\documentclass[journal,10pt,twocolumn,twoside]{IEEEtran}

\usepackage{mathtools}
\usepackage{empheq}
\usepackage{algpseudocode, algorithm}
\usepackage{algorithmicx}
\usepackage{subfig}
\usepackage[table]{xcolor}%
\usepackage{varwidth}
\usepackage{url}
\usepackage{multirow}
\usepackage{amssymb}
\usepackage{mathtools}
\usepackage{changes}
\usepackage[short]{optidef}
\usepackage{caption}
\usepackage{amsmath}
\usepackage{textcomp}
\usepackage{gensymb}
\usepackage{cite}
\usepackage{booktabs}
\usepackage{hyperref}
\usepackage{balance}
\usepackage{booktabs}
\usepackage{array}
\usepackage{balance}
\newcolumntype{P}[1]{>{\centering\arraybackslash}p{#1}}

\begin{document}
	
	\title{Enabling AI in Future Wireless Networks: \\ A Data Life Cycle Perspective}
	
	\author{Dinh C. Nguyen, Peng Cheng, Ming Ding, David Lopez-Perez, Pubudu N. Pathirana, Jun Li,  Aruna Seneviratne, Yonghui Li,~\IEEEmembership{Fellow,~IEEE,} and H. Vincent Poor,~\IEEEmembership{Fellow,~IEEE}

		\thanks{Dinh C. Nguyen and Pubudu N. Pathirana are with the School of Engineering, Deakin University, Australia (e-mails: \{cdnguyen, pubudu.pathirana\}@deakin.edu.au).}
		\thanks{Peng Cheng is with the Department of Computer Science and Information Technology, La Trobe University, Melbourne, VIC 3086, Australia, and also with the School of Electrical and Information Engineering, the University of Sydney, Sydney, NSW 2006, Australia (e-mails: p.cheng@latrobe.edu.au; peng.cheng@sydney.edu.au).}
		\thanks{Ming Ding is with the Data61, CSIRO, Australia (e-mail: ming.ding@data61.csiro.au). }	
		\thanks{David Lopez-Perez is with Huawei France R\&D, Boulogne-Billancourt, France (e-mail: dr.david.lopez@ieee.org).}
		\thanks{Jun Li is with the School of Electronic and Optical Engineering, Nanjing University of Science and Technology, Nanjing, 210094, China. He is also with the School of Computer Science and Robotics, National Research Tomsk Polytechnic University, Tomsk, 634050, Russia (e-mail: jun.li@njust.edu.cn).}
		\thanks{Aruna Seneviratne is with the School of Electrical Engineering and Telecommunications, University of New South Wales (UNSW), NSW, Australia (e-mail: a.seneviratne@unsw.edu.au).}
		\thanks{Yonghui Li is with the School of Electrical and Information Engineering, the University of Sydney, Australia, (e-mail: yonghui.lig@sydney.edu.au).}
		\thanks{H. Vincent Poor is with the Department of Electrical Engineering, Princeton University, Princeton, NJ 08544 USA (e-mail: poor@princeton.edu).}
		
		\thanks{This work was supported in part by the CSIRO Data61, Australia, and in part by U.S. National Science Foundation under Grant CCF-1908308. The work of P. Cheng was supported by ARC under Grant DE190100162. The work of Jun Li was supported in part by National Key R\&D Program under Grants 2018YFB1004800 and in part by National Natural Science Foundation of China under Grants 61727802, 61872184. \textit{(Corresponding Authors: Dinh C. Nguyen and Peng Cheng)}.}
		
	}
	
	\markboth{Accepted at IEEE Communications Surveys \& Tutorials}%
	{}

	\maketitle
	
	\begin{abstract}
		Recent years have seen rapid deployment of mobile computing and Internet of Things (IoT) networks, which can be mostly attributed to the increasing communication and sensing capabilities of wireless systems. Big data analysis, pervasive computing, and eventually artificial intelligence (AI) are envisaged to be deployed on top of the IoT and create a new world featured by data-driven AI. In this context, a novel paradigm of merging AI and wireless communications, called \textit{Wireless AI} that pushes AI frontiers to the network edge, is widely regarded as a key enabler for future intelligent network evolution. To this end, we present a comprehensive survey of the latest studies in wireless AI from the data-driven perspective. Specifically, we first propose a novel Wireless AI architecture that covers five key data-driven AI themes in wireless networks, including {Sensing AI, Network Device AI, Access AI, User Device AI and Data-provenance AI}. Then, for each data-driven AI theme, we present an overview on the use of AI approaches to solve the emerging data-related problems and show how AI can empower wireless network functionalities. Particularly, compared to the other related survey papers, we provide an in-depth discussion on the Wireless AI applications in various data-driven domains wherein AI proves extremely useful for wireless network design and optimization. Finally, research challenges and future visions are also discussed to spur further research in this promising area.
	\end{abstract}
	
	
	\IEEEpeerreviewmaketitle

\section{Introduction}
\label{Sec:Introduction}
Digitization and automation have been widely recognized as the next wave of technological revolution, which is envisaged to create a connected world and fundamentally transform industry, business, public services, and our daily life. In this context, recent years have seen rapid advancements in the deployment of Internet of Things (IoT) networks, which can be mostly attributed to the increasing communication and sensing capabilities combined with the falling prices of IoT devices \cite{1}. According to the Cisco forecast, by 2030 there will be more than 500 billion IoT devices connected to the Internet \cite{3}. Moreover, annual global data traffic is predicted to increase threefold over the next 5 years, reaching 4.8 ZB per year by 2022. Fueled by soaring mobile data and increasing device communication, wireless communication systems are evolving toward next generation (5G) wireless networks, by incorporating massive networking, communication, and computing resources. Especially, future wireless networks should provide many innovative vertical services to satisfy the diverse service requirements, ranging from residence, work, to social communications. The key requirements include the support of up to 1000 times higher data volumes with ultra-low latency and massive device connectivity supporting 10-100x number of connected devices with ultra-high reliability of 99.999\% \cite{5}. Such stringent service requirements associated with the increasing complexity of future wireless communications make traditional methods to network design, deployment,  operation, and optimization no longer adequate. Past and present wireless communications, regulated by mathematical models, are mostly derived from extant conventional communication theories. Communication system design  significantly depends on initial network conditions and/or theoretical assumptions to characterize real environments. However, these techniques are unlikely to handle complex scenarios with many imperfections and network nonlinearities \cite{7}. The future wireless communication networks, where the quality of services (QoS) is far beyond the capabilities and applicability of current modeling and design approaches, will require robust and intelligent solutions to adapt to the high network dynamicity for different services in different scenarios \cite{8}.

\begin{figure*}
	\centering
	\includegraphics[width=0.95\linewidth]{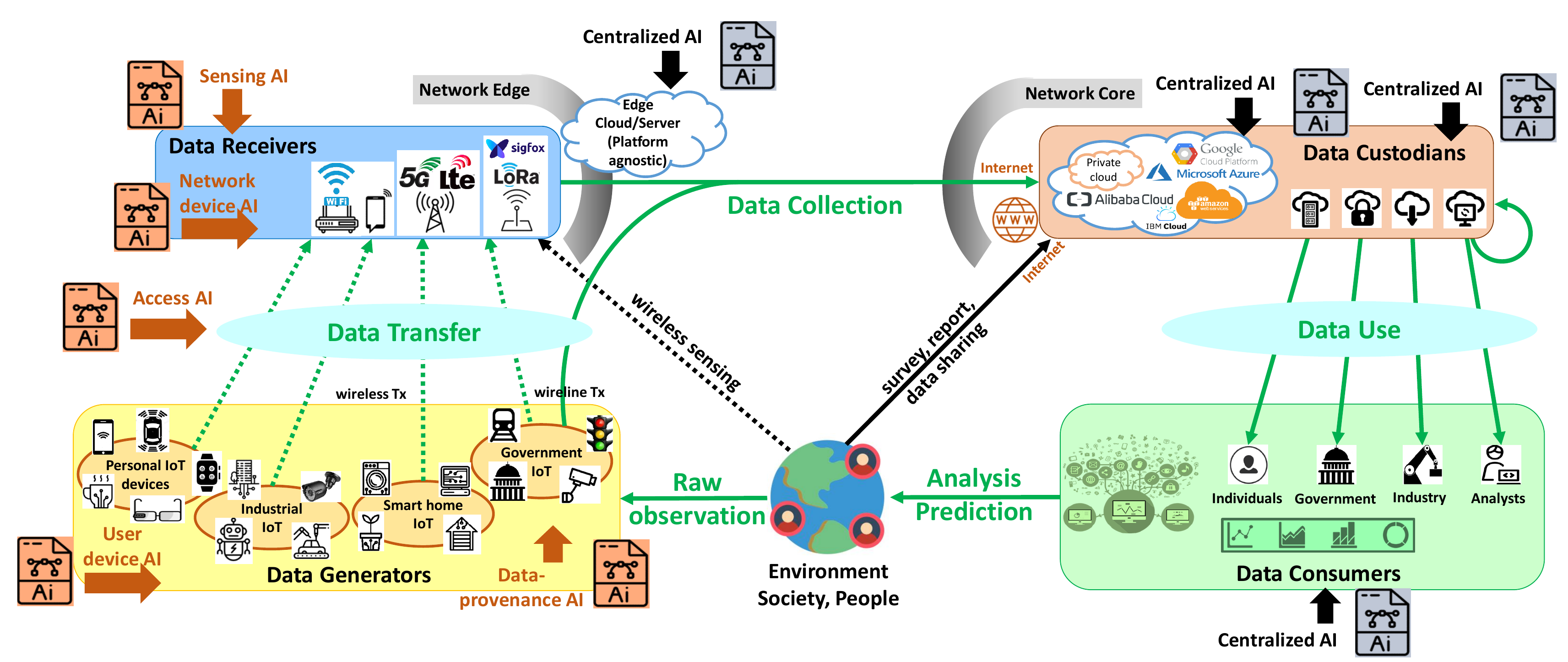}
	\caption{An illustration of the Wireless AI paradigm.  }
	\label{Fig:WirelessAI_Overview}
	\vspace{-0.17in}
\end{figure*}
Among the existing technologies, artificial intelligence (AI) is one of the most promising drivers for wireless communications to meet various technical challenges. {AI with machine learning (ML) techniques well known from computer science disciplines,} is beginning to emerge in wireless communications with recent successes in complicated decision making, wireless network management, resource optimization, and in-depth knowledge discovery for complex wireless networking environments \cite{9}. It has been proved that AI techniques can achieve outstanding performances for wireless communication applications thanks to their online learning and optimization capabilities in various complex and dynamic network settings. AI also lowers the complexity of algorithm computations enabled by data learning and interactions with the environment, and hence accelerates the convergence in finding optimal solutions compared to conventional techniques \cite{10}. By bringing the huge volumes of data to AI, the wireless network ecosystem will create many novel application scenarios and fuel the continuous booming of AI. Therefore, the integration of AI techniques into the design of intelligent network architectures becomes a promising trend for effectively addressing technical challenges in wireless communication systems. 

Indeed, the marriage of wireless communications and AI opens the door to an entirely new research area, namely \textit{Wireless AI} \cite{11}. Instead of heavily relying on data centres, i.e., cloud servers, to run AI applications, Wireless AI takes advantage of resources at the network edge to gain AI insights in wireless networks. More specifically, AI functions can be deployed over every corner of the wireless IoT networks, including in data generators (i.e., IoT devices), data receivers (e.g., mobile users, edge devices) and during on the wireless data transmissions (i.e., wireless links). This would make the wireless networks fully self-automating all operations, controls and maintenance functions with limited human involvement. Wireless AI thus solves effectively the heterogeneity and dynamicity of future wireless networks by leveraging its intelligent, self-adaptive, and auto-configurable capabilities. Notably, Wireless AI has attracted much attention from both the academia and industry. For example, according to Ericsson, wireless AI will have a significant role in reshaping next-generation wireless cellular networks, from intelligent service deployment to intelligent policy control, intelligent resource management, intelligent monitoring, and intelligent prediction. Major enterprises, such as Apple and Google, have been designing and developing many wireless AI-empowered applications for mobile users (e.g., Face ID and Siri apps as Apple products \cite{14} or Google Assistant tool of Google \cite{15}). These effort have boosted a wide spectrum of wireless AI applications, from data collection, data processing, massive IoT communication to smart mobile services. Therefore, Wireless AI is potentially a game changer not only for wireless communication networks, but also for emerging intelligent services, which will impact virtually every facet of our lives in the coming years. 

\subsection{Wireless AI Architecture}
In this paper, we propose a novel Wireless AI architecture for wireless networks as shown in Fig.~\ref{Fig:WirelessAI_Overview}. This new paradigm is positioned to create cognition, intelligence, automation, and agility for future wireless communication system design. As shown in Fig.~\ref{Fig:WirelessAI_Overview}, there are mainly four entities involved in this architecture. 

\begin{itemize}
	\item Firstly, our environment, society, and people are monitored and sensed by a variety of data generators, e.g. sensors and wearable devices to transform the observations and measurements of our physical world into digital via smart IoT devices.
	\item Secondly, such data will be transferred via wireless or wire-line links to wireless data receivers or directly to the Internet. Typical wireless data receivers include 4G/5G networks, Wi-Fi, LoRa, etc. 
	\item Thirdly, such data will be collected by data custodians via the Internet using cloud services, private cloud or enterprise cloud. A data custodian usually maintains a platform that organizes, stores and provides private and secure data access to data consumers. Note that a data custodian can also assume the role of a data consumer. 
	\item Fourthly, data consumers such as government, individuals, analysts, etc., will access the data so that they can conduct big data mining and harvest useful information to analyze and predict our physical world. {Besides, wireless service operators such as base station controllers can also exploit such data to optimize their network for seamless service delivery across data utilities in wireless networks. }
\end{itemize}

\indent It is important to note that in Fig.~\ref{Fig:WirelessAI_Overview} there are two shortcuts originated from the physical world as follows,
\begin{itemize}
	\item The first shortcut goes from the physical world to the data receivers, skipping the data generators. This framework is also known as wireless sensing because the data receivers directly see the physical world using radio frequency spectrum. Typical applications include people counting, human activity detection and recognition, etc.
	\item The second shortcut goes from the physical world to the data custodians, skipping both the data generators and the data receivers. This framework allows the data custodians to directly collect data using survey, report, and cross-organization data sharing, etc. Typical applications include national population censuses, tax return lodging, authorized medical record sharing, etc.
\end{itemize}

\indent Conventionally, AI functions sit in the cloud, or in powerful data center owned by data custodians, or in data consumers. In some cases, AI functions may even exist in an edge cloud, close to data receivers, performing edge computing tasks. Generally speaking, those AI functions are mainly designed for big data analysis and most of them follow a traditional paradigm of centralized AI, e.g., bringing data to computation functions for processing. Recently, many concerns begin to rise regarding this centralized AI paradigm, such as user privacy, power consumption, latency, cost, availability of wireless links, etc. {Among them is the user data privacy bottleneck caused by the centralized network architecture where IoT devices mainly rely on a third party, e.g., a cloud service provider. Although this model can provide convenient computing and processing services, the third party can obtain illegally personal information, leading to serious information leakages and network security issues.} For example, due to the Facebook data privacy scandal in Year 2018, privacy has become an obstacle for many applications involving personal data, such as living habits, healthcare data, travel routes, etc. Hence, enabling useful data collection/analysis without revealing individuals sensitive information has become an urgent task for the research community. {Moreover, the traditional centralized AI paradigm shows critical limitations in optimizing wireless networks \cite{19,20}. In the 5G era with ubiquitous mobile devices and multi-channel communications, the complexity of wireless networks is extremely high. The solution of using only a centralized AI controller to manage the whole network is obviously inefficient to optimize network operations and ensure high-quality service delivery.}

In this context, pushing AI to the network edge, i.e., bringing computation functions to data, close to data generators and receivers, has been recognized as a promising solution to protect privacy, reduce latency, save power and cost, and increase reliability by less-communication. In this paper, we identify four categories of wireless AI at the network edge and discuss their technologies, recent developments and future trends. More specifically, the four categories of wireless AI are shown in Fig.~\ref{Fig:WirelessAI_Overview} and briefly explained as follows.

\begin{itemize}

	\item \textit{Sensing AI at the Data Receivers:} AI supports well sensing tasks such as people and object detection over wireless networks. By analyzing Wi-Fi CSI characteristics, AI-based people counting and sensing systems can be designed to train the CSI values and generate the model so that people detection can be implemented. For example, convolutional neural networks (CNNs) are particularly useful in learning human images and recognizing humans within a crowd at different scales \cite{236, 242}. Besides, AI can be used for human motion and activity recognition using wireless signal properties \cite{274}. In fact, the relation between the CSI value and the dynamicity of human motion has been exploited to build the learning model using AI algorithms \cite{278}. 
	
	\item \textit{{User and Network Device AI} at the Data Generators and Receivers:} At the data generators such as sensors and mobile devices, AI functions have been embedded into end devices to create a new paradigm: on-device AI. This would open up new opportunities for simplifying AI implementations in wireless networks by eliminating the need of external servers, e.g., clouds. For example, deep neural network (DNN) inference now can be run on mobile devices such as Android phones for local client applications such as image classification \cite{199} and activity recognition \cite{209}. At the data receivers like edge servers, AI has been applied in the network edge to learn and detect signals, by embedding an AI-based intelligent algorithm on the base stations or access points. Specially, content catching at the network edge is now much easier thanks to the prediction capability of AI. Moreover, AI also has the potential in edge computing management such as edge resource scheduling \cite{187} and orchestration management \cite{190}. 
	
	\item \textit{Access AI in the Data Transfer Process:} AI can be applied to facilitate the data transmission from three main layers, including physical (PHY) layer, media access control (MAC) layer, and network layer. For example, in PHY layer, AI has been applied to facilitate CSI acquisition and estimation \cite{67} that are vitally important for improving data transfer process in terms of channel allocation, traffic congestion prediction. Further, AI also supports channel allocation, power control and interference mitigation functions in MAC layer. In fact, the learning capability of AI is particularly useful in estimating traffic channel distribution to realize flexible channel allocation \cite{93} and channel power management \cite{100}, aiming at minimizing channel interference in wireless networks. Specially, AI is able to empower spectrum sharing, data offloading, multi-connectivity, and resource management services from the network layer perspective. For example, AI has proved its effectiveness in spectrum resource management by online adaptive learning to achieve a reliable and efficient spectrum sharing in wireless networks \cite{404}.
	
	\item \textit{Data-provenance AI in the Data Generation Process:} In the data generation process, AI can provide various useful services, such as AI data compression, data privacy, and data security. As an example, AI has been applied for supporting sensor data aggregation and fusion, by classifying useful information from the collected data samples \cite{279}. Also, the classification of AI helps monitor the data aggregation process, aiming to analyze the characteristics of all data and detect data attacks \cite{281}. In addition to that, AI has been an ideal option to recognize legitimate users from attackers by training the channel data and performing classification \cite{503}. Meanwhile, DRL has proved extremely useful in building attack detection mechanisms \cite{509}, \cite{510} that can learn from the user behaviours and estimate the abnormal events for appropriate security actions.

\end{itemize}
\subsection{Comparison and Our Contributions}
\begin{table*}
	\centering
	\caption{\textcolor{black}{Existing surveys on AI and wireless networks.} }
	\label{Table:Comparisons}
	{\color{black}
	\setlength{\tabcolsep}{5pt}
	\begin{tabular}{|P{1.1cm}|p{2.3cm}|p{6.5cm}|p{6.6cm}|}
		\hline
		\centering \textbf{Related Surveys}& 
		\centering \textbf{Topic}&	
		 \textbf{Key contributions}&
		\textbf{Limitations}
		\\
		\hline
		\cite{16}    &	Edge computing for AI &	A survey on the edge intelligence enabled by  edge computing and AI integration.	&The applications of AI in wireless networks have not been presented. 
		\\
		\hline
		\cite{17}&	Edge computing for ML&	A review on the convergence of edge computing and ML.&	The applications of ML in wireless networks have not been presented.
		\\
		\hline
		\cite{18} &	AI for heterogeneous networks &	A survey on the use of AI in heterogeneous networks. &
		The potential of AI in data-driven wireless networks has not been discussed. 
		\\
		\hline
		\cite{19} & DL for edge computing &	A review on the convergence of edge computing and DL. &	The applications of DL in data-driven wireless networks have not been presented. 
		\\
		\hline
		\cite{20} &	DL for mobile networks &	A survey on the integration of DL and mobile networks in some domains such as mobile data analysis, user localization, sensor networks, and security. &	The roles of DL in data-driven wireless applications, such as intelligent network sensing, data access, have not been discussed. 
		\\
		\hline
		\cite{21}& 	DL for IoT data analytics&	A survey on the applications of DL in IoT data analytics and fog/cloud computing. &	The use of DL for wireless sensing, network access, and data-provenance services has not been explored. 
		\\
		\hline
		\cite{22}& 	DL for network traffic control&	A survey on the use of DL for network traffic control. &	The discussion of the DL applications for data-driven wireless networks has not been provided. 
		\\
		\hline
		\cite{23}&	DL for wireless networks &	A survey on the applications of DL for the design of wireless networks. &	The advantages of DL for data-driven wireless applications have not been discussed. 
		\\
		\hline
		\cite{24}&	DRL for communication networks &	A survey on the integration of DRL  and communication networks in some domains, i.e., network access control, content catching, and network security. &	The use of DRL in wireless networks for data-driven applications has been overlooked. 
		\\
		\hline
		\cite{25}&	ML for IoT wireless communications &	A survey on the use of ML in IoT wireless applications such as interference management, spectrum sensing. &	The benefits of ML for solving data-driven wireless issues such as data sensing, data access,  and user device intelligence have not been explored. 
		\\
		\hline
		\cite{26}& 
		ML for wireless networks&	A survey on the use of ML in  some domains in wireless networks, i.e., resource management, networking, mobility management, and localization. &	The potential of ML for data-driven wireless applications has not been discovered. 
		
		\\
		\hline
		\cite{27}&	ML for wireless networks &	A survey on the roles of AI in in wireless networks from the physical layer and the application layer. &	The potential of AI for data-driven wireless applications has not been discovered.
		\\
		\hline
		\cite{28}&	ML for optical networks &	A survey on the use of ML in optical networks. &	The potential of ML for data-driven wireless applications has not been discovered.
		\\
		\hline
		\cite{ref1} &	ML in data-driven 5G networks&	An introduction of the use of ML in online data-driven proactive 5G networks, including big data and edge-clouds. &	There is the lack of the discussion of AI applications in sensing networks and user device at the data generators and receivers. 
		\\
		\hline
		\cite{ref2}	&	AI for data-driven radio access networks&	A survey of AI applications in data-driven intelligent radio access networks, including transmission control and mobility management tasks. &	The analysis of data-driven AI-wireless applications at the data generators and data users has not been provided. 
		\\
		\hline
		\cite{ref3}	&	Data-driven explainable AI	&	A survey on the basic concepts of data-driven and knowledge-aware explainable AI. &	The roles of AI in wireless networks have not been analyzed. 
		\\
		\hline
		\cite{ref4}	&	Data-driven AI with security&	A survey of data-driven AI in security domains. &	There is the lack of the discussion on AI-based wireless network applications, such as network access, network devices and data provenance.
		\\
		\hline
		\textit{Our paper} &Data-driven AI for wireless networks &The applications of AI in data-driven wireless networks are comprehensively surveyed with five key data-driven AI domains, including Sensing AI, Network Device AI, Access AI, User Device AI and Data-provenance AI.   Particularly, we provide an in-depth discussion of Wireless AI applications with various network use cases.& -
		\\
		\hline
	\end{tabular}}
\end{table*}
Driven by the recent advances of AI in wireless networks, some efforts have been made to review related works and provide useful guidelines. Specifically, the work in \cite{16} presented a survey on the basics of AI and conducted a discussion on the use of edge computing for supporting AI, including edge computing for AI training, and edge computing for AI model inference. \textcolor{black}{The authors in \cite{17} analyzed edge ML architectures and presented how edge computing can support ML algorithms. The use of AI in heterogeneous networks was explored in \cite{18}. However, in these survey articles, the potential of AI in wireless networks from the data-driven perspective such as data processing in edge network devices, data access at edge devices, has not been elucidated.}   Meanwhile, Deep learning (DL) which has been an active research area in AI, is integrated in wireless systems for various purposes. For example, the authors in \cite{19} discussed the roles of DL in edge computing-based applications and then analyzes the benefits of edge computing in supporting DL deployment. The study in \cite{20} presented the discussion on the integration of DL and mobile networks, \textcolor{black}{but the roles of DL in data-driven wireless applications, such as intelligent network sensing, data access, have not been considered.} The authors in \cite{21}, \cite{22} studied the potentials of DL in enabling IoT data analytics (including surveys on the use of DL for IoT devices, and the performance of DL-based algorithms for fog/cloud computing) and network traffic control, while the authors in \cite{23} mainly concentrated on the quantitative analysis of DL applications for the design of wireless networks. Deep reinforcement learning (DRL), an emerging AI technique recent years, has been studied for wireless networks in \cite{24}. This survey focused on the discussion on the integration of DRL and communication networks, with a holistic review on the benefits of DRL in solving communication issues, including network access and control, content catching and offloading, and data collection. Moreover, the benefits and potentials of ML in wireless communications and networks were surveyed in recent works \cite{25,26,27,28} in various use cases, such as resource management, IoT data networks, mobility management and optical networking. \textcolor{black}{However, the use of ML for data-driven wireless applications in sensing networks and user device has not been analyzed.}

{\color{black}From the data-driven perspective, some AI-based works have been presented. For example, the survey in \cite{ref1} discusses the roles of ML in online data-driven proactive 5G networks, with the focus on physical social systems enabled by big data and edge-clouds. The authors in \cite{ref2} study the application of AI in data-driven intelligent radio access networks where an AI-based neural network proves its efficiency in transmission control and mobility management tasks. However, these works \cite{ref1}, \cite{ref2} lack the discussion of applications of AI in sensing networks and user devices at the data generators and receivers. Another study in \cite{ref3} introduces the basic concepts of data-driven and knowledge-aware explainable AI. However, the roles of AI in wireless networks are not analyzed. Moreover, a survey of data-driven AI has been provided in \cite{ref4}, but it is only limited to the security domain and does not cover wireless network applications, such as network access, network device and data provenance. The comparison of the related works and our paper is summarized in Table~\ref{Table:Comparisons}.

Despite the significant body of work of surveying wireless AI applications, some key issues remain to be considered:
\begin{itemize}
	\item Most current AI surveys focus on the use of AI in edge computing \cite{16, 17} or heterogeneous networks \cite{18}, while the potential of AI in data-driven wireless networks has not been explored.
	\item	Most existing works mainly focus on specific AI techniques (ML/DL or DRL) for wireless networks \cite{19,20,25,26}. Furthermore, the literature surveys on ML/DL only focus on some certain network scenarios in wireless communication from either data transmitters' \cite{19,20} or data receivers' \cite{26,27,28} perspectives.
	\item  Some recent data-driven AI surveys \cite{ref1, ref2, ref3, ref4} only provide a brief introduction to the use of AI in some certain wireless domains such as edge cloud computing \cite{ref1}, radio access networks \cite{ref2}, explainable AI \cite{ref3}, and security \cite{ref4}, while a comprehensive survey on the application of AI in wireless networks from the data life cycle perspective is still missing. 
	\item	 Moreover, the roles of wireless AI functions from the data-driven viewpoint, i.e., data sensing AI, data access AI, data device AI, and data-provenance AI, have not been explored well in the literature. 
\end{itemize}

Motivated by these limitations, we here present an extensive review on the applications of AI/ML techniques in wireless communication networks from the data-driven perspective. In particular, compared to the existing survey papers, we provide a more comprehensive survey on the key data-driven wireless AI domains in the data life cycle, from Sensing AI, Network Device AI, Access AI to User Device AI and Data-provenance AI. } To this end, the main contributions of this survey article are highlighted as follows:
\begin{enumerate}
	\item {We provide an overview of the state-of-the-art AI/ML techniques used in wireless communication networks.}
	\item We propose a novel Wireless AI architecture that covers five key data-driven AI themes in wireless networks, including {Sensing AI, Network Device AI, Access AI, User Device AI and Data-provenance AI}.
	\item  For each data-driven AI theme, we present an overview on the use of AI approaches to solve the emerging data-related problems and show how AI can empower wireless network functionalities. Particularly, we provide an in-depth discussion on the Wireless AI applications in various data-driven use cases. The key lessons learned from the survey of each domain are also summarized. 
	\item We provide an elaborative discussion on research challenges in wireless AI and point out some potential future directions related to AI applications in wireless networking problems.
\end{enumerate}

\subsection{	Organization of the Survey Paper}
The structure of this survey is organized as follows. Section~\ref{Sec:State-of-Art} presents an overview of state-of-the-art AI/ML techniques including ML, DL, and DRL. Based on the proposed wireless AI architecture outlined in the introduction section, we start to analyze the wireless AI applications in communication networks with five key domains: Sensing AI, Network Device AI, Access AI, User Device AI and Data-provenance AI. More specifically, we present a state-of-the-art review on the existing literature works in the Sensing AI area in Section~\ref{Sec:Sensing_AI}. We classify into three main specific domains, namely people and objection detection, localization, and motion and activity recognition. Section~\ref{Sec:Network_Device_AI} presents a state-of-the-art survey on the Network Device AI domain, highlighting the development of AI for supporting wireless services on edge servers (e.g., base stations). Next, Section~\ref{Sec:Access_AI} provides an extensive review for the Access AI domains, with a focus on the discussion on the roles of AI in PHY layer, MAC layer, and network layer.  We then analyze the recent developments of the User Device AI domain in Section~\ref{Sec:User_Device_AI}, by highlighting the development of AI for supporting wireless services on user devices. Meanwhile, Section~\ref{Sec:Data-provenance_AI} presents a discussion on the recent use of Data-provenance AI applications in the data generation process in various domains, namely AI data compression, data clustering, data privacy, data security. Lastly, we point out the potential research challenges and future research directions in Section~\ref{Sec:Challenges}. Finally, Section~\ref{Sec:Conclusions} concludes the paper. A list of key acronyms used throughout the paper is presented in Table~\ref{Table:Acronyms}.
\begin{table}
	\caption{List of key acronyms.}
	\label{Table:Acronyms}
	\scriptsize
	\centering
	\captionsetup{font=scriptsize}
	\setlength{\tabcolsep}{5pt}
	\begin{tabular}{p{1.5cm}|p{5.5cm}}
		\hline
		\textbf{Acronyms}& 
		\textbf{Definitions}
		\\
		\hline
		AI& Artificial Intelligence
		\\
		ML & Machine Learning
		\\
		DL & Deep Learning
		\\
		DRL & Deep Reinforcement Learning 
		\\
		SVM &Support vector machine
		\\
		SVR &  Support vector regression 
		\\
		k-NN & k-neural networks
		\\
		DNN &Deep Neural Network
		\\
		CNN & Convolutional Neural Network
		\\
		LSTM& Long-short Term Memory 
		\\
		RNN & Recurrent Neural Networks
		\\
		DQN & Deep Q-network 
		\\
		FL & Federated learning 
		\\
		PCA &Principal component analysis 
		\\
		MDp & Markov decision processes
		\\
		MAC & Media Access Control 
		\\
		IoT& Internet of Things
		\\
		MEC & Mobile edge computing 
		\\
		UAV &  Unmanned aerial vehicle
		\\
		MIMO & Multiple-input and multiple-output
		\\
		NOMA & Non-orthogonal multiple access
		\\
		OFDM & Orthogonal Frequency-Division Multiplexing
		\\
		CSI & Channel state information
		\\
		V2V & Vehicle-to-vehicle
		\\
		D2D& Device-to-Device
		\\
		M2M & Machine-to-Machine 
		\\
		MTD & Machine type device
		\\
		BS & base station
		\\
		MCS & Mobile Crowd Sensing
		\\
		QoS & Quality of Service
		\\
		SINR & Signal-to-interference-plus-noise ratio
		\\
		\hline
	\end{tabular}
	\label{tab1}
\end{table}

\section{AI Technology: State of the Art}
\label{Sec:State-of-Art}
\textcolor{black}{In this section, we review the state-of-the-art AI/ML technology used in wireless communication applications. Motivated by the AI tutorial in \cite{goodfellow2016deep}, we here classify AI into two domains, including conventional AI/ML techniques and advanced AI/ML techniques, as shown in Table~\ref{Table:AIclassify}}.

	\begin{table}[ht]
		\centering
		\caption{Classification of key AI techniques used in wireless networks. }
		\label{Table:AIclassify}		
		\setlength{\tabcolsep}{5pt}
		\begin{tabular}{|P{3.95cm}|P{4cm}|}
			\hline
			 \centering
			\textbf{Conventional AI/ML techniques}&	
			\textbf{Advanced AI/ML techniques}
			\\
			\hline
			\begin{itemize}
				\item Supervised learning
				\item Unsupervised learning 
				\item Semi-supervised learning
			 	\item	Reinforcement learning	
			\end{itemize}	&
			\begin{itemize}
				\item Convolutional neural networks (CNNs)
				\item Recurrent neural networks (RNNs)
				\item Long-short term memory (LSTM) networks
				\item Deep reinforcement learning (DRL)
			\end{itemize}
			\\
			\hline
		\end{tabular}
	\end{table}
\subsection{\textcolor{black}{Conventional AI/ML Techniques}}
\textcolor{black}{This sub-section surveys some of the most researched and applied conventional AI/ML algorithms for wireless communications,} including supervised learning, semi-supervised learning, unsupervised learning, and reinforcement learning. 

\subsubsection{Supervised Learning}
Supervised learning, as the name implies, aims to learn a mapping function representing the relation of the input and the output under the control of a supervisor with a labelled data set. In this way, the data model can be constructed so that each new data input can be learned by the trained model for making decisions or predictions \cite{34}. Supervised learning is a very broad ML domain. Here we pay attention to some techniques that have been employed in wireless AI systems, including support vector machine, K-nearest neighbors, and decision trees. 

- \textit{Support vector machine (SVM):} The SVM utilizes a part of dataset as support vectors, which can be considered as training samples distributed close to the decision surface. The main idea behind SVM is to use a linear classifier that maps an input dataset into a higher dimensional vector. The aim of its mapping concept is to separate between the different classes by maximizing their distance. {In SVMs, a nonlinear optimization problem is formulated to solve the convex objective function, aiming at determining the parameters of SVMs. It is also noting that SVMs leverage kernel functions, i.e. polynomial or Gaussian kernels, for feature mapping.  Such functions are able to separate data in the input space by measuring the similarity between two data points. In this way, the inner product of the input point can be mapped into a higher dimensional space so that data can be separated \cite{35}.} 

-	\textit{K-nearest neighbors (k-NN):}  Another supervised learning technique is K-NN that is a lazy learning algorithm which means the model is computed during classification. In practical applications, k-NN is used mainly for regression and classification given an unknown data distribution. It utilizes the local neighbourhood to classify a new sample by setting the parameter K as the number of nearest neighbours. Then, the distance is computed between the test point and each training point using the distance metrics such as Euclidean or Chebyshev \cite{36}. 

-	\textit{Decision trees:} Decision trees aim to build a training model that can predict the value of target variables through decision rules. Here, a tree-based architecture is used to represent the decision trees where each leaf node is a class model, while the training is set as the root. From the dataset samples, the learning process finds the outcome at every leaf and the best class can be portioned \cite{37}.  
\subsubsection{	Unsupervised Learning}
Unsupervised learning is a ML technique that learns a model to estimate the output without the need of labelled dataset. One of the most important unsupervised learning algorithms used in wireless networks is K-means that aims to find the clusters from the set of unlabelled data. The algorithm implementation is simple with two required parameters, including original data set and the target number of clusters. {Another popular unsupervised learning algorithm is Expectation-Maximization (EM) that is used to estimate the values of latent variables (which cannot be observed directly) under the known variable probability distribution. The importance of EM algorithms is shown via some use cases, such as estimation for parameters in Hidden Markov Models, finding the missing data in the data sample space, or support for clustering tasks \cite{38}.}
\subsubsection{Semi-supervised Learning} As the intermediate technique between supervised learning and unsupervised learning, semi-supervised learning uses both labelled and unlabelled samples. The key objective is to learn a better prediction rule using the unlabelled data than that based on the data labelling in supervised learning that is always time-consuming and easy to bias on the model. Semi-supervised learning is mainly used for classification tasks in computer visions and image processing \cite{goodfellow2016deep}. 
\begin{figure*}
	\centering
	\includegraphics[width=0.95\linewidth]{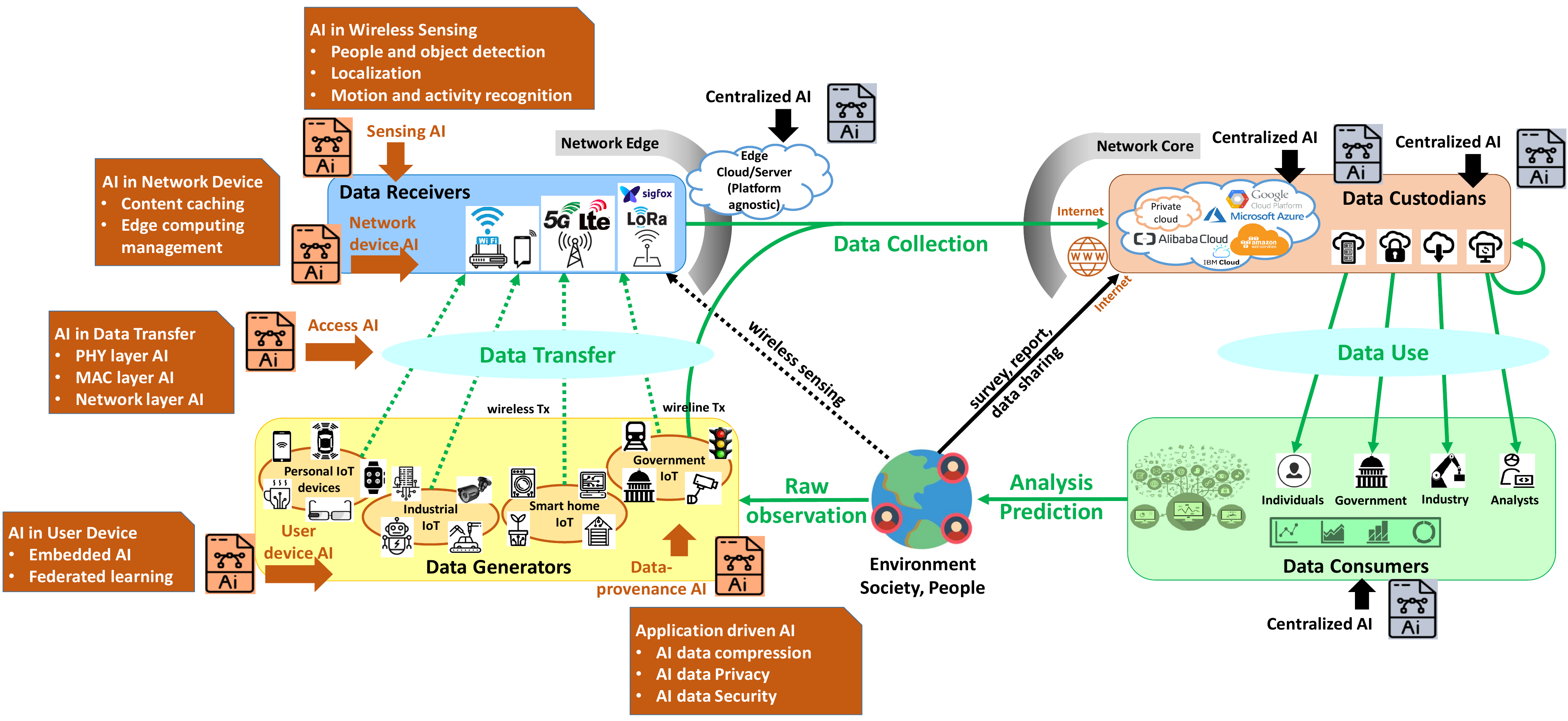}
	\caption{Future AI functions in wireless networks.  }
	\label{Fig:WirelessAI_Function}
	\vspace{-0.17in}
\end{figure*}
\subsubsection{Reinforcement Learning} In Reinforcement Learning (RL), there is no need to define the target outcomes. This method is based on the concept of environment interaction which means that an agent directly interacts with the surrounding environment, sense the states and select appropriate actions \cite{40}. As a result, the agent can get a reward after an action is taken. The aim of the agent is to maximize the long-term reward that corresponds an objective in wireless networks, such as system throughput optimization. The most popular RL technique is Q-learning that uses the Q-functions to adjust its policy to seek an optimal solution via trial and error without the need of prior environment knowledge. This concept can be applied to wireless scenarios where the system model is hard to find due to the lack of mathematical formulation, such as real time network optimization and mobile computation offloading \cite{55}. 
\subsection{\textcolor{black}{Advanced AI/ML Techniques}}
\textcolor{black}{In this sub-section, we summarize some key advanced AI/ML techniques used in wireless networks, including CNNs, RNNs, LSTM networks, and DRL.}

\subsubsection{	Convolutional Neural Networks}
CNN is a type of neural network used mainly for image processing and recognition with large pixel datasets \cite{44}.  Basically, the CNN architecture is similar to a DNN, but adding two new convolutional and pooling layers in between hidden layers. The convolution layer is regarded as the core component of a CNN, responsible to process the input samples based on a convolution operation to extract the features during the sample compression thanks to a set of filters \cite{45}. Another block of a CNN is the pooling layers that aim to make the spatial size of the representation smaller for reducing the computing overheads and mitigating overfitting. In this fashion, the output of previous layer can be combined with a perceptron of the next layer. Lastly, the extracted features are further processed in the final layer until reaching the output layer. Deep CNNs have been proved with many successes in wireless communication applications, such as massive MIMO \cite{46}, modulation classification in wireless networks \cite{47}. 
\subsubsection{Recurrent Neural Networks}
RNNs can be regarded as a deep generative architecture since they can be considered as several non-linear neuron layers between the input layer and the output layer in an unfolded fashion \cite{48}. The length of an RNN is highly correlated with the length of the data sequence and thus RNNs are well suitable for modelling text data or audio time series. This is based on the RNN training procedure where the training process is performed from the interactions between multilayer perceptrons in a feedback loop. That means the hidden layer is chained with the others via a loop to create a chained training structure. This operation is based on the memory capability of each neuron to store information of computation from the previous data input. Some RNN-based wireless applications can be fault detection in wireless sensor networks \cite{49} and wireless channel modelling \cite{50}. 
\subsubsection{	Long Short Term Memory Networks} LSTM networks are extended versions of RNNs. Unlike the conventional RNN, LSTM operates based on the gate concept \cite{51}. It keep information of the recurrent network in the neuron or gated cell that has different types of gates such as read, write and erasure gate. Different from digital logic gate in computers, these gates are analog and formulated using element-wise multiplication by sigmoids ranging from 0 to 1. Analog gates are differentiable to be well suitable for backpropagation tasks. These gates are operated similar to the neurons of neural networks. The cells perform data learning via an iterative process to adjust the weights via gradient descent. LSTM networks have been applied in several wireless scenarios such as orthogonal frequency division multiplexing (OFDM) \cite{52} and IoT data analysis \cite{53}.
\subsubsection{Deep Reinforcement Learning}
Reinforcement Learning is a strong AI technique, which allows a learning agent to adjust its policy to seek an optimal solution via trial and error without the need of prior environment knowledge. However, when the dimension of state and action space explodes, the RL-based solutions can also become inefficient. DRL methods such as deep Q-network (DQN) have been introduced and demonstrated their superiority for solving complex problems \cite{24}. Specially, instead of storing the variables in the table as the RL technique, DRL leverages a deep neural network as the approximator to store high-space actions and states. DRL is mainly used for model optimization, classification, and estimation where a decision making problem is formulated by using an agent and interactive environment. At present, DRL in wireless is a very active research area where some popular DRL techniques such as deep Q-learning, double DRL, and duelling DRL have been used widely for specific applications such as data offloading and resource allocation \cite{56}.

\subsection{{Visions of Wireless AI}}
The aforementioned AI techniques would have the great potential in supporting wireless networks from two perspectives: signal processing and data processing. Here, signal processing aims to increase the data throughput, while data processing focuses on producing data utility for specific applications. In fact, AI functions are particularly useful for signal processing tasks through signal learning and estimation. In general, signal processing takes advantages of signal parameters to recognize and classify types of signal in involved applications. Some AI functions for signal processing have been recently explored, such as channel interference signal detection, frequency and bandwidth estimation, and modulation recognition \cite{19, 20}. AI is able to facilitate  feature selection, an important domain in signal processing, for extracting instantaneous time features, statistical features or transform features. In practice, feature selection from the received signals is particularly challenging due to the lack of prior information of signal parameters and the uncertainty of signal time series. Feature engineering could be left out by using AI architectures to recognize effectively important signal features such as amplitude, phase, and frequency that are important for many signal-related domains, i.e. spectrum sharing \cite{405}, channel coding \cite{701} or edge signal detection \cite{300}. 

Moreover, AI functions are also particularly useful for data processing. In the era of IoT, data exhibits unique characteristics, such as large-scale streaming, heterogeneity, time and space correlation. How to obtain hidden information and knowledge from big IoT data is a challenge. AI can come as a promising tool for recognizing and extracting meaningful patterns from enormous raw input data, aiming to create a core utility for big data processing in wireless applications such as object detection, big data analysis or security. For example, big data collected from ubiquitous sensors can be learned using DL for human/object recognition \cite{274}. In such scenarios, AI is able to extract useful features which represent most the target object, and then classified using DL architectures, e.g., CNNs or DNNs. The classification capability of AI is also beneficial to data processing tasks for network security enhancements. Recent works \cite{503,504} show that a DL network is able to detect intrusions and threats in IoT datasets so that modified or unauthorized data can be recognized and eleminated for better security of data-driven applications.

The migration of computation power from the core to the edge has already begun. Wireless AI will further lead such an exodus beyond the edge, illuminating a future with pervasive and end-to-end AI. In the next sections, we conduct a detailed survey on AI functions in wireless networks following a migration trail of computation functions from the edge to the source data. Such AI functions include Sensing AI, Network Device AI, Access AI, User Device AI and Data-provenance AI as shown in Fig.~\ref{Fig:WirelessAI_Function}.

\section{Sensing AI}
\label{Sec:Sensing_AI}
\begin{figure}
	\centering
	\includegraphics [width=0.95\linewidth]{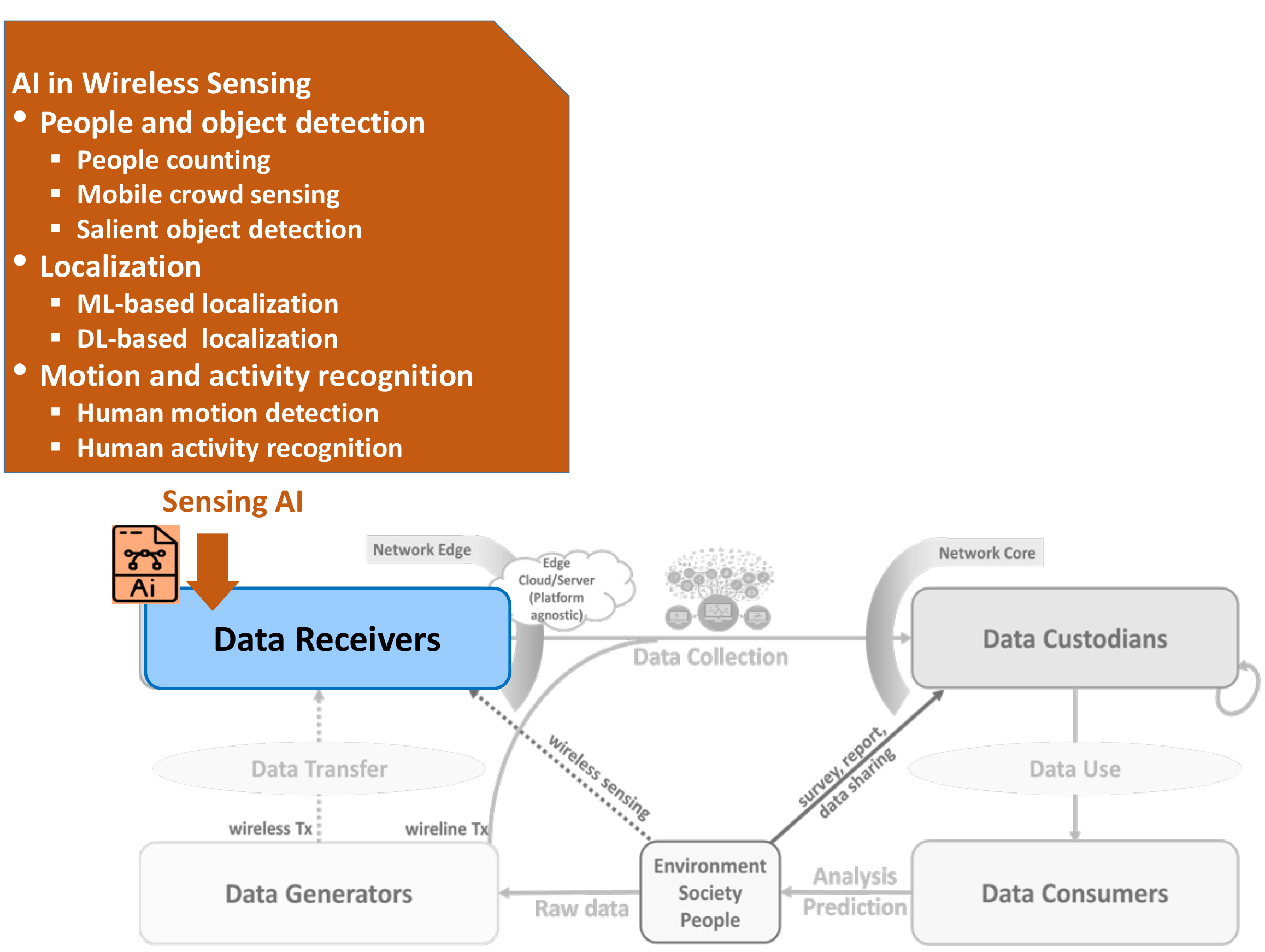}
	\caption{Sensing AI function. }
	\label{Fig:Sensing_AI}
	\vspace{-0.1in}
\end{figure}

{As shown in the Wireless AI architecture in Fig.~\ref{Fig:WirelessAI_Function}, the first Wireless AI function of the data life cycle is Sensing AI that is responsible to sense and collect data from the physical environments using sensors and wearable devices, followed by smart data analytics using AI techniques. In this section, we present a state-of-the-art review on the existing literature studies working around the Sensing AI function for wireless networks as shown in Fig.~\ref{Fig:Sensing_AI}. Here, Sensing AI focuses on three main domains, namely people and objection detection, localization, and motion and activity recognition.}
\subsection{	People and Object Detection}
\subsubsection{	People Detection}
Here we focus on discussing the use of AI in two people detection domains, including people counting and mobile crowd sensing. 

\textit{1.1) People counting:}

\textbf{Problem formulation: }{People counting provides essential information for various pervasive applications and services, e.g., crowd control for big events, smart guiding in events, indoor monitoring. It is important to support crowd analytics for human crowd size estimation and public space design.  
	
	\textbf{Drawbacks of conventional methods: }Most of the existing work uses computer vision, infrared sensors and devices attached on people such as RFID and smart phones. However, infrared sensors deployed at the doorway count the entering or exiting people based on the assumption that people are moving through one by one under certain time interval. Vision-based approaches count people by image object detection, while its performance can be weakened by object overlapping, poor lighting conditions and dead zones \cite{222}. 
	
	\textbf{Unique advantages of using wireless AI techniques: }The use of AI would enable smart people counting with light-weight and better accurate human information analytics \cite{223}. 
	
	\textbf{Recent works: } We here summarize some representative liturature studies working on AI/ML techniques for people counting. 
	\begin{itemize}
		\item Linear regression and SVR methods: In order to improve the accuracy of the people counting, ML methods such as linear regression and SVR have been proposed for estimating the number of people using the existing Wi-Fi access points in indoor environments \cite{224}. The experiment result indicates that SVR yields the better accuracy in estimating people numbers. SVR is also applied for pedestrian counting from captured video images. To improve the efficiency of feature extraction, a Histogram of Oriented Gradient technique is designed that can capture more meaningful information from the raw images for better human movement recognition. \textcolor{black}{The proposed model is able to generate a 98.2\% motion detection accuracy with less computational complexity.} SVR is also used to learn the estimation function modelled from the extracted features for people counting tasks \cite{226}. Radar sensor BumbleBee is employed to detect the human movement images, and features extracted from the images are classified into three domains, i.e., time, frequency, and joint time-frequency domains. Here, an experiment is set up with 60 sessions of data collected in room settings with 43 people hired to do a trial. K-fold cross validation is used to evaluate the feasibility of the proposed scheme, showing a high correlation and low error with baseline methods.
		\item Crowdsourcing methods: Another AI-based approach for people counting is proposed in \cite{227} that implements a people counting model called Wi-Fi-counter using smartphone and Wi-Fi. The Wi-Fi-based counter scheme employs a crowdsourcing technique to gather the information of people and Wi-Fi signals. However, the useful information between of them that is necessary to estimate people count may be hidden. Then a five-layer neural network architecture is built, including two main phases: the offline phase for computing the eigenvalues from Wi-Fi signals at the access point and the online phase for collecting Wi-Fi data that is then used for training the model. The performance of Wi-Fi-counter model is mainly evaluated via counting accuracy and system power usage overheads. Meanwhile, the people counting project in \cite{228} investigates the efficiency of Random Forest technique. The number of people at train stations is estimated from images captured from CCTV videos. Instead of using features extracted randomly that is time-consuming, each captured image is divided into sub-windows for fast human recognition during the learning. 
		\item {Advanced AI/ML methods:} In general, ML has been applied and achieved certain successes in people counting over wireless networks. However, in fact the performance in terms of accuracy, efficient large-scale training and the ability to solve complex counting tasks with variety of data samples still need to be improved. DL can be an answer for facilitating future people counting applications. For example, the authors in \cite{230} work in developing a new crowd counting model, call Wicount, using Wi-Fi devices. The dynamicity of human motions makes CSI values changeable over time. Based on this concept, the CSI characteristics are captured, including amplitude and phase information. A critical challenge is to find a mathematical function which represents the relationship between CSI values and people count due to the varying number of people. To solve this issue, a DL solution is proposed to train the CSI values and generate the model so that people counting can be implemented. A cross-scene crowd counting framework is presented in \cite{231}. The key focus is on mapping the images captured by camera sensors and people counts. To solve the issue in people counting in the case of surveillance crowd scenes unseen in the training, a CNN architecture is proposed to train the samples iteratively and a data-driven technique is used to fine-tune the trained CNN model. 
	\end{itemize}
	
	\textit{1.2) 	Mobile crowd sensing:}
	
	\textbf{Problem formulation: }Mobile Crowd Sensing (MCS) is a large-scale sensing paradigm empowered by the sensing capability of mobile IoT devices, such as smartphones, wearable devices. MCS enables mobile devices to sense and collect data from sensors and share with cloud/edge servers via communication protocols for further processing. The high mobility of mobile devices makes MCS flexible to provide different ubiquitous services, such as environment monitoring, human crowd monitoring, and traffic planning \cite{235}. 
	
	\textbf{Drawbacks of conventional methods: } Existing approaches like dynamic programming have tried to assume MCS as a static model where the motion of mobile devices are assumed to follow in a predefined trajectory in the static environement \cite{235}. In fact, MCS tasks should be always considered in time-varying environment and the uncertainty of mobile devices, which make current MCS solutions inefficient. 
	
	\textbf{Unique advantages of using wireless AI techniques: }AI helps to realize the potential of MCS and overcome challenges in crowd sensing applications in terms of high MCS system dynamicity, sensing network complexity and mobile data diversity. 
	
	\textbf{Recent works: } Several AI approaches have been used to faciliate MCS applications.
	\begin{itemize}
		\item MCS using ML approaches: For example, the work in \cite{236} considers a robust crowd sensing framework with mobile devices. The key objective is to solve the uncertainty of sensing data from the mobile user participants and optimize the sensing revenue under a constrained system budget. In practical scenarios, a sensing task owner often selects randomly the member who can gather the most information among all participants for maximizing the profits. However, it may be infeasible in the context where data quality is uncertain which makes crowdsensing challenging. To solve this uncertainty issue, an online learning approach is applied to obtain the information from sensing data thanks to a selection step without knowing the prior statistics of the system. \textcolor{black}{The unique benefit of this ML-based model lies in its low sensing cost, e.g., low training computation, but this scheme has not been tested in practical sensor crowdsensing networks.}
		The authors in \cite{237} suggest to use ML for feature selection from similar metrics in mobile sensing data (i.e., Wi-Fi signal strength and acceleration data). To detect human movement patterns, two techniques: individual following and group leadership are proposed that use an underlying ML algorithm for identifying features that can classify human groups. Meanwhile, unsupervised ML is investigated in \cite{238} to estimate the number of people from audio data on mobile devices. An application called Crowd++ is implemented on Android phones to collect over 1200 minutes of audio from 120 participants. To estimate the number of target object, namely active speakers in the crowd, an ML process is taken with three steps: speech detection, feature extraction and counting. 
		\item {Advanced ML methods:} Recently, some advanced ML techniques such as DL have been investigated to facilitate MCS ecosystems \cite{241}. How to achieve high-quality data collection in terms of optimizing collected data with lowest energy consumption is important for MCS applications in smart cities where mobile terminals equipped with sensors to gather ubiquitous data. Therefore, a DRL approach is proposed to create a movement and sensing policy for mobile terminals so that the efficiency of data collection is maximized. Here, each mobile terminal acts as an agent to interact with the MCS environment to find the best trajectory for data sensing and collection. Another DL approach is introduced in \cite{242} where a joint data validation and edge computing scheme is proposed for a robust MCS. Stemming from the fact that the current MCS models suffer from several limitations, lack of incentive mechanism for attracting more members for data sensing, lack of validation of collected data, and network congestion due to high transmit data volumes. Thus, DL based on a CNN network for pattern recognition is adopted to detect unimportant information and extract useful features for lightweight data sensing. \textcolor{black}{However, this CNN-based model is vulnerable to data loss and malware bottlenecks due to the lack of learning security mechanisms.}
	\end{itemize}
	
	\subsubsection{	Object Detection} In this subsection, we study the roles of AI in object detection.  
	
	\textbf{Problem formulation: } As one of the key computer vision areas, object detection plays an important role in realizing the knowledge and understanding of images and videos, and is used in various applications, such as image/video classification, object analytics or object motion tracking. 
	
	\textbf{Drawbacks of conventional methods: } Traditional object detection approaches mostly use a three-step model, with the adoption of informative region selection, feature selection, and classification using mathematical models \cite{zhao2019object}. For example, a multi-scale sliding window is often used to scan the image to detect target objects, but it is computationally expensive and time-costing. Further, due to the high similarity of visual features, current solutions may be unlikely to represent exactly the detection models and mis-classify objects. 
	
	\textbf{Unique advantages of using wireless AI techniques: }Recent years, many research have made attempts to use AI for object detection thanks to its higher prediction accuracy but low computational complexity.
	
	\textbf{Recent works: } Object detection (i.e., salient object detection) employs deep neural networks to recognize the distinctive regions in an image. The work in \cite{245} presents a recurrent fully convolutional network for better accurate saliency detection. This is enabled by an iterative correction process during learning, which refines better saliency maps and brings a better performance in terms of higher prediction accuracy. \textcolor{black}{This object detection solution can help achieve accurate learning inference, but requires complex multi-layer data training.} The authors in \cite{246} focus on designing a cascaded partial decoder mechanism which helps to remove shallower features and refine the features of deeper layers in the CNN network for obtaining precise saliency maps. Based on that, the proposed scheme can achieve higher prediction accuracy but lower algorithm running time. To reduce the training complexity and improve the evaluation speed in traditional neural network in salient object detection, a simplified CNN network is introduced in \cite{247}. Instead of minimizing the boundary length as the existing Mumford-Shah approaches have implemented, the focus of this work is to minimize the intersection over union among the pixels of the images. This non-reliance on super-pixels makes the method fully convolutional and thus enhances the estimation rate.  
	
	\subsection{Localization}
	\begin{figure*}
		\centering
		\includegraphics [width=0.95\linewidth]{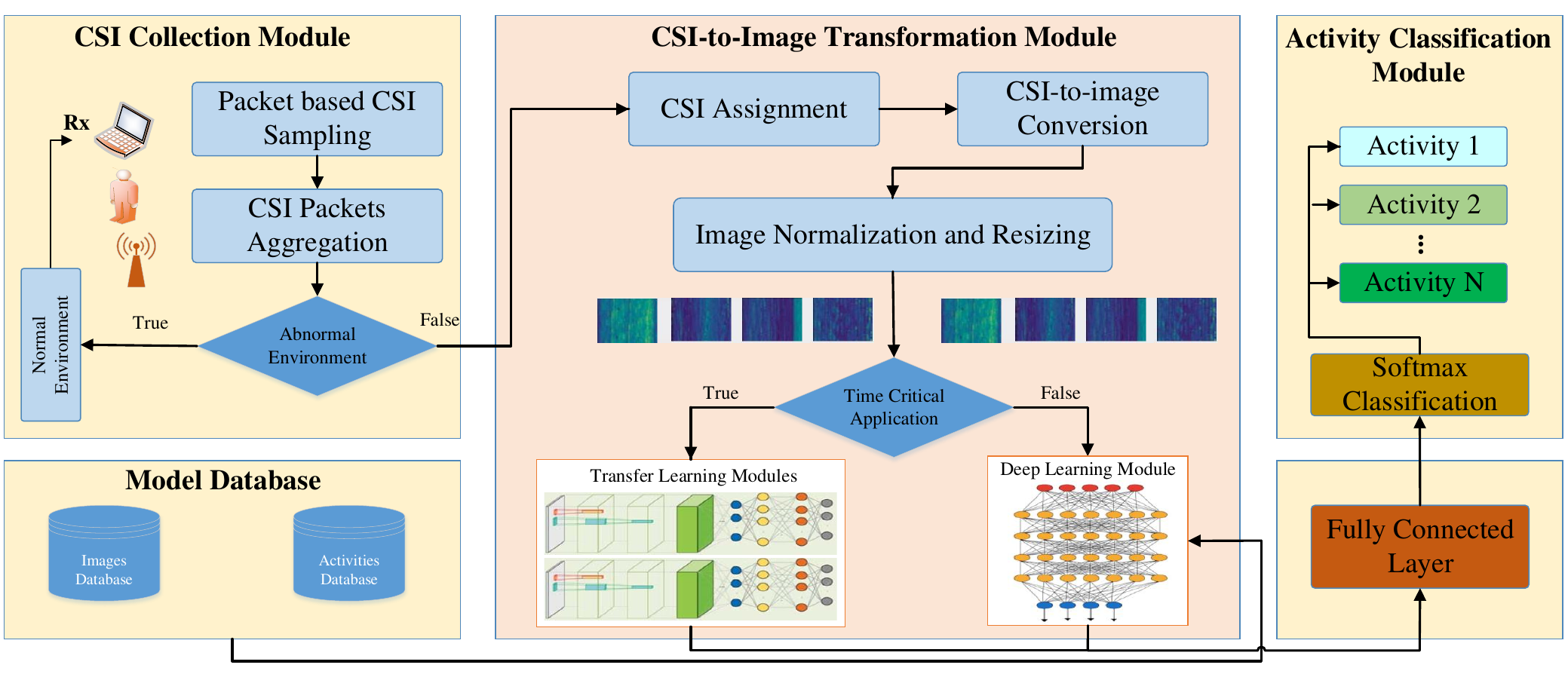}
		\caption{DL for CSI-based human activity classification.  }
		\label{Fig:DRL_ActivityClassification}
		\vspace{-0.17in}
	\end{figure*}
	
	\textbf{Problem formulation: }Localization aims to determine the geographic coordinates of network nodes and objects \cite{254}. It is important to various applications, such as object tracking, human recognition or network resource optimization. 
	
	\textbf{Drawbacks of conventional methods: }In many scenarios, it is infeasible to use global positioning system (GPS) hardware to specify where a node is located, especially in indoor environments. Distance of nodes can be calculated using mathematical techniques such as RSSI, TOA, and TDOA \cite{255}. However, the location of nodes may be changeable over time due to movements. 
	
	\textbf{Unique advantages of using wireless AI techniques: }AI comes as a promising solution to effectively estimate locations of objects/nodes due to some benefits. First, AI can approximate the relative locations of nodes to absolute locations without the need of physical measurements. Second, in the complex dynamic network, AI can help estimate accurate localization without mathematical formulation which reduces the complexity of localization problem. 
	
	\textbf{Recent works: } Some important works are highlighed with the exploitation of AI for localization tasks, {by using AI/ML techniques.}
	\begin{itemize}
		\item {Conventional AI/ML-based localization:} The work in \cite{256} presents a sensor node localization scheme using ML classifiers. Data is collected with time-stamp from four capacitive sensors with a realistic room setting. A variety of ML classifiers such as Bayes network, random forest, and SVM are used to evaluate the performance of indoor localization. \textcolor{black}{The simulation results confirm that random forest performs best overall in terms of low training time, lower computational complexity and high accuracy (93\%). However, this ML scheme is only verified in a simple and small sensor network.}  The authors in \cite{257} propose a non-linear semi supervised noise minimization algorithm for sensor node localization via iterative learning. The collected labelled and unlabelled data are expressed as a weighted graph. The localization of wireless sensor node can be considered as semi supervised learning. Another work in \cite{258} implements a neural network on Android phones for indoor localization. The system is a two-layer model, where the first layer is a neural network coupled with random forest, while the second one is a pedometer for accelerating the data sampling process. \textcolor{black}{Its drawback is the high risks of electronic eavesdropping for data collection in mobile networks.}
		
		\item {Advanced AI/ML-based localization:}  The work in \cite{263} explores a solution for device-free wireless localization in wireless networks. The discriminative features are first extracted from wireless signals and learned by a three-layer deep neural network. Using the extracted features from the DL model, a ML-based softmax regression algorithm is then employed to estimate the node location in indoor lab and apartment settings. Meanwhile, a deep extreme learning machine technique is presented in \cite{264} for fingerprint localization based on signal strength of wireless sensor networks. Instead of using random weight generation, neural network-based autoencoder is used for sample training in the encoding and decoding manner. This allows to extract better features and then localize the node better. Recently, DRL has been applied to facilitate localization tasks \cite{268}. To formulate the localization problem as a MDP, which is the key structure of DRL approaches, a continuous wireless localization process is derived. The localization task is divided into time slots where the agent inputs the location information at the previous time slot and outputs the new location in the next slot. Therefore, the computation of the location at each step only depends on the location at the previous step, which makes it appropriate to formulate a MDP problem. Then, a deep Q-learning algorithm is built to solve the proposed problem, aiming to minimize the error of localization estimations. 
	\end{itemize}
	
	\subsection{Motion and Activity Recognition}
	\textbf{Problem formulation: }In social networks, human motion data may be very large and activity parterns are changable over time. How to deal with a large-scale datasets for motion and activity recognition tasks is a challenge. 
	
	\textbf{Drawbacks of conventional methods: } Many traditional techniques have assumed to have prior knowledge of the CSI or environment information for motion recognition \cite{269}, \cite{270}, but it is hard to achieve in practice. Moreover, the samples collected from human motion activities are often very large associated with the complexity of multiple motion patterns, which make current approaches such as dynamic programming inefficient to model exactly the motion estimation task so that the problem can be solved effectively.

	\textbf{Unique advantages of using wireless AI techniques: }
	Thanks to high classification accuracy and efficient online learning based on scalable big datasets, AI has been employed for human motion and activity recognition based on wireless signal properties for better accuracy of activity recognition. 
	
	\textbf{Recent works: }In \cite{269}, the authors study the CSI retrieved from the host computer to estimate and classify the four body motions: normal gait, abnormal gait (ataxia) \cite{kashyap2020automated}. The main idea is to analyze the relation between the CSI value and the change of human motion. Then, a system setting with an access point and network interface for CSI detection is considered so that gait abnormality and hand tremors can be estimated using an ML algorithm. The authors in \cite{271} develop a motion detection system using Wi-Fi CSI data. The system relies on the variance of CSI to detect the human motion that is empowered by a moving filter in supervised ML algorithm. Compared to traditional motion recognition approaches using wearable sensors, radars that require hardware settings and management efforts, motion detection based on Wi-Fi signals is much more flexible, cheaper due to using the available infrastructure. \textcolor{black}{However, the proposed scheme is sensitive to CSI variance that can adversely affect the motion detection performance.} 
	
	Moreover, a multiple access point-based CSI aggregation architecture is introduced in \cite{274}, aiming to estimate human motion. A large-scale Wi-Fi CSI datasets from the data collection process are fed to a multi-layer CNN model for training and analyzing the relations between CSI strengths and levels of human motion. In comparison to SVM-based training, CNNs provide better accuracy in activity recognition. As an investigation on the influence of human motions on Wi-Fi signals, the work in \cite{275} suggest a DL network for image processing. From the measures on multiple channels, CSI values are transformed to a radio image, which is then extracted to generate features via a DNN learning process and trained for motion classification with ML. A DL model for CSI-based activity recognition can be seen in Fig.~\ref{Fig:DRL_ActivityClassification}. As an further effort for improving the feature selection from CSI datasets in human activity recognition, an enhanced learning model is suggested in \cite{276}. In fact, the data collected from Wi-Fi signals contains redundancy information that may not be related to human motion, and thus needs to be remoted for better estimation. Motivated by this, a background reduction module is designed to filter unrelated motion information from the original CSI data. Then, correlation features are extracted from the filtered CSI data. To this end, a DL-based RNN is employed to extract the deeper features via offline training. Specially, LSTM is integrated with an RNN to achieve better recognition accuracy with lower computational complexity. Moreover, a Wi-Fi CSI collection framework based on IoT devices is introduced in \cite{278} for recognizing human activity. To learn effectively the features extracted from CSI datasets, a joint DL model of autoencoder, CNN and LSTM is designed. Here, autoencoder aims to eliminate the noise from CSI data, the CNN extracts the important features from the output of autoencoder, while LSTM extracts inherent dependencies in the features.  \textcolor{black}{Although this framework can achieve high detection accuracy (97.6\%), it suffers from slow learning rates and incurs high computational costs. }
	
	\subsection{Lessons Learned} The main lessons acquired from the survey on the Sensing AI domain are highlighted as the following.
	\begin{itemize}
		\item 	AI has been applied and achieved certain successes in people and object detection over wireless networks. By analyzing Wi-Fi CSI characteristics, AI-based people counting and sensing systems can be built to train the CSI values and generate the model so that people detection can be implemented. We have also observed that CNNs are particularly useful in learning human images and recognizing humans among the crowd at different scales. We also find that most studies leverage deep CNNs as classifiers for better accurate object recognition thanks to their fully convolutional networks \cite{247}. 
		\item	Moreover, AI helps estimate effectively locations of objects/nodes due to some benefits. First, AI can approximate the relative locations of nodes to absolute locations without the need of physical measurements. Second, in the complex dynamic network, AI can help to divide into sub-class using classifiers, which reduces the complexity of localization problem for better node estimation. Many research efforts have been made to apply both ML and DL in supporting localization applications in mobile IoT networks \cite{264},  \cite{268}.
		\item	 Lastly, the use of AI for human motion and activity recognition using wireless signal properties has been realized. The relation between the CSI value and the change of human motion has been exploited to build the learning model using AI algorithms. Especially, to learn effectively the features extracted from CSI datasets for human detection tasks, a joint DL model consisting of an autoencoder, CNN and LSTM is designed \cite{278}. Here, the autoencoder aims to eliminate the noise from CSI data, the CNN extracts the important features from the output of autoencoder, while the LSTM network extracts inherent dependencies in the features. This triple AI combination is promising to improve recognition accuracy and reduce the training latency, which is important for wireless AI deployments in realistic scenarios. 
		\item \textcolor{black}{In general, in the Sensing AI domain, SVR demonstrates high efficiency in learning tasks, such as people counting \cite{224}, with high  learning accuracy and easy implementation, compared to other AI schemes. CNNs have also emerged as a very promising AI technique in data sensing models, such as human activity recognition \cite{278} for low training latency, improved learning efficiency, and enhanced data estimation rate, while other AI techniques such as DNNs \cite{275} remain high learning latency. Other AI techniques such as SVM \cite{256} and neural networks \cite{258} also prove their potential in supporting sensing tasks in wireless networks. However, some important issues including high energy consumption for learning, learning security and complex learning structures need to be solved for the full realization of sensing AI.} 
	\end{itemize}
	\textcolor{black}{In summary, we show the classification of the Sensing AI domain in Fig.~\ref{Fig:01_ Taxonomy_Sensing} and summarize the AI techniques used along with their pros and cons in the taxonomy Table~\ref{Table:Sensing_AI}. }

\begin{figure}
	\centering
	\includegraphics [width=0.95\linewidth]{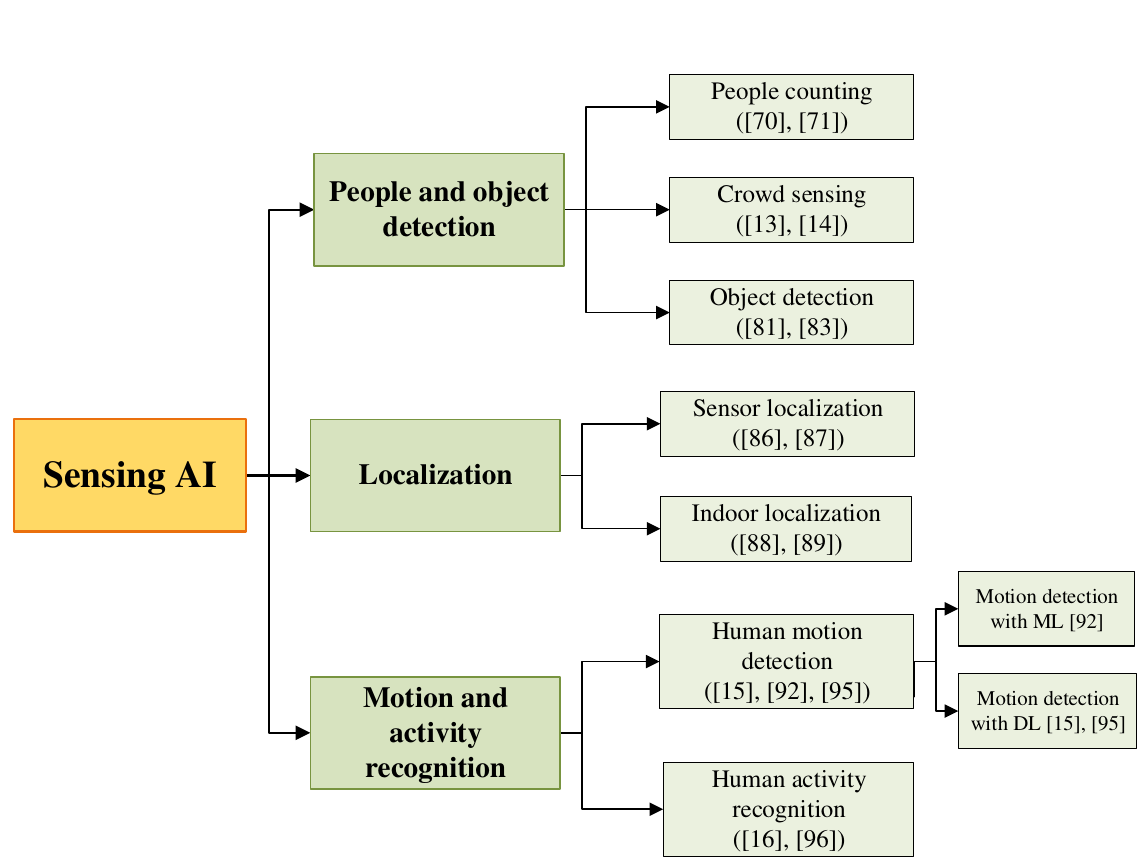}
	\caption{\textcolor{black}{Summary of Sensing AI domain.}}
	\label{Fig:01_ Taxonomy_Sensing}
	\vspace{-0.1in}
\end{figure}

\begin{table*}
	\centering
	\caption{{\color{black}Taxonomy of Sensing AI use cases.} }
	\label{Table:Sensing_AI}
	{\color{black}
	\begin{tabular}{|P{0.4cm}|P{0.3cm}|P{1.8cm}|P{1.3cm}|P{1.2cm}|P{1cm}|P{1cm}|P{1cm}|P{1.3cm}|P{2.2cm}|P{2cm}|}
		\hline
		\textbf{Issue}& 	
		\textbf{Ref.} &	
		\textbf{Use case}&	
		\textbf{AI Algorithm}& 
		\textbf{Complexity}& 	
		\textbf{Training Time }&
		\textbf{Training Data}&
		\textbf{Accuracy}&
		\textbf{Network}&
		\textbf{Pros}&
		\textbf{Cos}
		\\
		\hline
		\parbox[t]{1.5cm}{\multirow{11}{*}{\rotatebox[origin=c]{90}{People and object detection}}} & 
		\cite{224}&	People counting	&SVR&	Low &	Low &	Low &	High (98.2\%)&	Wi-Fi networks&Easy implementation&	Less operational flexibility  
		\\ \cline{2-11}&
		\cite{226}&	People counting	&SVR&	Low &	Low &	Low &	Fair	&Radar networks&	Low estimation error&	Learning bias with small data
		\\ \cline{2-11}&
		\cite{236}&	Crowd sensing&	Online ML&	Fair &	Low &	Low &	Fair&	Sensor networks	&Low sensing cost &Lack of simulating practical datasets
		\\ \cline{2-11}&
		\cite{242}&	Crowd sensing&	CNN	&Fair&	High&	High&	High &	Sensing networks&	Robust learning	&Unsuitable for real-time crowd sensing 
		\\ \cline{2-11}&
		\cite{245}&	Object detection&	RNN&	High&	Fair&	Fair	&High&	Wireless networks&	Accurate learning inference&	Complex learning implementation 
		\\ \cline{2-11}&
		\cite{247}&	Object detection&	CNN&	Fair&	Low&	High &	Fair&	Wireless networks&	Less computation time&	Lack of realistic detection tests
		\\ \cline{2-11}
		\hline
		\parbox[t]{1.5cm}{\multirow{11}{*}{\rotatebox[origin=c]{90}{Localization}}}& 
		\cite{256}	&Sensor localization&	SVM&	Low	&Low&	Low	&High (93\%)&	Sensor networks&	Low localization error; require low computational resources	&Lack of large-scale deployment 
		\\ \cline{2-11}&
		\cite{257}&	Sensor localization&	Semi supervised ML&	Fair&	Low &	Low&	High&	Sensor networks&	Reduce noise; high accuracy&Risks of learning eavesdropping
		\\ \cline{2-11}&
		\cite{258}&	Indoor localization&	NN&	Low	&Low&	Low&	High
		(95\%)	&Indoor networks &	Easy implementation; low localization error&	Data loss risks for online training
		\\ \cline{2-11}&
		\cite{263}&	Wireless localization&	DNN&	Fair&	Fair &	Fair&	Fair 
		(85\%)	&Wireless networks	&Energy efficiency &	Small-scale settings
		\\ \cline{2-11}
		\hline
		\parbox[t]{1.5cm}{\multirow{13}{*}{\rotatebox[origin=c]{90}{Motion/activity recognition}}} & 
		\cite{269}&	Human motion classification&	Online ML& Low&	Fair&	Low	&High (93\%)&	Sensor networks	&Implement in clinical settings; low cost&	Need specific hardware
		\\ \cline{2-11}&
		\cite{271}	&Human motion detection&	LSTM, Naive Bayes&	Low	&Low&	Low&	Fair&	Wi-Fi networks&	Comprehensive learning tests&	Sensitive to CSI variance
		\\ \cline{2-11}&
		\cite{274}&	Human motion detection&	CNN&	Fair &	High&	High&	High&	Wi-Fi networks&	Learn well complex data&	Sensitive to CSI variance
		\\ \cline{2-11}&
		\cite{275}	&Human activity recognition&	DNN&	High&	High&	High&	Fair&	Wi-Fi networks&	Characterize well human actions via Wi-Fi signals	&Training privacy concerns
		\\ \cline{2-11}&
		\cite{278}	&Human activity recognition&	CNN&	Fair&	Low&	Fair&	High 
		(97.6\%)&	IoT networks&	Improved learning efficiency; real experiments&	High energy cost with on-device learning
		\\ \cline{2-11}
		\hline
	\end{tabular}}
\end{table*}
\section{Network Device AI}
\label{Sec:Network_Device_AI}
{The next Wireless AI function of the data life cycle in Fig.~\ref{Fig:WirelessAI_Function} is Network Device AI where AI has been applied in the network edge to learn and detect signals, by embedding an AI-based intelligent algorithm on the base stations or access points. We here focus on two main applications of the Network Device AI function, including content caching and edge computing management as shown in Fig.~\ref{Fig:NetworkDeviceAI}.}
\begin{figure}
	\centering
	\includegraphics [width=0.95\linewidth]{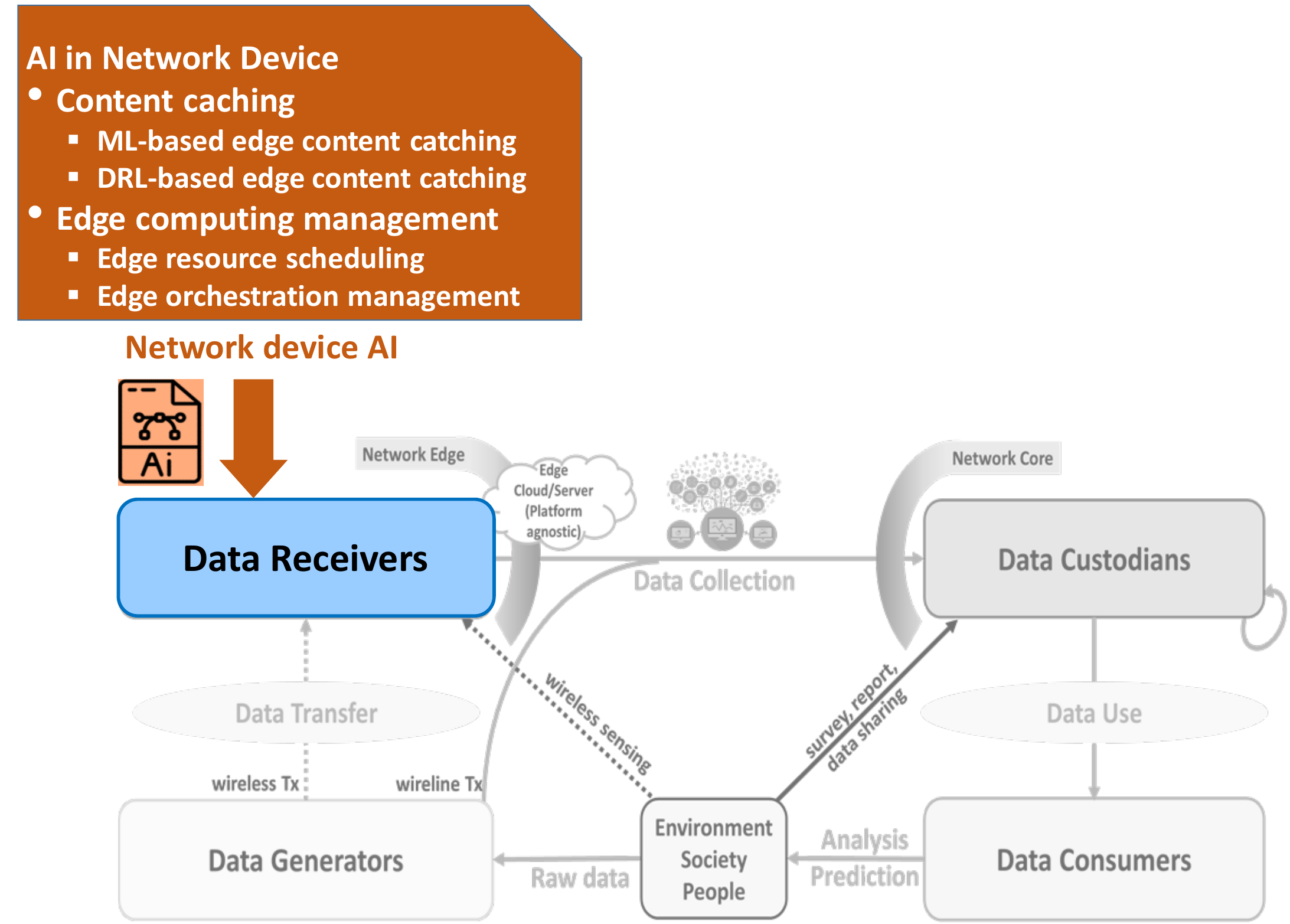}
	\caption{Network Device AI function. }
	\label{Fig:NetworkDeviceAI}
	\vspace{-0.1in}
\end{figure}
\subsection{	Content Caching}
\textbf{Problem formulation: }Caching contents at base stations/access points at the edge of the network has emerged as a promising approach to reduce traffic and enhance the quality of experience (QoE) \cite{173}. 

\textbf{Drawbacks of conventional methods: } Most existing studies assume that the distribution of content caching popularity is known, but in practice, it is complex and  non-stationary because of high content dymanics \cite{173}. 

\textbf{Unique advantages of using wireless AI techniques: }Consider the complexity and dynamicity of contents, recent studies have investigated the integration of AI at BSs to facilitate content caching policies. 

\textbf{Recent works: } Many solutions have been proposed to support content caching, using ML and DL approaches.
\begin{itemize}
	\item {Conventional AI/ML-based edge content caching:} For example, the authors in \cite{174} consider optimization of caching efficiency at small cell base stations. They focus on minimizing the backhaul load, formulate it as an optimization problem and leverage a K-means clustering algorithm to identify spatiotemporal patterns from the content requests at the BS. As a result, the similar content preferences are grouped into different classes so that the BS can cache the most important content with high accuracy and low caching latency. \textcolor{black}{The experiments indicate that the proposed caching method with K-means can reduce data noise by 70\% with high learning reliability and low training latency, but the learning parameters need to be optimized further to enhance the caching performance.}
	RL has recently been applied to edge-based content caching architectures. With an assumption that content popularity and user preference are not available at MEC servers, a multi-agent RL is an appropriate option for learning and training the caching decisions \cite{177}. Each MEC server as an agent learns on how to coordinate its individual caching actions to maximize the optimal caching reward (i.e., the downloading latency metric). 
	\item {Advanced AI/ML-based edge content caching:} Unlike aforementioned solutions, the work in \cite{180} suggests a DRL approach for content caching at the edge nodes, e.g., at the base stations, to solve the policy control problem. Indeed, content delivery is always complex with a huge number of contents and different cache sizes in realistic scenarios, which pose critical challenges on estimating content popularity distribution and determining contents for storage in caches. DRL which is well known in solving complex control problems is suitable for this caching decisions at the BS. User requests to the BS are considered as the system state and caching decisions (cache or not) are action variables in the RL formulation, while cache hit rate is chosen as the reward to represent the goal of the proposed scheme. \textcolor{black}{Through the simulation, DRL has proven its efficiency with better cache hit rate in various system settings (e.g., edge cache capability) and achieved low runtime and action space savings. A drawback of this scheme is to only consider a simple edge caching architecture with a single base station, which thus needs to be extended to multiple base stations-based edge networks in 5G scenarios.}
	For content caching in fog computing-empowered IoT networks, an actor–critic deep RL architecture is provided in \cite{181} and deployed at the edge router with the objective of reducing caching latency. Instead of deriving a mathematical formulation which is hard to obtain in practice, the authors propose a model-free learning scheme, where the agent continuously interacts with the fog caching environment to sense the states (i.e., numbers of contents and user requests). Since the edge router prefers to cache those contents with better popularity, the content transmission delay is formulated as a reward function. Thus, the content with higher popularity is more likely to have lower time cost, which motivates the agent to take the action of caching for long-term reward. \textcolor{black}{ However, the proposed DRL algorithm has a  high computational cost due to its multi-layer learning structure with the complex state-action handling model that makes it challenging to integrate into ultra low-latency 5G content caching systems. }
	
\end{itemize}

\subsection{Edge Computing Management}
\textbf{Problem formulation: }With the assistance of mobile edge computing, mobile devices can rely on edge services for their computation and processing tasks. Due to the constrained resource of edge servers and the unprecedented mobile data traffic, resource scheduling and orchestration management are required to achieve robust edge management and ensure long-term service delivery for ubiquitous mobile users. 

\textbf{Drawbacks of conventional methods: }Most of current strategies rely on mixed integer nonlinear programming or dynamic programming to find the edge computing management policies, but it is unlikely to apply to complex wireless edge networks with unpredictable data traffic and user demands [16], [17]. Further, when the dimension of the network increases due to the increase of mobile users, the traditional techniques are unable to solve the increasingly computational issue and then hard to scale well to meet QoS of all users. 

\textbf{Unique advantages of using wireless AI techniques: }AI would be a natural option to achieve a scalable and low-complex edge computing management thanks to its DL and large-scale data optimization. In this sub-section, we investigate the use of AI for edge computing management with some representative recent works in two use case domains: edge resource scheduling and orchestration management.

\textbf{Recent works: } {Many studies have leveraged AI/ML techniques} to support edge computing management via two important services: edge resource scheduling and orchestration management. 
\begin{itemize}
	\item Edge resource scheduling: AI has been applied widely to facilitate edge resource scheduling.
	The authors in \cite{185} present a DL-based resource scheduling scheme for hybrid MEC networks, including base stations, vehicles and UAVs connecting with edge cloud. A DNN network is proposed with a scheduling layer aimed to perform edge computing resource allocation and user association. \textcolor{black}{A unique advantage of the proposed model is its low computational complexity inherited by online learning, but it potentially raises high security risks, e.g., malware, due to the UAV-ground communication.} 
	To provide an adaptive policy for resource allocation at the MEC nodes in mobile edge networks with multi-users, a deep Q-learning scheme is introduced in \cite{186}. In the multi-user MEC network, it is necessary to have a proper policy for the MEC server so that it can adjust part of edge resource and allocate fairly to all users to avoid interference and reduce latency of users for queuing to be served. Therefore, a smart resource allocation scheme using deep Q-learning is developed and deployed directly on the MEC server so that it can allocate effectively resource for offloaded tasks of network users under different data task arrival rates. Moreover, a multi-edge resource scheduling scheme is provided in \cite{187}. An edge computing use case for a protest crowd incident management model is considered where ubiquitous data including images and videos are collected and offloaded to the nearby MEC server for execution. Here, how to allocate edge resource to the offloaded tasks and adjust edge computation with varying user demands to avoid network congestion is a critical challenge. Motivated by this, an edge computing prediction solution using ML is proposed, from the real data (transmission and delay records) in wireless network experiments. The ML algorithm can estimate network costs (i.e., user data offloading cost) so that efficient edge resource scheduling can be achieved. To further improve the efficiency of online edge computing scheduling, another solution in \cite{188} suggests a DRL-based schem. The authors focus on the mobility-aware service migration and energy control problem in mobile edge computing. The model includes three main entities: a network hypervisor for aggregating the states of input information such as server workloads, bandwidth consumption, spectrum allocation; a RL-based controller for producing the control actions based on the input information, and action executor for obtaining the control decisions from the RL controller to execute corresponding actions. \textcolor{black}{As a case study, the authors investigate the proposed DRL scheme with Deep Q-learning for an edge service migration example, showing that DRL can achieve superior performance, compared to a greedy scheme with lower running cost and adaptability to user mobility. However, this model is developed only for a specific DRL architecture and is thus hard to apply to other similar DL systems.}
	
	\item Edge orchestration management: In addition to edge resource scheduling, the potential of AI has been investigated in edge orchestration management. As an example, the work in \cite{190} considers an orchestration model of networking, caching and computing resources with MEC. The authors concentrate on the resource allocation problem that is formulated as an optimization problem with respect to the gains of networking, caching and computing. The triple model is highly complex that is hard to be solved by traditional optimization algorithms. Therefore, a deep Q-learning approach is proposed to achieve the high convergence performance with the best system utility. Meanwhile, in \cite{191}, a joint scheme of virtual edge orchestration with virtualized network functions and data flow scheduling is considered. Motivated by the fact that real-world networks may be hard to be modelled due to the complex and dynamic nature of wireless networks, a model-free DRL algorithm using deep deterministic policy gradients is proposed, aiming to minimize system delay and operation costs. \textcolor{black}{A remaining issue with this scheme is the relatively high running time due to the complex learning architecture and unoptimized training parameters. }
	Specially, an edge-Service Function Chain (SFC) orchestration scheme based on blockchain \cite{nguyen2019integration} is introduced in \cite{192} for hybrid cloud-edge resource scheduling. Blockchain is adopted to create secure transmission between users and edge service providers, while the SFC orchestration model is formulated as a time-slotted chain to adapt with the dynamicity of IoTs. Based on that, a joint problem of orchestration control and service migration is derived and solved by a DRL approach using deep Q-learning. The numerical simulation results show that the learning-based method can achieve high performance with low latency of orchestration algorithm, and system cost savings.
\end{itemize}

\subsection{Lessons Learned} The deployment of AI functions at edge servers (e.g., base stations) instead of remote clouds has emerged as a new direction for flexile network designs with low latency and privacy protection. AI has been applied in the network edge learn and detect signals, by embedding an AI-based intelligent algorithm on the base stations or access points. Specially, content caching at the network edge is now much easier thanks to the prediction capability of AI. Due to the huge number of contents and different cache sizes in realistic network scenarios, how to accurately estimate content popularity to achieve a high caching rate at the edge is highly challenging. AI comes as a natural choice to provide smart caching policies for implementing content caching with respect to latency constraints and user preference \cite{174}. Moreover, the potential of AI in edge computing management has been also investigated in mainly two use-case domains: edge resource scheduling \cite{185} and orchestration management \cite{190}.

\textcolor{black}{ In general, in the Network Device AI domain, K-means learning \cite{174} can provide simple and efficient solutions for device intelligence. In particular, DRL techniques such as Deep Q-learning \cite{186}, \cite{190} exhibit superior performance in learning accuracy and convergence reliability with limited computing resource use due to recent optimization advances, including optimized learning parameters and efficient learning-layer structure organization \cite{190}.}
	
\textcolor{black}{In summary, we show the classification of the Network Device domain in Fig.~\ref{Fig:02TaxonomyNetworkDevice} and summarize the AI techniques used along with their pros and cons in the taxonomy Table~\ref{Table:NetworkDeviceAI}.}

\begin{figure}
	\centering
	\includegraphics [width=0.95\linewidth]{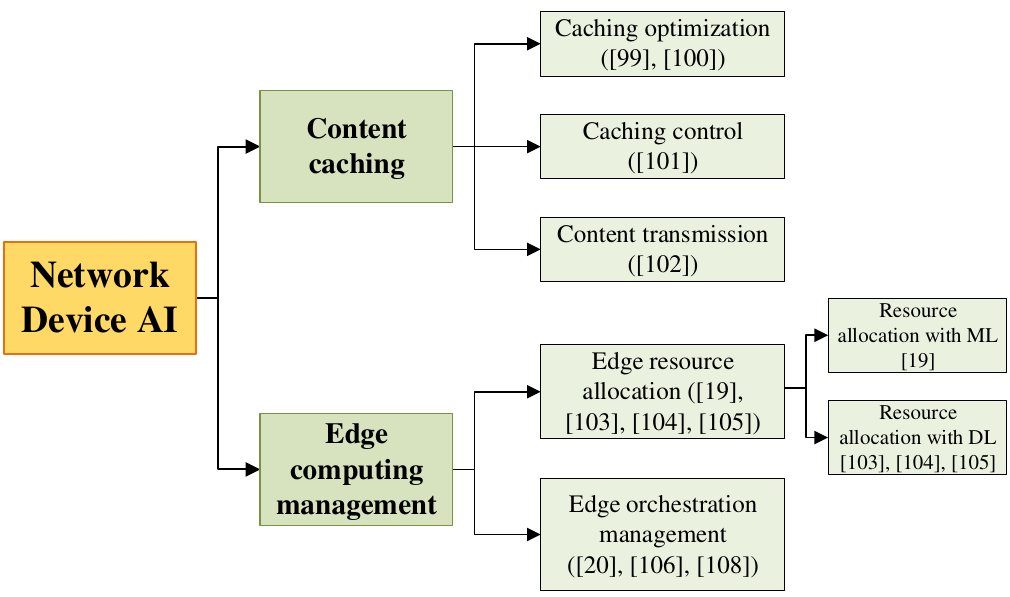}
	\caption{\textcolor{black}{Summary of Network Device AI domain.}}
	\label{Fig:02TaxonomyNetworkDevice}
	\vspace{-0.1in}
\end{figure}

\begin{table*}
	\centering
	\caption{{\color{black}Taxonomy of Network Device AI  use cases. } }
	\label{Table:NetworkDeviceAI}
	{\color{black}
	\begin{tabular}{|P{0.4cm}|P{0.4cm}|P{1.7cm}|P{1.3cm}|P{1.2cm}|P{1cm}|P{1cm}|P{1cm}|P{1.3cm}|P{2.2cm}|P{2cm}|}
		\hline
		\textbf{Issue}& 	
		\textbf{Ref.} &	
		\textbf{Use case}&	
		\textbf{AI Algorithm}& 
		\textbf{Complexity}& 	
		\textbf{Training Time }&
		\textbf{Training Data}&
		\textbf{Accuracy}&
		\textbf{Network}&
		\textbf{Pros}&
		\textbf{Cos}
		\\
		\hline
		\parbox[t]{1.5cm}{\multirow{11}{*}{\rotatebox[origin=c]{90}{Content caching}}} & 
		\cite{174}&	Caching optimization&	K-means&	Fair&	Low&	Fair&	Fair&	Sensor networks&	Reduce noise by 70\%; high learning reliability &	No optimization of learning parameters
		\\ \cline{2-11}&
		\cite{177}&	Caching decisions&	Multi-agent RL&	Low&	Low&	Low&	-&	Edge computing&	Improved content cache hit rate&	No accuracy evaluation  
		\\ \cline{2-11}&
		\cite{180}&	Caching control&	DRL&	Fair&	Low&	Fair&	-&	Wireless networks&	Low runtime, action space savings&	Lack of multiple base stations settings
		\\ \cline{2-11}&
		\cite{181} &	Content transmission&	DRL&	Low&	Fair&	Fair&	Fair&	IoT networks&	Efficient large-scale data learning &	Hard to obtain wireless datasets
		\\ \cline{2-11}
		\hline
		\parbox[t]{1.5cm}{\multirow{19}{*}{\rotatebox[origin=c]{90}{Edge computing management}}}& 
		\cite{185}	&Resource scheduling&	DNN&	Low&	Fair&	High&	Fair&	UAV networks&	Online learning; low computing complexity&	Data loss risks on UAV-ground links
		\\ \cline{2-11}&
		\cite{186}	&Resource allocation&	Deep Q-learning&	Fair&	Low&	Fair&	High (99\%)&	Edge computing&	Low-latency learning, high task success ratio&No training optimization 
		\\ \cline{2-11}&
		\cite{187}&	Multi-edge resource scheduling&	Online ML&	Fair&	Fair&	Fair&	Fair&	Mobile user networks&	Reduce scheduling cost by 70\%&	Only consider small-scale settings
		\\ \cline{2-11}&
		\cite{188}&	Edge computing&	DRL&	High&	High&	High&	-&	Edge computing&	Low running cost; adaptive to user mobility&	Not general for all DRL models
		\\ \cline{2-11}&
		\cite{190} 	&Edge orchestration management&	Deep Q-learning&	High&	High&	Fair&	Fair&	Vehicular networks&	Enable dynamic learning &	High computing resources on edge devices
		\\ \cline{2-11}&
		\cite{191}	&Virtual edge orchestration&	Model-free DRL&	Fair&	Fair&	Low&	-&	Virtual edge networks&	Optimized learning parameters&Lack of practical edge settings
		\\ \cline{2-11}&
		\cite{192}	&Edge service orchestration&	DRL&	Fair&	Low&	Low&	High&	IoT networks&	Low learning costs&	Require specific edge networks 
		\\ \cline{2-11}
		\hline
	\end{tabular}}
\end{table*}
\section{Access AI}
\label{Sec:Access_AI}
{The third Wireless AI function of the data life cycle in Fig.~\ref{Fig:WirelessAI_Function} is Access AI where AI can be applied to facilitate the data transmission from three main layers, including PHY layer, MAC layer, and network layer. Here, we focus on discussions on the roles of AI in such layers as shown in Fig.~\ref{Fig:AccessAI}.}
\begin{figure}
	\centering
	\includegraphics [width=0.95\linewidth]{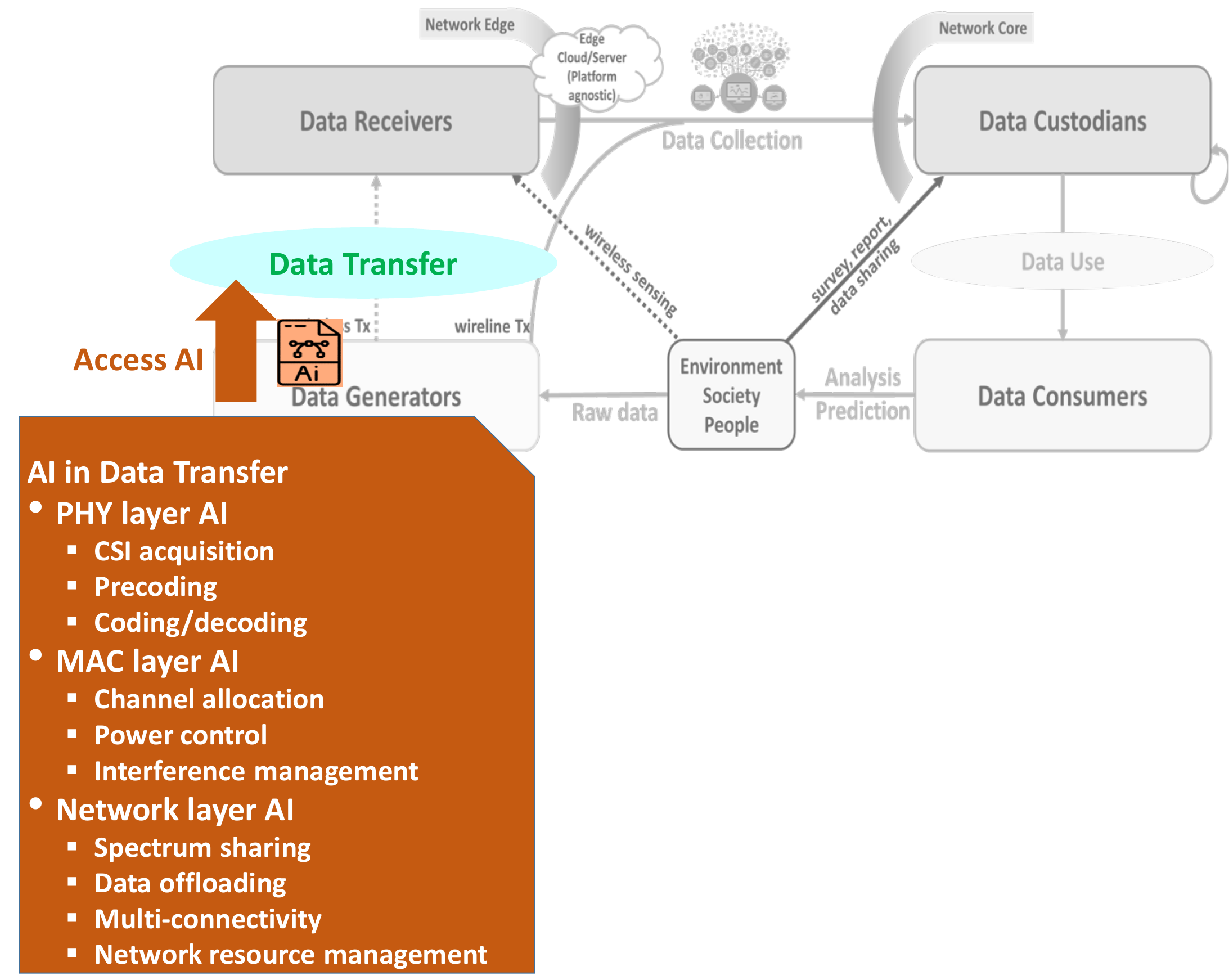}
	\caption{Access AI function. }
	\label{Fig:AccessAI}
	\vspace{-0.1in}
\end{figure}
\subsection{PHY Layer AI}
AI has been applied in physical layer communications and shown impressive performances in recent years \cite{62}. Here, we focus on discussing the roles of AI in three important domains, namely CSI acquisition, precoding, and coding/decoding. 
\subsubsection{CSI Acquisition} In this sub-section, we analyze the application of AI in CSI acquisition. 

\textbf{Problem formulation: }The massive multiple-input multiple-output (MIMO) system is widely regarded as a major technology for fifth-generation wireless communication systems. By equipping a base station (BS) with hundreds or even thousands of antennas in a centralized or distributed manner, such a system can substantially reduce multiuser interference and provide a multifold increase in cell throughput. This potential benefit is mainly obtained by exploiting CSI at BSs. In current frequency division duplexity (FDD) MIMO systems (e.g., long-term evolution Release-8), the downlink CSI is acquired at the user equipment (UE) during the training period and returns to the BS through feedback links. 

\textbf{Drawbacks of conventional methods: }Vector quantization or codebook-based approaches are usually adopted to reduce feedback overhead. However, the feedback quantities resulting from these approaches need to be scaled linearly with the number of transmit antennas and are prohibitive in a massive MIMO regime. 

\textbf{Unique advantages of using wireless AI techniques: }Recently, AI techniques such as ML and DL have been adopted as a promising tool to facilitate CSI acquisition in wireless communications. For example, AI potentially lowers the complexity and latency for CSI acquisition tasks such as CSI feedback and CSI sensing. Particularly, it is shown that a deep neural network with online learning is able to facilitate the parameter tuning and expedite the convergence rate, which improves the estimation of CSI without occupation of uplink bandwidth resource \cite{62}. 

\textbf{Recent works: } {There are a number of studies working on AI/ML techniques for CSI acquisition.}
\begin{itemize}
	\item {Conventional AI/ML-based CSI acquisition:} The work in \cite{64} introduces a ML-based CSI acquisition architecture for massive MIMO with frequency division duplex (FDD). The learning algorithm is divided into three steps: offline regression model training, online CSI estimation, and online CSI prediction by using linear regression (LR) and support vector regression (SVR). The proposed ML solution enables each user to estimate the downlink channel separately and then feed back the estimated low dimension channel to the base station based on the simple codebook method, which reduces the overhead of uplink feedback. Another work in \cite{67} is also related to CSI estimation. It provides a ML-based time division duplex (TDD) scheme in which CSI can be estimated via a ML-based predictor instead of conventional pilot-based channel estimator. Here, two ML-based structures are built to improve the CSI prediction, namely, a CNN combined with autoregressive (AR) predictor and autoregressive network with exogenous inputs. \textcolor{black}{This advanced ML-CNN model can improve the prediction accuracy and save 77\% in learning overhead, but its training may become unstable in dynamic scenarios with high user mobility and varying CSI datasets.}
	\item {Advanced AI/ML-based CSI acquisition:}  The authors in \cite{69} introduce CsiNet, a novel CNN-enabled CSI sensing and recovery mechanism that learns to effectively use channel structure from training samples. In CsiNet, the recent and popular CNNs are explored for the encoder (CSI sensing) and decoder (recovery network) so that they can exploit spatial local correlation by enforcing a local connectivity pattern among the neurons of adjacent layers. To compress CSI in massive MIMO system, a multiple-rate compressive sensing neural network framework is proposed in \cite{70}. The key focus is on the investigation of CSI reconstruction accuracy and feasibility at the user equipments (UE) for the CSI feedback problem by using a DL-based network architecture, called CsiNet+. This framework not only enhances reconstruction accuracy but also decreases storage space at the UE, thus enhancing the system feasibility. The authors in \cite{71} also propose a convolutional DL architecture, called DeepCMC, for efficient compression of the channel gain matrix to alleviate the significant CSI feedback load in massive MIMO systems. The proposed scheme is composed of fully convolutional layers followed by quantization and entropy coding blocks for better CSI estimation under low feedback rates. \textcolor{black}{ However, the proposed DL algorithm requires lengthy learning parameter estimation and requires huge training datasets (80000 realizations).}

\end{itemize}

\subsubsection{Precoding} In this sub-section, we discuss the roles of AI in supporting precoding.

\textbf{Problem formulation: }Hybrid precoding is a significant issue in millimeter wave (mmWave) massive MIMO system. To cope with a large number of antennas, hybrid precoding is required where signals are processed from both analog and digital precoders. Typically, the estimation of MIMO channels is neccessary to establish the hybrid precoding matrices where AI has emerged as a promising tool to support precoding.

\textbf{Drawbacks of conventional methods: }Due to the large-scale antennas as well as the high resolution analog-to-digital converters (ADCs)/digital-to analog converters (DACs), current digital precoding frameworks \cite{77} suffer from high power consumption and hardware costs. Besides, the radio frequency (RF) chains are considerable that may make traditional solutions inefficient to achieve a desired throughput rate. 

\textbf{Unique advantages of using wireless AI techniques: }Recently, AI has attracted much attention in precoding system designs thanks to its low complexity, online learning for better estimation of hybrid precoding matrices, and its adaptability to time-varying mmWave channels. 

\textbf{Recent works:} There are many AI/ML techniques used to facilitate precoding tasks in the literature. 
\begin{itemize}
	\item {Conventional AI/ML approaches:} For example, ML has been adopted in \cite{80} to build a hybrid precoding architecture with two main steps. First, an optimal hybrid precoder is designed with a hybrid precoding algorithm enabled by an enhanced cross-entropy approach in ML. The second step is to design the optimal hybrid combiner with an approximate optimization method, aiming to obtain the optimal precoding matrix and combining matrix to maximize the spectrum efficiency of mmWave massive MIMO system with low computational complexity. To support hybrid precoding in mmWave MIMO-OFDM systems, a ML approach using cluster analysis is introduced in \cite{82}. The key purpose is to group dynamic subarray with the help of PCA that is used to extract frequency-flat RF precoder from the principle component of the optimal frequency-selective precoders. Similarly, the study in \cite{83} incorporates an unadjusted probability and smoothing constant to across-entropy method in ML to perform a hybrid precoding implementation to solve the issue of energy loss in mmWave massive MIMO systems with multiple RF chains.
	\item {Advanced AI/ML approaches:} The work in \cite{85} applies DNNs to obtain a decentralized robust precoding framework in multi-user MIMO settings, aiming to cope with the continuous decision space and the required fine granularity of the precoding. To this end, the authors develop a decentralized DNN architecture, where each TX is responsible to approximate a decision function using a DNN, and then all the DNNs are jointly trained to enforce a common behaviour for all cooperative TXs. \textcolor{black}{The proposed DNN architecture is simple and easy to implement with a flexible learning process, but it has low convergence reliability that may degrade the training throughput.}
	To reduce the large computational complexity and local-minimum problems faced by the traditional design of hybrid precoders for multi-user MIMO scenarios, the work in \cite{86} considers a DL scheme using a CNN for a new mmWaves hybrid precoding architecture. More specific, the precoder performs an offline training process by using the channel matrix of users as the input to train the CNN, which generates the output labelled as the hybrid precoder weights. After that, based on the trained model, the CNN-MIMO can predict the hybrid precoders by feeding the network with the user channel matrix. A significant advantage of the CNN architecture is a capability to handling the imperfections of the input data and the dynamicity of the network, which contributes a stable and better performance in terms of high throughput rates.  Recently, DRL has been appeared as a new research direction to further improve the performance of current ML/DL approaches in precoding design. The work in \cite{88} also focuses on precoding issues, but it uses DRL for hybrid precoding in MIMO under uncertain environment and unknown interference. The benefit of DRL is the ability to learn from the environment using an agent to force a policy for optimal actions and then obtain rewards without a specific model. Using this theory, the authors develop a game theoretic DRL-based algorithm to learn and update intelligently a dynamic codebook process for digital precoding. \textcolor{black}{The simulation results show that DRL can effectively maximize multi-hop signal coverage and data rate for MIMO systems, enabled by a well-trained architecture with data uncertainty. However, its computational cost is relatively high due to the existing multiple DNNs in the actor-critic structure. }
	Similarly, the study in \cite{89} also analyzes a mmWave hybrid precoding model. This work introduces a mmWave point-to-point massive MIMO system where the hybrid beamforming design is considered using DRL. In this scheme, channel information is considered as the system state, precoder matrix elements are actions, while spectral efficiency and bit error rate are regarded jointly as a reward function. Based on that, a mmWave hybrid precoding problem is formulated as a Markov decision process (MDP) that is solved by a deep Q-learning algorithm. The authors concluded that the use of DRL can achieve high performance in terms of high spectral efficiency and minimal error rate, compared to its algorithm counterparts. 
\end{itemize}
\subsubsection{Coding/decoding}
In this sub-section, we present a discussion on the use of AI in supporting coding/decoding in wireless networks.

\textbf{Problem formulation: }Channel coding and decoding are significant components in modern wireless communications. Error correcting codes for channel coding are utilized to achieve reliable communications at rates of the Shannon channel capability. For instance, low-density parity-check (LDPC) codes are able to exhibit near Shannon capability with decoding algorithms such as belief propagation (BP) \cite{700}.

\textbf{Drawbacks of conventional methods: } Many coding and decoding solutions have been proposed in the literature but they still remain some limitations. For example, LDPC codes may not achieve desired performance due to channel fading and coloured noise that also introduces high coding complexity \cite{700}. Another approach is whitening that aims to transform coloured noise to white noise, but it requires matrix multiplication that is known to highly complex for long-length codes. 

\textbf{Drawbacks of conventional methods: } The advances of AI open up new opportunities to address coding/decoding issues. Instead of relying on a pre-defined channel model, AI is able to learn the network channel via learning with low computational complexity, better coding accuracy and improved decoding performances.

\textbf{Recent works:} The work in \cite{701} uses a CNN to develop a smart receiver architecture for channel decoding. Stemming from the fact that the difficulty of accurate channel noise estimation with presence of decoding error, neural layers are exploited to estimate channel noise. In this process, training is particularly important to eliminate estimation errors and capture useful features for the BP decoder. The continuous interaction between the BP decoder and CNN would improve decoding SNR with much lower complexity and better decoding accuracy, as confirmed from simulation results. The authors in \cite{702} concern about the stability of channel coding in terms of block lengths and number of information bits. This can be achieved by using neural networks as replacement of an existing polar decoder. In this case, the large polar encoding graph is separated into sub-graphs so that the training of codeword lengths is improved for better decoding performances with low computation latency. An improvement of BP algorithms for decoding block codes using AI is shown in \cite{703}. An RNN decoder is designed to replace the conventional BP decoder which desires to enhance the decoding efficiency on parity check matrices. The ability of AI in assisting coding and decoding is also demonstrated in \cite{704} that employs a parametrized RNN decoder for decoding. The main objective is to train weights of the decoder using a dataset that is constructed from noise collections of a codeword. This training process is performed using stochastic gradient descent, showing that the bit error rate is reduced significantly without causing computational complexity.

\subsection{MAC Layer AI }
In this sub-section, we survey the recent AI-based solutions for MAC layer in three main domains, including channel allocation, power control and interference mitigation.
\subsubsection{Channel Allocation} In this sub-section, we present a discussion on the use of AI in channel allocation.

\textbf{Problem formulation: }Channel allocation is a major concern in densely deployed wireless networks wherein there has a large number of base stations/ access points but limited available channels. Channel allocation is possible to lower network congestion, improve network throughput and enhance user experience. 

\textbf{Drawbacks of conventional methods: } In existing studies \cite{90}, the major challenge lies in accurately modelling the channel characteristics to make effectively channel decisions for long-term system performance maximization under the complex network settings. 

\textbf{Unique advantages of using wireless AI techniques: }AI with its learning and prediction capabilities shows great prospects in solving wireless channel allocation, e.g., better solving high dimensions of states, improving channel network learning for optimal channel allocation. 

\textbf{Recent works:} Some AI/ML approaches are applied to empower channel allocation applications. 
\begin{itemize}
	\item {Conventional AI/ML approaches:} An AI-based solution is proposed in \cite{90} for channel allocation in Wireless Networked Control Systems (WNCSs) where each subsystem uses a timer to access the channel. To formulate a channel allocation for subsystems, an online learning is considered so that the subsystems can learn the channel parameters (i.e., Gilbert-Elliott (GE) channel) to adjust their timers according to channel variation for gaining optimal channel resource. To support channel assignment for multi-channel communication in IoT networks, the work in \cite{91} introduces a ML-based algorithm running on IoT devices. Through the learning process, each IoT device estimates the channel with higher resources (i.e., bandwidth) to make decisions for selecting the appropriate channel in the dynamic environment. Another solution for channel allocation is \cite{93} where a RL-based ML approach is designed for optimizing channel allocation in wireless sensor networks. The RL agent is used to interact with the environment (e.g., sensor networks) and select the channel with high frequency for sensor nodes, aiming to reduce the error rate during the transmission. 
	\textcolor{black}{The key benefit of the proposed RL scheme is its low learning error, but it lacks the consideration of dynamic networks with time-varying channel information for adaptive online learning. }
	
	\item {Advanced AI/ML approaches:} Another approach in \cite{94} solves the channel allocation issue by taking advantage of pilot symbols prior known by both transmitters and receivers in OFDM systems. In fact, the estimation of sub-channels relies on the number of pilots and pilot positions. Motivated by this, the authors propose a autoencoder-based scheme with a DNN to find the pilot position. The autoencoder is first trained to encode the input (i.e., the true channel) into representation code to reconstruct it. Then, the pilots are allocated at nonzero indices of the sparse vector so that the channel allocation can be done only at the indices corresponding to the pilots. Besides, a conjunction of DRL and Graph Convolutional Networks is considered in \cite{96} for a channel allocation scheme in wireless local area networks. The high density makes the use of DRL suitable for solving high dimensions of states coming from the large number of mobile users and coordinators (e.g., access points). 
\end{itemize}
\subsubsection{Power Control}  In this sub-section, we present a discussion on the use of AI in power control.

\textbf{Problem formulation: }The energy-efficient power control is one of the most important issues for the sustainable growth of future wireless communication networks. How to ensure smooth communications of devices while efficient power control on the channels is critical. This topic is of paramount importance for practical communication scenarios such as MIMO systems, OFDM networks and D2D communications \cite{99}.

\textbf{Drawbacks of conventional methods: }Most traditional approaches follow a predefined channel of wireless communication to formulate the power control/allocation problem \cite{100}. However, these schemes only work well in the static wireless networks with predictable channel conditions (e.g., user demands, transmit data sizes), while it is hard to be applied in the networks without a specific channel model. 

\textbf{Unique advantages of using wireless AI techniques: }Recently, AI has been investigated to solve power control issues in complex and dynamic wireless communications with low computational complexity and better throughput for power transmisison and power allocation. 

\textbf{Recent works: }
\begin{itemize}
	\item {Conventional AI/ML approaches:} The work in \cite{100} presents a power control scheme using ML for D2D networks. The major objective is to dynamically allocate power to the network of D2D users and cellular users for avoiding channel interference. This can be done by using a Q-learning algorithm which learns a power allocation policy such that the successful data transmission of each user is ensured with a certain level of power. In the complex wireless communication networks with nonstationary wireless channels, the control of transmit power for transmitters is a challenging task. The authors in \cite{101} propose a learning-based scheme with the objective of optimizing transmit power over the wireless channel under a certain power budget. The distribution of power allocation to channels is estimated through a learning process so that the use of power serving data transmissions is efficiently predicted and optimized in a long run.
	\item {Advanced AI/ML approaches:} Meanwhile, the study in \cite{102} investigates the potential of DL in addressing power control issues in wireless networks. As an experimental implementation, the authors employ a DNN architecture to perform a power control in large wireless systems without the need of a complex channel estimation procedure. The DNN network is trained using CSI values on all network links as inputs to capture the interference pattern over the different links. With the aid of learning, the proposed scheme can adapt well when the size of the network increases. Similar to this work, the authors in \cite{106} also leverage DL to propose a dynamic power control scheme for NOMA in wireless catching systems. They concentrate on minimizing the transmission delay by formulating it as an optimization problem with respect to transmit deadline between base station and users as well as total power constraint. Then, a DNN architecture is built to learn in an iterative manner, evaluate and find an optimal solution in a fashion the transmission time is minimized and optimal power allocation is achieved under a power budget. Recently, DRL also shows the high potential in solving power control problems. A multi-agent DRL-based transmit power control scheme is introduced in \cite{107} for wireless communications. Different from the existing power control frameworks which require accurate information of channel gains of all users, the proposed model only needs certain some channels with high power values. As a result, the complexity of the learning algorithm is low regardless of the size of the network, which shows a good scalability of the proposed scheme. DRL is also employed in \cite{108} to develop a joint scheme of resource allocation and power control for mobile edge computing (MEC)-based D2D networks. A model-free reinforcement learning algorithm is designed to update the parameters and rules of resource allocation and power control via a policy gradient method. In the training, each D2D transmitter acts as an agent to interact with the environment and take two appropriate actions: channel selection and power selection, aiming to find a near-optimal solution to maximize these two objective values. 
\end{itemize}
\subsubsection{	Interference Management}  In this sub-section, we present a discussion on the use of AI in interference management.

\textbf{Problem formulation: }In the ultra-dense networks, interference stems from the usage of the same frequency resource among neighbouring users or network cells. In OFDMA networks, for example, inter-cell interference can be caused by two or more neighbour cells using the same frequency band, which can degrade the overall performance of wireless networks. 

\textbf{Drawbacks of conventional methods: }Usually, different frequency reuse factors (FRF) or partial-frequency reuse approaches are employed to coordinate and manage interference among users. However, these strategies may be inefficiently in future wireless networks with variable network traffic and dynamic user patterns, leading to under-usage or lack of the spectrum in user/cell networks. 

\textbf{Unique advantages of using wireless AI techniques: }By using AI, it is possible that an adaptive and online approach is devised to enable minimum network interference for efficient cellular deployments in terms of spectral allocation efficiency and as well as QoS fulfilment. 

\textbf{Recent works: } Many AI-based studies have been done to solve interference management issues. 
\begin{itemize}
	\item {Conventional AI/ML approaches:} In \cite{112}, the authors consider uplink interference management in multi-user LTE cellular networks with the aid of data-driven ML. A strategy for power control is necessary to adjust the transmit power on the transmission channels for interference balance among users under the settings of cell-specific power control parameters in LTE networks. Then, a stochastic leaning-enabled gradient algorithm is developed to determine optimal power control parameters in a fashion interference among user channels is maintained in an acceptable level for improving the data rate in LTE networks. The work in \cite{113} solves the issue of inter-cell interference which is caused by two or more neighbour cells using the same frequency band in OFDMA networks through a decentralized RL approach. Different from the previous study, the authors focus on the spectrum assignment for interference control among cells. To formulate a learning problem, each cell is regarded as an agent which determines optimally an amount of spectrum resource based on obtained context information (i.e., average signal to interference ratio (SIR)/chunk in the cell and user throughput). Concretely, the proposed scheme demonstrates its high performance in cellular deployments in terms of spectral allocation efficiency and SINR enhancement as well as QoS fulfilment. 
	\item {Advanced AI/ML approaches:} To investigate the potential of DL in interference management for wireless communications, the authors in \cite{116} suggest a DNN architecture to develop a real-time optimization scheme for time sensitive signal processing over interference-limited channels in multi-user networks. By introducing a learning process, the relationship between the input and the output of conventional signal processing algorithms can be approximated and then \textcolor{black}{the DNN-based signal processing algorithm can be implemented effectively in terms of high learning accuracy (98.33\%), low computational complexity and better system throughput (high spectral efficiency and low user interference levels).} 
	Moreover, a joint DL-enabled scheme of interference coordination and beam management is presented in \cite{117} for dense mmWave networks. The ultimate objective is to minimize the interference and enhance the network sum-rate with respect to beam directions, beamwidths, and transmit power resource allocations. To do so, a beamforming training is required to establish directional links in IEEE 802.11ay. Then, the BM-IC is formulated as a joint optimization problem that is approximated by a DNN network through an iterative learning procedure.  Also, the interference management problem is considered in \cite{120} that concentrates on in the context of vehicle-to-vehicle (V2V) communications. In fact, in the complex vehicular networks, a large number of V2V links results in a high interference level, which degrades the overall performance of the involved network, such as high latency and less transmission reliability. By introducing an RL concept where each V2V link acts as an agent and spectrum usage and power allocation are determined using CSI and mutual information among channels, a DRL architecture is built to provide a unified network observation model for adjusting resource allocation among vehicles to avoid high interference on V2V links. Lastly, another solution for interference management is presented in \cite{121} from the QoS perspective. To evaluate QoS of massive MIMO cognitive radio networks in terms of transmission rate and interference, a user selection approach is proposed in  using a deep Q-learning. Two key cognitive radio scenarios are taken into consideration, including radio networks with available and unavailable CSI knowledge at the secondary base station. While the former case can be considered with a conventional programming tool to formulate and solve the optimization problem of power allocation, the later case is realized by using a Deep Q-learning scheme. The base station is regarded as an agent which senses the cognitive radio environment and makes decisions on how much power resource should be allocated to each secondary user so that interference in the network is minimal. 
\end{itemize}
\subsection{	Network Layer AI}
In this section, we discuss the roles of AI in supporting spectrum sharing, data offloading, multi-connectivity, and network resource management. 
\subsubsection{	Spectrum Sharing} In this sub-section, we discuss the roles of AI in supporting spectrum sharing.

\textbf{Problem formulation: }The ultra-dense networks in the emerging 5G era promises to support one million devices per $km^2$ in the future. However, with the full swing of IoTs reaching every corner, the anticipated device density could even surpass this target, particularly in large-scale cellular networks. To address this problem, consensus has been reached that spectrum sharing by various users and devices will play a key role, monitoring frequency-band use and dynamically determining spectrum sharing. With the available spectrum identified, secondary users (SUs) will share the opportunities efficiently and fairly, through an appropriate spectrum sharing strategy. Essentially, a distributed self-organized multi-agent system is highly desirable, where each SU acts in a self-organised manner without any information exchange with its peers. Since the complete state space of the network is typically very large, and not available online to SUs, the direct computation of optimal policies becomes intractable. 

\textbf{Drawbacks of conventional methods: }Most existing studies on the multi-agent case use rely on game theory and matching theory \cite{400} to obtain structured solutions. But these model-dependent solutions make many impractical assumptions, such as the knowledge of signal-to-interference-plus-noise ratio (SINR), transmission power, and price from base stations. Moreover, when the number of users is larger than that of channels, model-dependent methods (graph colouring in particular) will become infeasible. Therefore, a fundamentally new approach to spectrum sharing needs to be developed, handling the large state space and enabling autonomous operation among a large number of SUs. 

\textbf{Unique advantages of using wireless AI techniques: }AI has emerged as a highly efficient solution to overcome these challenges in spectrum sharing. Here, we focus on analyzing the applications of AI in spectrum sharing with recent successful works. Many AI-empowered solutions with ML, DL and DRL techniques have been proposed to spectrum sharing issues in wireless communications. 

\textbf{Recent works: } The work in \cite{403} presents an inter-operator spectrum sharing (IOSS) model that takes advantage of the benefits of spectral proximity between users and BSs. By executing a Q-learning scheme at each BSs, the network operators can achieve efficient spectrum sharing among users with high quality of experience and spectral resource utilization. To be clear, the network operators can share their licenced spectrum with others via an orthogonal spectrum sharing based business model. The BSs can make decisions to utilize the IOSS mode based on the amount of spectral resources requested from the neighbours, while inferences among users are taken to formulate the action (e.g., spectrum sharing parameters) to satisfy user demands in different load scenarios. \textcolor{black}{To this end, a Q-learning algorithm is derived so that the BS can adjust dynamically the spectrum resources based on the states and actions for self-organized sharing among different types of users, including users with high service demands and users with lower resource requirements. However, the learning structure needs to be improved to enhance the learning quality, e.g., combining with DNN training for better data feature selection.}

Database-based spectrum sharing is of the important solutions for wireless IoT communications  \cite{404}. In fact, the spectrum sharing policies of a spectrum can be aggregated as a database so that the SUs can refer for avoiding spectrum interference with primary users, especially in the primary exclusive region (PER). Therefore, updating the PER once spectrum interference occurs is necessary to make sure that the smooth spectrum exchange is ensured. To do that, a supervised learning algorithm is derived to update the PER and learn a sharing policy over time for allocating optimally spectrum over the network, aiming to increase the efficiency of spectrum usage among users. To support spectrum sharing in vehicular networks, the work in \cite{405} considers a multi-agent RL scheme where multiple vehicle-to-vehicle (V2V) links reuse the frequency spectrum occupied by vehicle-to-infrastructure (V2I) links. Here, the V2V links interconnect neighbouring vehicles, while the V2I network links each vehicle to the nearby BS. The process of spectrum sharing among V2V links and V2I connections is formulated as a multi-agent problem with respect to spectrum sub-band selection. The key objective is to optimize the capability of V2I links for high bandwidth content delivery, which is enabled by a reward design and centralized learning procedure. Then, a multi-agent deep RL is proposed where each V2V link acts as an agent to learn from the vehicular environment for gaining high spectrum usage. The implementation results indicate the advantage of the learning approach in terms of better system capacity of V2I links and enhanced load delivery rate of V2V links with spectrum savings.

\subsubsection{	Data Offloading} In this sub-section, we discuss the roles of AI in supporting data offloading. 

\textbf{Problem formulation: }Data offloading refers to the mechanism where the resource-limited nodes offload part or full of their data to the resourceful nodes (e.g., edge/cloud servers). Data offloading potentially mitigates the burden on resource-constrained nodes in terms of computation and storage savings. Worthy of mention, this solution would benefit both end users for better service processing and service providers (e.g., edge/cloud services) for network congestion control and better user service delivery. For example, the advances of mobile cloud/edge technologies open up opportunities to solve the computation challenges on mobile devices by providing offloading services. In this way, mobile users can offload their computation-extensive tasks (i.e., data and programming codes) to resourceful cloud/edge servers for efficient execution. This solution potentially reduces computation pressure on devices, and enhances service qualities to satisfy the ever-increasing computation demands of modern mobile devices. 

\textbf{Drawbacks of conventional methods: }Some literature works listed in \cite{144}  show that the traditional offloading strategies mainly leverage Lyapunov or convex optimization methods to deal with the data offloading problem. However, such traditional  optimization algorithms is only suitable for low-complexity tasks and often needs the information of system statistics that may not be available in practice. Besides, how to adapt the offloading model to the varying environments (e.g., varying user demands, channel availability) is a critical challenge to be solved. 

\textbf{Unique advantages of using wireless AI techniques: }In the data offloading process, AI plays an important role in supporting computation services, channel estimations and device power control for optimal offloading. 

\textbf{Recent works: } Many AI-based studies have been done to solve data offloading issues.
\begin{itemize}
	\item {Conventional AI/ML approaches:} An ML-enhanced offloading scheme is proposed in \cite{144} for industrial IoT where mobile devices can offload their data tasks to nearby edge servers or remote cloud servers for computation. The main performance indicator considered in offloading is service accuracy (e.g., accuracy of computation results). A compressed neural network is deployed on edge servers to learn model parameters from data across edge servers, instead of relying on a cloud server for model updates. Then, the predictive model is trained based on a specific requirement of service delay and accuracy of computation tasks, congestion status of the edge-IoT network, and available computing edge resource. \textcolor{black}{The learning-based offloading framework is able to estimate how much data should be offloaded for optimal service accuracy in real-time network settings with a simple transfer learning architecture. In the future, it is necessary to consider more realistic IoT settings, e.g., offloading in smart city services, to evaluate the efficiency of the proposed learning model.}
	\item {Advanced AI/ML approaches:} A DRL-based solution for supporting offloading in wireless networks is  evaluated in \cite{146}. To be clear, a vehicular traffic offloading scheme is considered to deal with the issue of computation burden on resource-limited vehicles by using MEC services. Transmission mode selection and MEC server selection are taken into consideration for building an optimization problem, which is then solved effectively by a deep Q-learning algorithm. Moreover, a cooperative offloading architecture for multi-access MEC is given in \cite{147}. In the complex environment with multiple MEC servers and varying user offloading requirements, how to offload tasks to support mobile users while preserve network resource and improve data catching on MEC servers is challenging. Inspired by these, a joint offloading scheme of MEC and D2D offloading is formulated and offloading decisions are formulated as a multi-label classification problem. To solve the proposed problem, a deep supervised learning-based algorithm is then derived which aims to minimize the overhead of data offloading. Considering the limitation of battery and computation capability of IoT devices, the study in \cite{148} suggests an offloading scheme which enable IoT devices to submit their heavy data tasks to nearby MEC servers or cloud servers. A challenge here is how to choose which computation platform (MEC or cloud) to serve IoT users for the best offloading latency efficiency. To tackle this problem, a DL mechanism is proposed to learn the offloading behaviours of users and estimate offloading traffic to reduce computing latency with respect to task size, data volume and MEC capability. Moreover, to learn a computation offloading process for wireless MEC networks, a DRL approach is described in \cite{150} as shown in Fig.~\ref{Fig:DRL_Offloading}. The key problem to be solved is how to optimize offloading decisions (i.e., offloading tasks to MEC servers or locally executing at mobile devices) and achieve network power resource preservation with respect to network channel dynamicity. Motivated by this, a joint optimization problem is derived that is then solved effectively by a deep Q-learning empowered by a DNN network, showing that DRL not only provides flexible offloading decision policies but also adjusts adaptively network resource to satisfy the computation demands of all network users. \textcolor{black}{The pros of this scheme include low training time and minimal CPU usage, but it should be  extended to more general edge computing settings, e.g., integrated content offloading-caching or offloading-resource allocation scenarios.}
	DRL is aslo useful in vehicular data offloading for Internet of vehicle networks \cite{151}. To achieve an optimal offloading decision for tasks with data dependence, a RL agent can be neccessary to sense the vehicular networks to collect necessary information, including user mobility and resource demands, and then train offline the collected data on the MEC nodes. Besides, thanks to the online training process, the vehicular service transactions can well adapt with the changes of vehicular environments for efficient offloading. 
\end{itemize}

\begin{figure}
	\centering
	\includegraphics [width=0.95\linewidth]{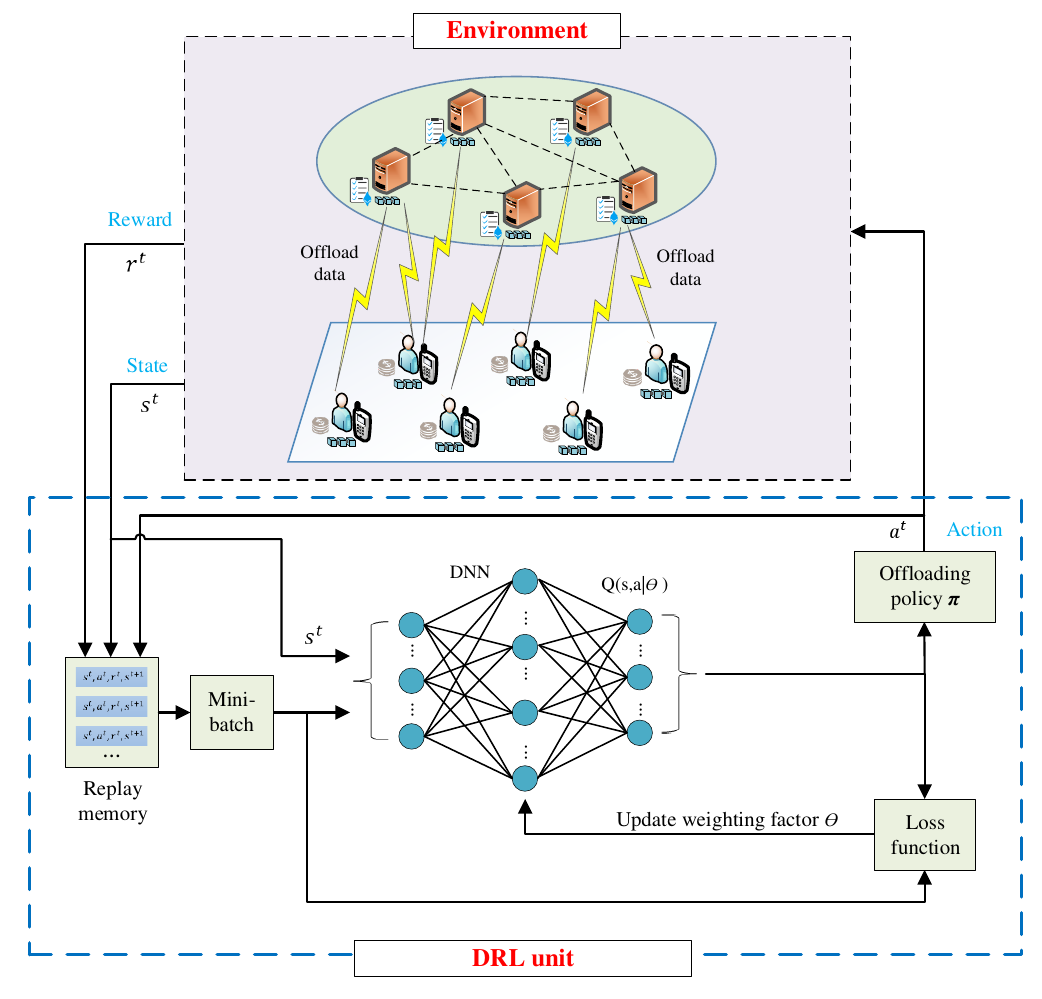}
	\caption{Deep reinforcement learning for data offloading. }
	\label{Fig:DRL_Offloading}
	\vspace{-0.1in}
\end{figure}
\subsubsection{Multi-connectivity}
Due to the increasing number of base stations and mobile users in 5G ultra-dense networks, the demands of multi-connectivity with high reliability and spectrum efficiency is growing. In such as a context, AI has come as a promising solution to cope with the enormous state space caused by high network complexity and well support interconnection among users \cite{154}. Communication techniques such as duel connectivity can realize the data transmission between base station and mobile users such that each station can select a codeword from a code book to establish an analog beam for assisting downlink communication with users. ML approaches such as SVM can support classification for codeword selection by a learning process where the inputs include user transmit power and CSI data. By using trained model, users can select effectively channels for uplink transmission and base stations can select codeword for downlink transmission with low complexity, which achieves fast and reliable ultra-dense network communications. To solve the mobility management issue caused by frequent handovers in ultra-dense small cells, a LSTM-based DL algorithm is proposed in \cite{155}. This aims to learn the historical mobility patterns of users to estimate future user movement, which is important for hanover estimation for regulate user connectivity with base stations. Further, to improve the scalability of wireless communications in 5G vehicular networks, a traffic workload prediction scheme using neural networks is considered in \cite{156}. The model is able to learn connectivity behaviour based on data rate, numbers of users and QoS requirements in order to make decisions for adjusting scalability for efficient vehicular communications in terms of low latency and high service request rates.
\subsubsection{Network Resource Management}
In this sub-section, we present how AI can enable smart network resource management.

\textbf{Problem formulation: }The current cellular technology and wireless networks may not satisfy the strides of wireless network demands. Resource management becomes more complex to maintain the QoE and achieve the expected system utility. The development of advanced technologies in 5G such as software defined networking (SDN) \cite{800} and network slicing makes network resource management much more challenging that thus urgently requires new innovative and intelligent solutions.

\textbf{Drawbacks of conventional methods: }Many solutions have been proposed to solve network resource management issues, such as generic algorithms or dynamic programming approaches, but they have high complexity in terms of computation and optimization \cite{25}. Further, these conventional approaches have been demonstrated to be inefficient under presence of dynamicity of network resources and user demands. 

\textbf{Drawbacks of conventional methods: }AI can achieve smart and reliable network resource management thanks to its online learning and optimization capability. An important feature of AI is its smart making decision ability (e.g., RL approaches) to adaptively allocate network resources and control intelligently resource so that the network stability is guaranteed.   

\textbf{Recent works: }The work in \cite{801} presents an AI-based network resource management scheme for SDN-empowered 5G vehicular networks. Both LSTM networks and CNNs are used to classify and detect the resource demands at the control plane. To mimic the resource usage behaviour, learning is necessary to train the virtualization manager placed on the SDN controller. It is able to adjust the resource (e.g., bandwidth) based on the user demand and QoS requirements so that network fairness among users can be achieved. In the line of discussion, DL has been used in \cite{802} to manage SDN resources in vehicular networks. To do that, a flow management scheme is designed and integrated in the SDN controller so that resource utilization is optimized. Due to the different priority levels and data requests of different vehicular users, a dynamic resource allocation algorithm is vitally important to avoid over resource usage as well as resource scarcity at any users. A CNN is deployed on the network controller to manage the flow control for multiple users, aiming to optimize resource usage. Based on the network information such as bandwidth, available resource, user requests, the network controller can learn the new resource usage patterns for better allocation.

\subsection{Lessons Learned} The main lessons acquired from the survey on the Access AI domain are highlighted as the following.
\begin{itemize}
	\item AI well support the designs of three key network layers: PHY layer, MAC layer, and Network layer. For PHY layer, AI has been applied to facilitate CSI acquisition tasks in wireless communications such as CSI estimation \cite{67}, CSI sensing and recovery \cite{69}. AI provides promising solutions by its learning and prediction capabilities to simplify the precoding design under uncertain environments and unknown network traffic patterns \cite{88}.	Further, AI also facilitates channel coding/decoding for wireless systems \cite{703}.  
	\item Many AI-based solutions have been proposed for MAC layer, mainly in three domains, namely channel allocation, power control and interference mitigation. In the multi-channel communications with less CSI and high channel variation, AI especially DL is able to estimate dynamically the channel allocation by establishing the adaptive channel assignment policy with low computational complexity. AI has been also investigated to solve power control issues in complex and dynamic wireless communications with recent successes \cite{100}, \cite{106}. Specially, some emerging research results demonstrate that AI is particularly useful for interference management in small cell networks or mmWave networks \cite{117}.
	\item	AI also gain enormous interests in network layer designs from four fundamental domains, spectrum sharing, data offloading, multi-connectivity, and network resource management. For instance, many AI-based approaches have been introduced to solve the critical spectrum sharing issues in wireless communications, ranging from wireless spectrum sharing \cite{403} to IoT spectrum exchange \cite{404}. Meanwhile, the learning capability of AI is also helpful to estimate data traffic and channel distribution to realize flexible data offloading \cite{144}, multi-connectivity \cite{155}, and resource management \cite{801} in wireless networks.
	\item \textcolor{black}{In general, in the Access AI domain, CNNs have been the preferred choice for many network access use cases, e.g., CSI estimation \cite{67} and channel precoding \cite{86} thanks to their low learning complexity and robust learning, compared to their counterparts such as DNNs. Moreover, DRL \cite{88} with its deep learning structure can solve well the large-scale learning issues remained in RL schemes \cite{93} in network access to achieve a better learning performance in the context of large datasets. LSTM networks have emerged as highly efficient learning methods for network intelligence, such as in smart multi-connectivity \cite{155}, with high learning accuracy and low computational complexity.}
\end{itemize}
\textcolor{black}{In summary, we show the classification of the Access AI domain in Fig.~\ref{Fig:03AccessAI} and summarize the AI techniques used along with their pros and cons in the taxonomy Table~\ref{Table:AccessAI}. }
\begin{figure}
	\centering
	\includegraphics [width=0.95\linewidth]{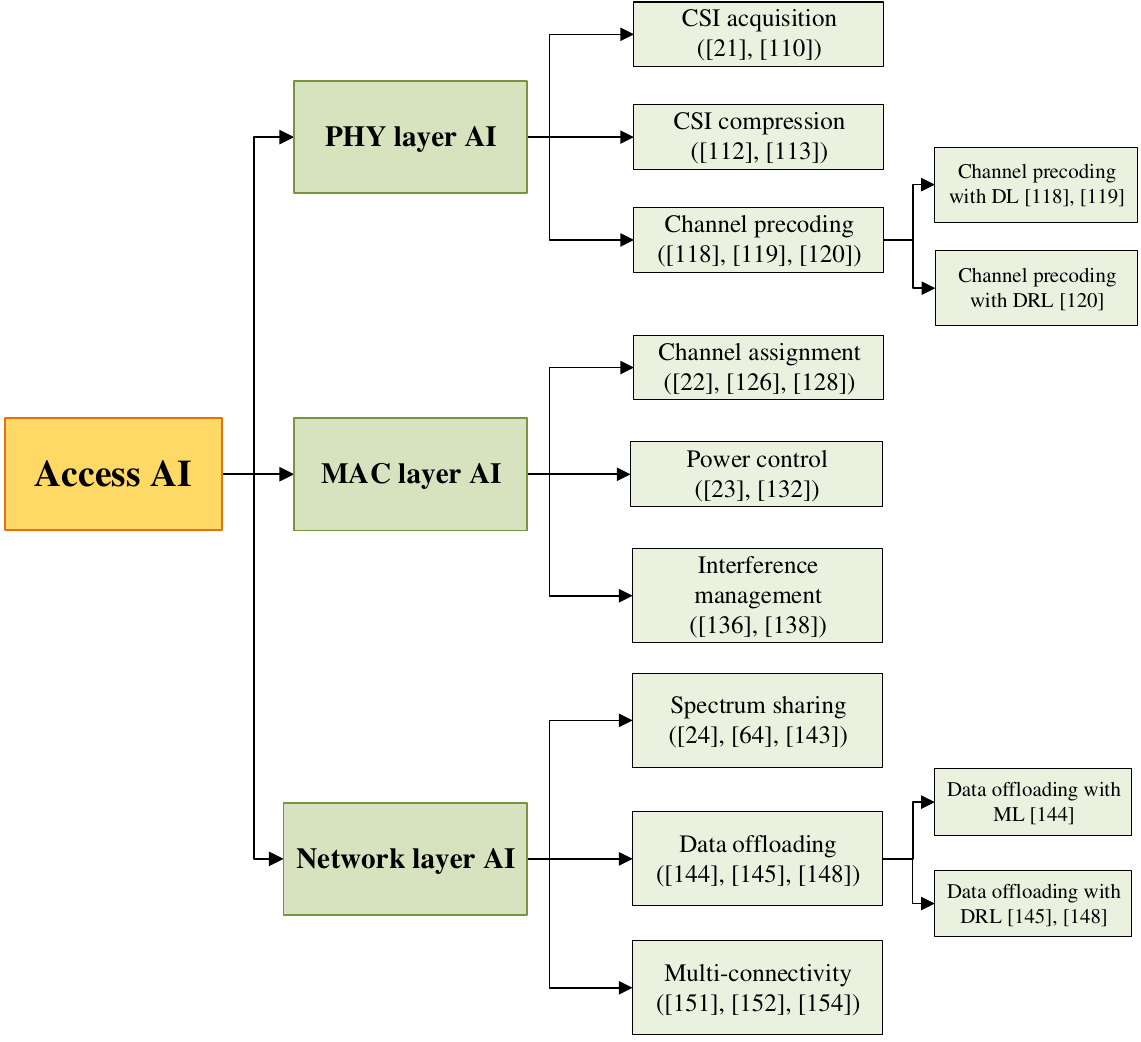}
	\caption{\textcolor{black}{Summary of Access AI domain.}}
	\label{Fig:03AccessAI}
	\vspace{-0.1in}
\end{figure}

\begin{table*}
	\centering
	\caption{{\color{black}Taxonomy of Access AI use cases.}  }
	\label{Table:AccessAI}
	{\color{black}
	\begin{tabular}{|P{0.4cm}|P{0.4cm}|P{1.8cm}|P{1.3cm}|P{1.2cm}|P{1cm}|P{1cm}|P{1cm}|P{1.3cm}|P{2.2cm}|P{2cm}|}
		\hline
		\textbf{Issue}& 	
		\textbf{Ref.} &	
		\textbf{Use case}&	
		\textbf{AI Algorithm}& 
		\textbf{Complexity}& 	
		\textbf{Training Time }&
		\textbf{Training Data}&
		\textbf{Accuracy}&
		\textbf{Network}&
		\textbf{Pros}&
		\textbf{Cos}
		\\
		\hline
		\parbox[t]{1.5cm}{\multirow{16}{*}{\rotatebox[origin=c]{90}{PHY layer AI}}} & 
		\cite{64}&	CSI acquisition	&SVR&	Low&	Low&	Low&	Fair&	Massive MIMO&	Low computational complexity&	Sensitive to channel uncertainty
		\\ \cline{2-11}&
		\cite{67}&	CSI estimation	&CNN &	High&	Fair&	High&	High&	Massive MIMO&	Reduce learning overhead by 77\%&Risks of unauthorized access
		\\ \cline{2-11}&
		\cite{70}	&CSI compression&	DL (CsiNet+)&	High&	High&	Fair&	High&	Massive MIMO&	High CSI estimation accuracy&	Complex training strategies
		\\ \cline{2-11}&
		\cite{71}&	CSI compression	&CNN&	Fair&	High&	High&	Fair&	Massive MIMO&	Flexible learning architecture &	Require huge datasets (80000 CSI realizations)
		\\ \cline{2-11}&
		\cite{85}&	Channel precoding&	DNN&	Low&	Fair&	Fair&	Fair&	Wireless networks&	Simple training procedure&	Low convergence reliability 
		\\ \cline{2-11}&
		\cite{86}&	mmWaves hybrid precoding&	CNN&	Low&	Low&	Fair&	Fair&	Massive MIMO&	Less computation time; robust training&	Ignore continuous channel data 
		\\ \cline{2-11}&
		\cite{88}&	Channel precoding&	DRL&	Fair&	Fair&	Fair &	Fair&	Massive MIMO&	Good training with data uncertainty &	No optimization of learning parameter 
		\\ \cline{2-11}
		\hline
		\parbox[t]{1.5cm}{\multirow{16}{*}{\rotatebox[origin=c]{90}{MAC layer AI}}}& 
		\cite{90}	&Distributed channel access&	Online ML&	Fair&	High&	Low&	Low&	Wireless networks&	Simple training process&	Low convergence reliability 
		\\ \cline{2-11}&
		\cite{93}&	Channel allocation&	RL&	Low&	Low &	Fair&	Fair&	Wireless networks&	Low training error&	Only consider static channels
		\\ \cline{2-11}&
		\cite{94}&	Channel allocation	&DNN&	Fair&	Fair&	Fair&	-&	OFDM systems&	Enable dynamic learning&	Lack of accuracy evaluation 
		\\ \cline{2-11}&
		\cite{100}	&Power control	&Q-learning&	Low&	Low&	Low&	Fair&	Wireless networks&	Adaptive learning procedure&	Simple model design
		\\ \cline{2-11}&
		\cite{102}&	Power control &	DNN&	Low&	Fair&	Fair&	Low&	Wireless networks&	High learning high speed &	Lack of channel estimation
		\\ \cline{2-11}&
		\cite{112}	&Interference management&	Stochastic gradient leaning&	Low&	Low&	Low&	High&	LTE cellular systems&	Learning improvement &	Only simulations with parameter assumptions
		\\ \cline{2-11}&
		\cite{116}&	Channel interference control&	DNN&	Fair&	High &	High&	High (98.33\%)&	Wireless networks&	High learning accuracy and stability&	Complex learning settings
		\\ \cline{2-11}
		\hline
		\parbox[t]{1.5cm}{\multirow{22}{*}{\rotatebox[origin=c]{90}{Network layer AI}}} & 
		\cite{403}&	Spectrum sharing&	Q-learning&	Low&	Low&	Low&	Fair&	Mobile networks&	Enable Self-organizing spectrum sharing&	No Q-learning improvement
		\\ \cline{2-11}&
		\cite{404}	&Spectrum sharing	&Supervised learning&	Low&	Low&	Low&	Fair&	Cognitive networks&	Require small learning iterations&	Simple machine learning design
		\\ \cline{2-11}&
		\cite{405}	&Spectrum sharing&	RL&	Low&	Low&	Low&	High
		(95\%)&	Vehicular networks&	Low-cost training &	Less robust learning
		\\ \cline{2-11}&
		\cite{144}&	Data offloading&	Neural networks&	Low&	Low&	Low&	High (89\%)&	IoT networks&	Simple transfer learning design&	Lack of practical IoT settings
		\\ \cline{2-11}&
		\cite{146}&	Traffic offloading&	Deep Q-learning&	Fair&	Fair&	High&	Fair&	Vehicular networks&	Optimized Q-learning &	Hard to obtain traffic datasets
		\\ \cline{2-11}&
		\cite{150}	&Computation offloading&	DRL &	Fair&	Low&	Low&	High&	Edge computing&	Low CPU training latency& Only apply to specific networks
		\\ \cline{2-11}&
		\cite{155}	&Multi-connectivity &	LSTM network &	Fair&	Fair&	Fair&	High
		(95\%)&	Mobile networks&	High energy efficiency&	No consideration of real datasets
		\\ \cline{2-11}&
		\cite{156} & 	Multi-connectivity&	Neural networks &	Fair&	Fair&	Low&	High
		(90\%)&	Mobile networks&	Low deployment latency&	Less scalable to large networks
		\\ \cline{2-11}&
		\cite{801} &	Resource management&	LSTM and CNN &	High &	High&	High&	Fair&	Vehicular networks&	Multiple learning settings&	Require repeated model updates.
		\\ \cline{2-11}
		\hline
	\end{tabular}}
\end{table*}

\section{User Device AI}
\label{Sec:User_Device_AI}
The next Wireless AI function of the data life cycle in Fig.~\ref{Fig:WirelessAI_Function} is User Device AI where AI have been embedded into end devices to create a new on-device AI paradigm. In this section, we study the User Device AI function for wireless networks by analyzing the integration of AI in mobile user devices as shown in Fig.~\ref{Fig:DeviceAI}. 
\subsection{Embedded AI Functions}
\begin{figure}
	\centering
	\includegraphics [width=0.95\linewidth]{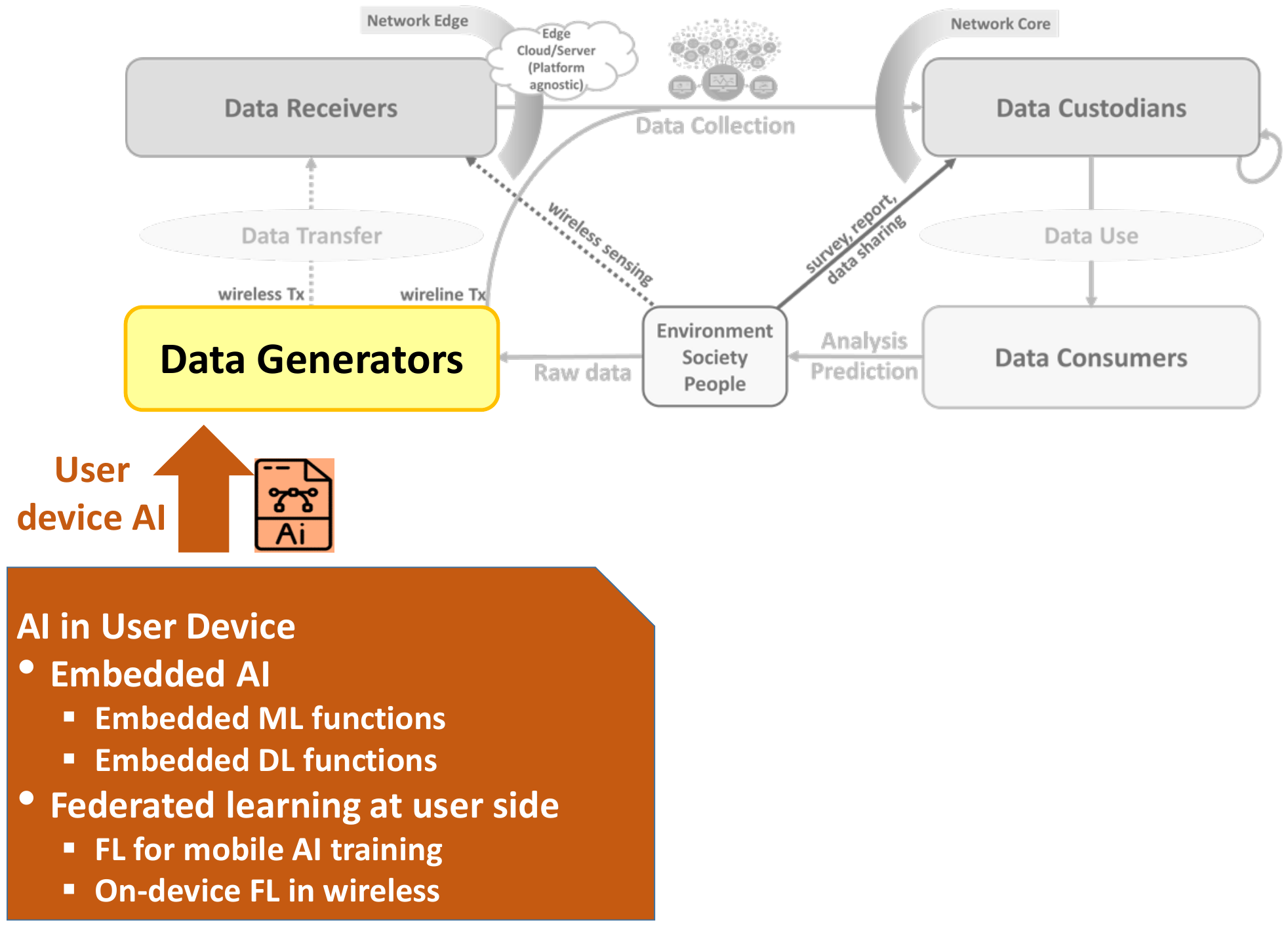}
	\caption{User Device AI function. }
	\label{Fig:DeviceAI}
	\vspace{-0.1in}
\end{figure}
In this sub-section, we present how AI is integrated into mobile devices to develop on-develop AI functions.

\textbf{Problem formulation: }With the rapid growth of smart mobile devices and improved embedded hardware, there is interest in implementing AI on the device \cite{193}, for both ML and DL functions. In the future networks with higher requirements in terms of low-latency data processing and reliable intelligence, pushing AI functions to mobile devices with privacy enhancement would be a notable choice to wireless network design. 

\textbf{Drawbacks of conventional methods: } Traditionally, AI functions are placed on remote cloud servers, which results in long latency for AI processing due to long-distance transmissions. Further, the reliance on a central server faces a serious scalable problem when processing data from distributed mobile devices. The use of external servers to run AI functions is also vulnerable to security and data privacy issues due to the third parties.

\textbf{Unique advantages of using wireless AI techniques: } AI can be run directly on mobile devices to provide smart and online learning functions for mobile networks at the edge. The key benefits of embedded AI functions consist of low-latency learning and improve the intelligence of mobile devices, which can open up new intesting on-device applications, such as mobile smart object detection, mobile human recognition. 

\textbf{Recent works: }Some important works are highlighed with the integration of AI on mobile devices, including embedded ML functions and embedded DL functions. 
\begin{itemize}
	\item Embedded ML functions: The work in \cite{194} introduces nnstreamer, an AI software system running on mobile devices or edge devices. This software is able to execute directly real-time ML functions, such as neural networks, with complex mobile data streams by simplifying ML implementations on mobile devices without relying on cloud servers. The feasibility of nnstreamer has been investigated via real-world applications, such as event detection through neural networks with sensor stream samples as the input data for training and classification. In \cite{195}, a GPU-based binary neural networks (BNN) engine for Android devices is designed, that optimizes both software and hardware to be appropriate for running on resource-constrained devices, e.g., Android phones. More specific, the authors exploit the computation capability of BNN on mobile devices, decoupled with parallel optimizations with the OpenCL library toolkit for enabling real-time and reliable neural network deployments on Android devices. \textcolor{black}{The experimental results indicate that the mobile BNN architecture can yield 95.3\% learning accuracy and enhance storage and CPU usage. However, it still exhibits high operational costs from on-device data training. } 
	Another research effort in buiding AI on devices is in \cite{196} that applies ML to Android malware detection on Android smart devices. They build an Application Program Interface (API) call function and integrate an ML classification algorithm to detect abnormal mobile operations or malicious attack on the device. As a trial, some typical ML classifiers are considered, including SVM, decision tree, and bagging to characterize the designed API scheme, showing that the ML-based bagging classifier achieves the best performance in terms of higher malicious application detection rate. Moreover, a ML-based API module is designed in \cite{197} to recognize malware on Android phones. There are 412 samples including 205 benign app and 207 malware app used in the test for classification by integrating with SVM and tree random forest classifiers. \textcolor{black}{The trial results show that ML classifiers can achieve high malware detection rates, with over 90\% for cross validation examination along with low training time. In the future, it is possible to extend the designed model to others mobile platforms, such as iOS.} 
	
	\item Embedded DL functions: Software and hardware support for performing DL on mobile devices is now evolving extremely fast. In \cite{199}, the authors run DNNs on over 10,000 mobile devices and more than 50 different mobile system-on-chips. The most complex task that can be run with the help of TensorFlow Mobile or TensorFlow Lite is image classification with MobileNet or Inception CNNs at the moment. Another example of mobile AI function is in \cite{201} that develops a MobileNet app running lightweight CNN on mobile devices for mobile localization applications. Images captured by phone cameras can be recognized by CNN to specify the centroid of the object (e.g., the human hand), which is useful to various applications in human motion detections. CNN is also applied to build classification functions for preserve user privacy on embedded devices \cite{202}. With the popularity of Android apps on the market, malicious apps can attack mobile operating systems to exploit personal information, which breaches mobile data over the Internet. By using a CNN-based training model, the mobile processor can recognize attack activities by integrating the trained model into the malicious apps. Therefore, all behaviour and data streams on mobile websites can be identified and detected. \textcolor{black}{The disadvantage of this mobile classification function includes the relatively high training complexity and high computational energy due to on-device multi-layer deep learning. }
	
	CNN inference is becoming necessary for improving user experience with recent applications, such as information extraction from mobile data streams, virtual reality. The authors in \cite{203} present a hardware design that considers to integrate CNN inference into embedded AMR devices. They implement a Kernel-levels scheme which represents the four-layer CNN across heterogeneous cores. Different cores would process different layers based on the consecutive images in a stream. The improvement of overall throughput of the CNN architecture stems from the computation resources of multi-cores and on-chip data memory. In addition to CNNs, DNNs have been explored for mobile learning \cite{204}. In this work, an on-chip DL model is designed for edge AI devices. Due to the overhead of computation on mobile CPUs, an improved DNN inference engine is considered with half precision support and optimized neural network layers to be suitable for resource-limited devices. This model potentially enhances the GPU throughput by 20 times and saves 80\% memory usage. 
	In particular, CNN can be integrated in edge devices to create an on-device CNN architecture \cite{209}. CNN layers can learn the features collected from the data samples on wearable sensor devices. The edge board such as SparkFun Apollo3 chip is used to run the DL algorithm. The novelty of this work is to use batch normalization for recognition tasks with CNN on mobile devices. To optimize the performance of on-chip DL-based system, a memory footprint optimization technique is also involved and compared with conventional statistical ML technique, showing a lower execution delay for running mobile DL models. Recently, an on-board AI model is designed for mobile devices for sensing and prediction tasks \cite{210}. To do this, a CNN architecture is considered, with two convolutional blocks, two linear blocks with a sigmoid block to perform learning and classification on embedded devices.  \textcolor{black}{A use-case using a CNN for detecting the seed germination dynamics in agriculture is taken with seed images as training data inputs, showing the high accuracy (97\%) in seed recognition with power savings for running DL algorithms. }
\end{itemize}
\begin{figure*}
	\centering
	\includegraphics [width=0.8\linewidth]{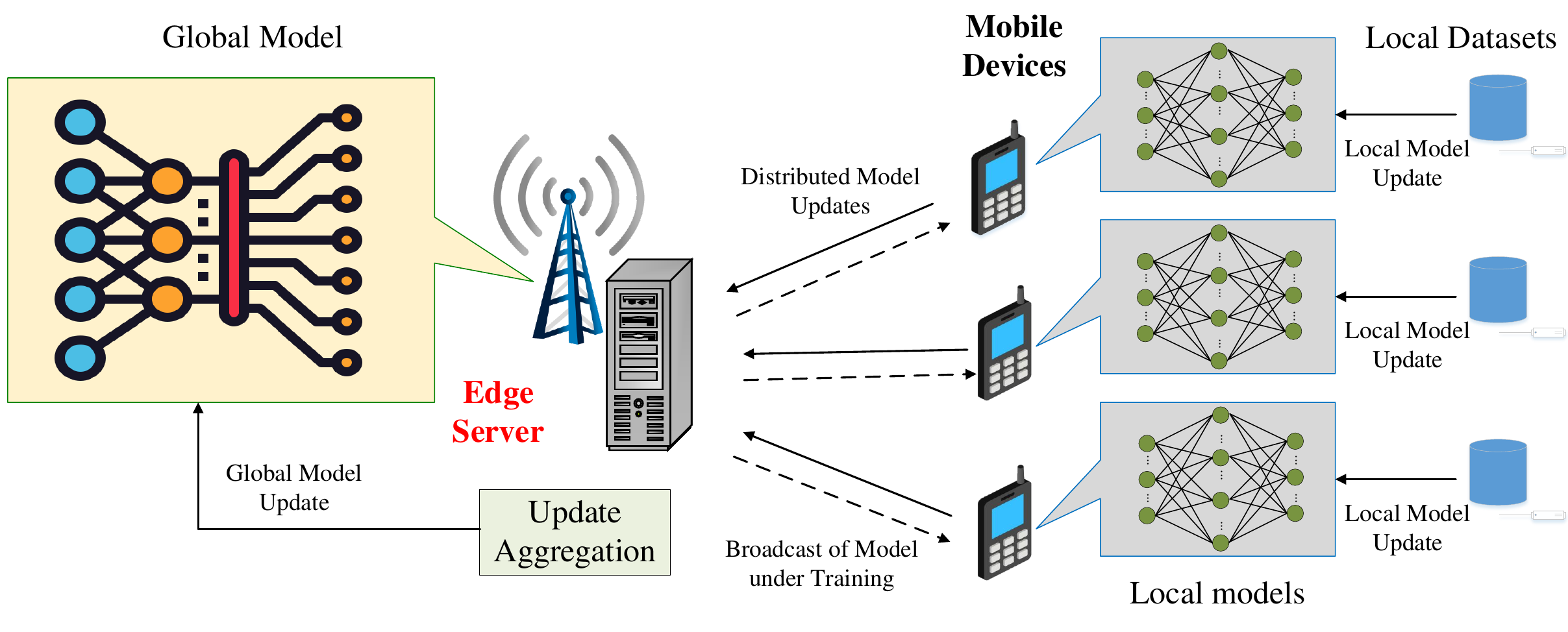}
	\caption{Federated learning for mobile wireless networks.}
	\label{Fig:FL}
	\vspace{-0.1in}
\end{figure*}

\subsection{Federated Learning at User Side}
In this sub-section, we present the recent development of FL in wireless networks. Conceptually, FL requires a cross-layer implementation involving user devices, network devices, and the access procedure. However, since almost all of the AI-related computations are carried out by user devices in FL, we put FL under the category of this section "User device AI".

\textbf{Problem formulation: } In the large-scale mobile networks, it is in-efficient to send the AI learning model to remote cloud or centralized edge servers for training due to high communication costs and AI data privacy issues. With widely distributed and heterogeneous user data located outside the cloud nowadays, centralized AI would be no longer a good option to run AI in wireless networks. FL can preserve privacy while training a AI model based on data locally hosted by multiple clients. Rather than sending the raw data to a centralized data-center for training, FL allows for training a shared model on the server by aggregating locally computed updates black \cite{211}.

\textbf{Drawbacks of conventional methods: } The traditional AI models rely on a cloud or edge server to run AI functions, which put high pressure on such a centralized data center in terms of high data storage and processing demands. Further, offloading to remote servers to run AI function may result in data loss due to the user reluctancy of providing sensitive data, which in return degrades data privacy. 

\textbf{Unique advantages of using wireless AI techniques: } 
FL allows to train AI models in a distributed manner where local data can be trained at mobile devices for low-latency learning and data privacy protection. It also reduce the flow of data traffic to avoid network congestion. 

\textbf{Recent works: } The work in \cite{213} presents a collaborative FL network for smart IoT users. The FL includes two steps, including model training on the smartphones using data collected from IoT devices and model sharing over the blockchain network \cite{nguyen2019blockchain}. To encourage more users to join the FL, an incentive mechanism is designed to award coins to users, which in return enhances the robustness of the mobile FL network. To support low-latency V2V communications, the authors in \cite{214} employ FL that enables to interconnect vehicular users in a distributed manner. In this context, each user transmits its local model information such as vehicle queue lengths to the roadside unit (RSU) where a centralized learning module is available for perform learning for maximum likelihood estimation. \textcolor{black}{Although the proposed scheme can achieve a reduction of exchanged data and low energy cost, it raises privacy concerns from distributed learning, e.g., malicious data training users.}

FL is also useful to DRL training in the offloading process of the MEC-based IoT networks \cite{215}. In the offloading scheme, each IoT device takes actions to make offloading decisions based on system states, e.g., device transmit power, CSI conditions. In such a context, FL allows each IoT user to use his data to train the DRL agent locally without exposing data to the edge server, which preserves the privacy of personal information. Each trained model on individual IoT user is then aggregated to build a common DRL model for. By using FL, the IoT data is trained locally that eliminates the channel congestion due to a large amount of data transmitted over the network. In addition to training support, FL is able to preserve the privacy of DL schemes, e.g., in the neural network training process \cite{216}. Here, the training model consists of centralized learning on the edge server and FL at the users. Specially, in FL, both users and servers collaborate to train a common neural network. Each user keeps local training parameters and synchronizes with the cloud server to achieve an optimal training performance. By implementing the FL, user privacy can be ensured due to the local training. Another solution introduced in \cite{218} uses FL for building a self-learning distributed architecture that aims to monitor security of IoT devices (e.g., detecting compromised devices). In this case, FL is used to aggregate detection profiles of the IoT network where the mobile gateways can train data locally without sharing with remote servers. This would not only mitigate the latency overhead of the training process, but also ensure IoT data privacy by using available resource for model training. Besides, to support low-latency communication in mobile wireless networks, a federated edge computing scheme is proposed in \cite{219} as depicted in Fig.~\ref{Fig:FL}. User data can be trained locally using local datasets and model updates can be transmitted over the multi-access channel to the edge server for aggregation of local models. The process is repeated until the global model is converged. \textcolor{black}{This model is able to achieve high training accuracy (90\%) with reduced aggregation error, but its energy consumed for the data learning in user devices is high which can make it inappropriate in practical low-latency mobile communication networks.} 
Further, an on-device FL model for wireless network is suggested by \cite{220}. The main focus is on a design for communication latency mitigation using a Newton method and a scheme for reducing error during the FL model update based on difference-of-convex functions programming.

\subsection{Lessons Learned} With the rapid growth of smart mobile devices and improved hardware technology, AI functions have been embedded into mobile devices to create a new paradigm: on-device AI. This would open up new opportunities in simplifying AI implementation in wireless networks by eliminating the need of external servers, e.g., clouds. In fact, the feasibility of on-device AI model has been demonstrated via recent successes with designs of mobile ML functions \cite{194} and mobile DL functions \cite{199}. For example, DNN inference now can be run on mobile devices such as Android phones for local client applications such as image classification \cite{201}, and object recognition \cite{209}. Developing deep AI models on mobile devices is predicted to revolutionize the way future intelligent networks operates and provides user services, especially in the booming era of the IoT. Moreover, AI can be implemented by the cooperation of distributed mobile devices, which gives birth to FL paradigms. Rather than sending the raw data to a centralized data-center for training, FL allows for training a shared model on the server by aggregating locally computed updates \cite{211}. FL can support training the collaborative model over a network of agents of users, service providers and agents with low training latency, traffic congestion mitigation and privacy preservation \cite{215}. 

\textcolor{black}{In general, in the User Device AI domain, conventional ML techniques such as NNs \cite{195} and SVM \cite{196} are well suitable for low-energy devices like smartphones due to their simple training in comparison with advanced ML techniques. Recently, much research effort has been devoted to improving the performance of mobile AI models such as CNNs \cite{209}, \cite{210}, by optimizing training and learning processes and advancing device hardware, which is very promising to not only improve learning accuracy but also reduce computational complexity in mobile devices.}

\textcolor{black}{In summary, we shows the classification of the User Device AI domains in Fig.~\ref{Fig:04TaxonomyUserDevice} and summarize the AI techniques used along with their pros and cons in the taxonomy Table~\ref{Table:Device_AI}. }

\begin{figure}
	\centering
	\includegraphics [width=0.95\linewidth]{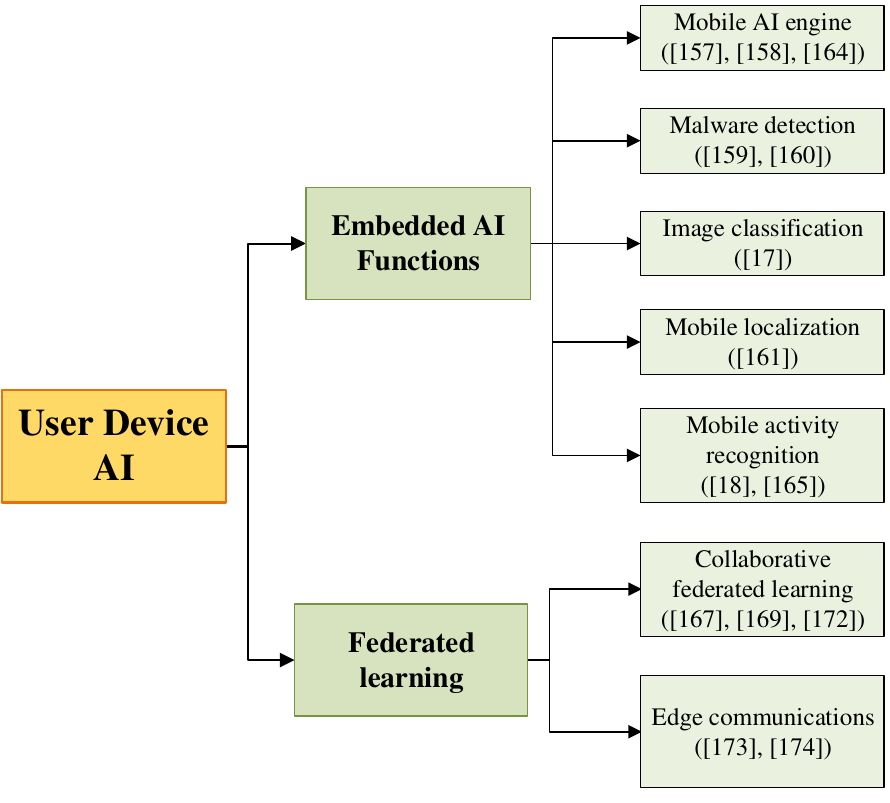}
	\caption{\textcolor{black}{Summary of User Device AI domain.}}
	\label{Fig:04TaxonomyUserDevice}
	\vspace{-0.1in}
\end{figure}

\begin{table*}
	\centering
	\caption{{\color{black}Taxonomy of User Device AI use cases. } }
	{\color{black}
	\label{Table:Device_AI}
	\begin{tabular}{|P{0.4cm}|P{0.4cm}|P{1.7cm}|P{1.3cm}|P{1.2cm}|P{1cm}|P{1cm}|P{1cm}|P{1.3cm}|P{2.2cm}|P{2cm}|}
		\hline
		\textbf{Issue}& 	
		\textbf{Ref.} &	
		\textbf{Use case}&	
		\textbf{AI Algorithm}& 
		\textbf{Complexity}& 	
		\textbf{Training Time }&
		\textbf{Training Data}&
		\textbf{Accuracy}&
		\textbf{ Network}&
		\textbf{Pros}&
		\textbf{Cos}
		\\
		\hline
		\parbox[t]{1.5cm}{\multirow{21}{*}{\rotatebox[origin=c]{90}{Embedded AI Functions}}} & 
		\cite{194}&	Embedded ML functions&	Neural networks&	High&	Fair&	Fair&	Fair&	Mobile networks&	Efficient resource utilization&	Complex software settings
		\\ \cline{2-11}&
		\cite{195}&	Mobile ML engine &	Bsinary neural networks&	Fair&	Fair&	Low&	High (95.3\%)&	Mobile networks&	Improve storage efficiency and CPU usage&	High computation for on-device learning
		\\ \cline{2-11}&
		\cite{196}	&Malware detection	&SVM, decision tree&	Fair&	Low&	Low&	Fair&	Mobile networks&	Simple app implementation&	Only specific to Android phones
		\\ \cline{2-11}&
		\cite{197}&	Malware detection&	SVM, tree random forest&	Low&	Low&	Low&	High (91.9\%)	&Mobile networks&	Simple app implementation&	Only specific to Android phones
		\\ \cline{2-11}&
		\cite{199}&	Image classification&	DNNs&	Fair&	Fair&	Low&	Fair (78\%)&	Mobile networks&	Low memory and GPU usage &	Hard to train complex data
		\\ \cline{2-11}&
		\cite{201}&	Mobile localization	&CNN&	Fair&	Low&	Fair&	High &	Mobile networks&	Handle well with complex datasets&	Relatively high on-device learning cost
		\\ \cline{2-11}&
		\cite{202}&	Mobile privacy protection &	CNN&	Fair&	Fair&	Fair&	High (93\%)&	Mobile networks&	Applicable to mobile security systems&	High computational energy
		\\ \cline{2-11}&
		\cite{204}&	DNN inference engine&	DNN&	Fair&	Fair&	Fair&	-&	Mobile networks&	Reduce memory usage by 80\%&	Lack of accuracy evaluation 
		\\ \cline{2-11}&
		\cite{209}&	Mobile activity recognition&	CNN&	Fair&	Low&	Low&	High (96.4\%)&	Mobile networks&	Execution time reduction&	Only apply to a specific application
		\\ \cline{2-11}&
		\cite{210}&	Mobile sensing and prediction&	CNN&	Fair&	Fair&	Fair&	High (97\%)&	Mobile networks&	Efficient RAM and CPU usage&	Lack of practical sensing deployment
		\\ \cline{2-11}
		\hline
		\parbox[t]{1.5cm}{\multirow{14}{*}{\rotatebox[origin=c]{90}{Federated learning}}}& 
		\cite{213}&	Collaborative federated learning	&FL&	Low&	Low&	Low&	High (90\%)&	IoT networks&	Flexible reward scheme; improved privacy &	Lack of learning cost evaluation 
		\\ \cline{2-11}&
		\cite{214}&	Vehicular communications&	FL&	Low&	Low&	Low&	High&	V2V networks&	Reduction of exchanged data; low energy cost&Privacy risks of distributed learning
		\\ \cline{2-11}&
		\cite{218}&	Security monitoring	&FL&	Low&	Low&	Low&	High (95.6\%)&	IoT networks&	Fast learning process; efficient attack detection&	Hard to deploy in realistic scenarios
		\\ \cline{2-11}&
		\cite{219}	&Edge communication&	FL&	Fair&	Fair&	Low&	High (95\%)&	Edge computing&	High latency-reduction ratio&Risks of adversarial learning
		\\ \cline{2-11}&
		\cite{220}&	Edge communication & 	FL&	Fair&	High&	Fair&	High&	Edge computing&	Reduced aggregation error; high prediction accuracy&	High energy and bandwidth for on-device training
		\\ \cline{2-11}
		\hline
	\end{tabular}}
\end{table*}
\section{Data-provenance AI}
\label{Sec:Data-provenance_AI}
The final Wireless AI function of the data life cycle in Fig.~\ref{Fig:WirelessAI_Function} is Data-provenance AI where AI can provide various useful services, such as AI data compression, data privacy, and data security. We will focus on discussing the roles of AI for such important services that are shown in Fig.~\ref{Fig:Data-provenance_AI} 
\begin{figure}
	\centering
	\includegraphics [height=7.2cm,width=8.8cm]{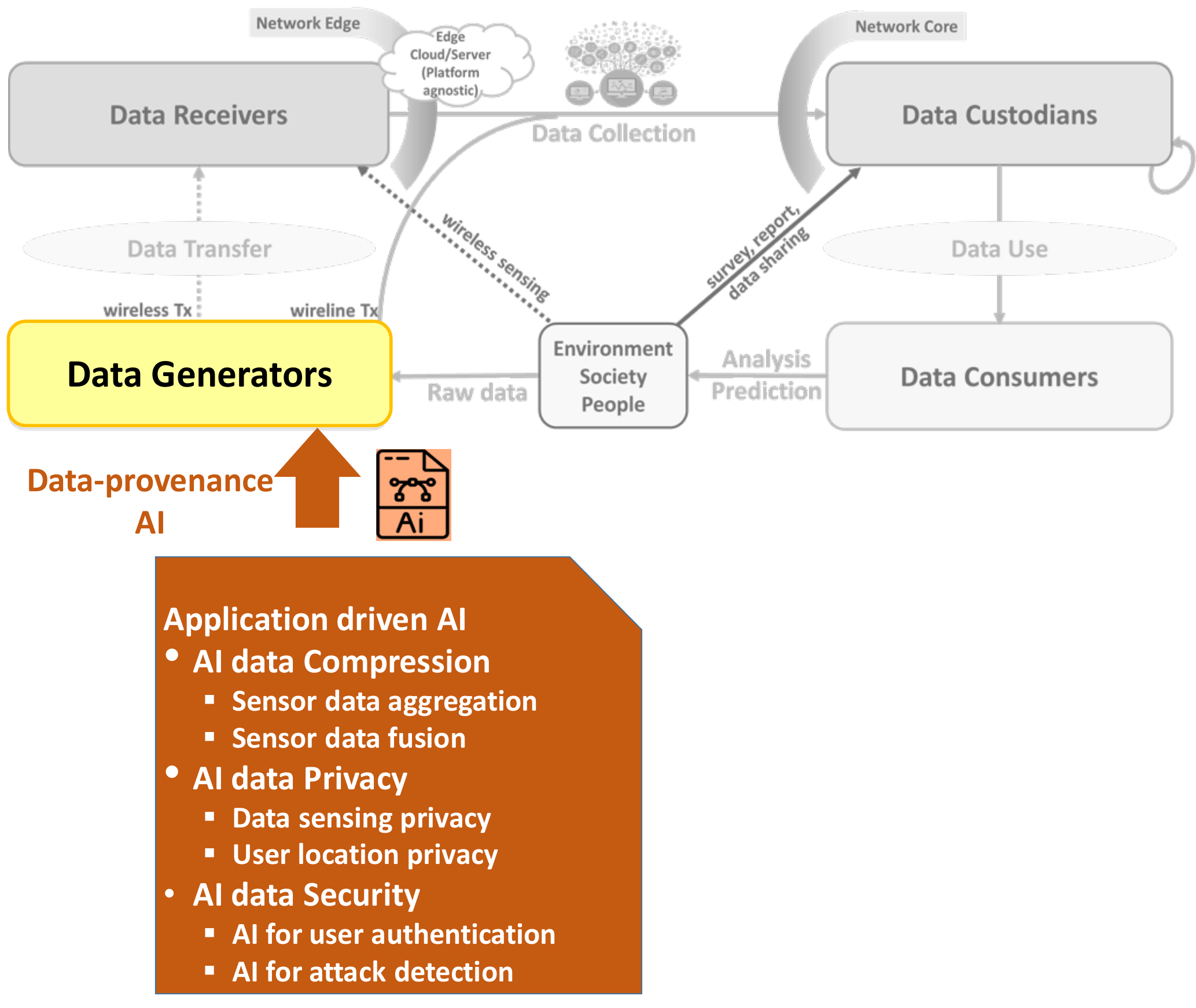}
	\caption{Data-provenance AI function. }
	\label{Fig:Data-provenance_AI}
	\vspace{-0.1in}
\end{figure}
\subsection{AI Data Compression }
We here analyze how AI can help data compression tasks in wireless networks.

\textbf{Problem formulation: }In the era of big data, e.g., big sensor data, sensor data needs to be compressed to mitigate network traffic and save resources of wireless networks. There are two important data compression tasks in wireless networks, including data aggregation and data fusion. Here, data aggregation is a process that aims to minimize the data packets from different sources (e.g., sensor nodes) to alleviate the transmit workload.  Due to the resource constraints of wireless sensor devices, data aggregation is of paramount importance for achieving efficient data collection while preserving energy resources on wireless networks. Meanwhile, in the data fusion, data from ubiquitous sensors can be fused to analyze and create a global view on the target subject. For example, in a smart home setting where sensors in each room collect data that represents the location information. 

\textbf{Drawbacks of conventional methods: } Many data compression algorithms have been proposed in \cite{23}, but they remain some unsolved issues. For example, a wireless sensor network often consists of heterogeneous sensor nodes with multi-media data associated with different statistical properties, which makes traditional data compression approaches using mathematical formulations inefficient. Further, in the future networks, the number of sensor nodes would increase rapidly with higher data complexity and high data volume, how to compress data with high scalability and robustness without loss of valuable data is a challenge.

\textbf{Unique advantages of using wireless AI techniques: } The major benefits of AI for data compression tasks in wireless networks are intelligence and scalability. Thanks to the online learning and prediction, AI can extract the useful information to compress the ubiquitous sensor data and identify the most important features that are necessary for later data processing. Moreover, AI can provide scalable data learning, data processing and data computation for a large dataset. The use of DNNs would overcome the limitations of current ML approaches to map and learn the big data collected from sensors.

\textbf{Recent works: }Some important works are highlighed with the exploitation of AI for data compression tasks, including sensor data aggregation and sensor data fusion. 
\begin{itemize}
	\item Sensor data aggregation: The work in \cite{279} presents a sensory data aggregation scheme for wireless sensor clusters. The main focus is on classifying the collected data using a supervised intrusion detection system. Here a hybrid ML system including a misuse detection and an anomaly detection system are designed to monitor the data aggregation process and detect data attacks. \textcolor{black}{To investigate the feasibility of the proposed approach, a knowledge discovery in data mining dataset is employed as the dataset for training and testing, which shows that ML can help achieve above 99\% detection rate and 99.8\% accuracy along with low training time. However, the analysis of energy consumption from the learning has not been provided.}
	 A new tree-based data aggregation framework is introduced in \cite{281} for sensor networks. Sensory data can be collected by an aggregator and then a part of data is transmitted to the sink associated with a function that represents the characteristics of all data, while the remaining data is processed in the sink for transmit energy savings. The classification of the aggregated data is implemented by a Bayesian belief network, which can achieve 91\% accuracy via numerical simulations. 
	
	The efficiency of data aggregation in wireless networks can be further improved with advanced ML/DL approaches \cite{282}. In this study, the focus is on privacy preservation for sensor data on the link between gateways and servers. Here, a DL autoencoder is used to ensure security for the sensor communication channel by reconstructing the obfuscated data from the original data, which can hidden data from unauthorized parties. \textcolor{black}{Although the proposed scheme can ensure sensor privacy preservation, it remains high computational complexity due to unoptimized learning parameters. }
	In some practical scenarios, aggregators may not be willing to collect data due to the lack of motivation (e.g., less aggregation revenues). To solve this issue, the work in \cite{283} implements a payment-privacy preservation scheme that enables aggregators to collect securely data while paying (e.g., rewards) to them. This process can formulated as a MDP problem and since the transition probabilities of payment is unknown to the aggregators, a model-free DRL approach using deep Q-learning is proposed to learn the model via trial and error with an agent. The reward here correlates with the optimal payment so that the agent explores the data aggregation network to find an optimal policy for maximizing revenues. 
	
	\item Sensor data fusion: The study in \cite{284} presents a multi-sensor data fusion that serves human movement classification based on ML. A network of heterogeneous sensors of a tri-axial accelerometer, a micro-Doppler radar and a camera is used to collect data that is then fused and classified by an ML algorithm with SVM and KNN. In \cite{285}, a sensor fusion architecture is proposed for object recognition. Parameters or information measured from sensors can be fused with those collected from other sensors to create a global view on the target subject, which helps evaluate it more comprehensively. To facilitate the fusion process, a processing model based on unsupervised ML is integrated that helps classify the features from the fused datasets. Experiments use camera images as the input data for training and classification, showing that ML can recognize the object well in different parameter settings. 
	
\end{itemize}

\subsection{	AI Data Privacy}
In this sub-section, we present how AI can help improve data privacy in wireless networks.

\textbf{Problem formulation: }At the network edge, data privacy protection is the most important service that should be provided. The dynamicity of the data collection, data transmisison and data distribution in future networks would put at risks of user's information leakage. For example, the data collection can expose sensor's information to third parties which makes privacy (e.g., privacy of user address, user name) vulnerable. Therefore, protecting data privacy is of paramount importance for network design and operations. By ensuring data privacy, a wireless network can appeal to a larger number of users which increases the utility of such a system. 

\textbf{Drawbacks of conventional methods: } Many existing approaches have been proposed to solve data privacy issues, mainly using encryption protocols. Although such techniques can keep the content of user's information private from the others and third parties, it is hard to apply to scalable wireless networks with large data. Further, they often apply a specific encryption protocol to a group of multi-media data collected from sensors and devices,  which is infeasible to future large-scale networks.

\textbf{Unique advantages of using wireless AI techniques: } AI potentially provides privacy functions for sensed data via classification and detection to recognize potential data threats. AI is also useful to user location privacy. In fact, AI can also support protection of user location privacy by building decision-making policies for users to adjust their location information and selecting privacy location preference.

\textbf{Recent works: }From the perspective of data generation process, we consider two AI data privacy domains, including data sensing privacy and user location privacy. 
\begin{itemize}
	\item Data sensing privacy: The work in \cite{290} concerns about privacy preservation for mobile data sensing. Differential privacy is used to protect the individual information during the data collection process. To authenticate the user identity based on their own data, a ML-based classifier approach using two-layer neural networks is designed. \textcolor{black}{It classifies the privacy sensitivity levels (0-9) and evaluates the privacy vulnerability with leakage possibility. However, no learning accuracy evaluation is provided.} 
	To overcome the limitations of existing encryption solutions in terms of computation complexity and the lack of ability to solving large-scale domains, a multi-functional data sensing scheme with awareness of different privacy is proposed in \cite{291}. The key objective is to provide flexible sensing functions and aggregation properties to serve different demands from different users. Besides, it is important to preserve the collected sensor data against adversarial attacks for better user information protection. In such contexts, the aggregation queries from sensors are collected as the dataset for training that is implemented by an ML algorithm, while the multifunctional aggregation can be performed by a Laplace differential privacy technique. Then, learning helps to estimate the aggregation results without revealing user privacy.  Meanwhile, the work in \cite{293} solves the data collection issue with privacy awareness using K-mean clustering. Indeed, to ensure data privacy, a technique of adding Laplace noise to the data compression process. Also, aggregated data is distributed over the mobile device network where each device holds a differentially-private sketch of the network datasets. 
	In the multimedia IoT networks, it is important to protect data privacy during the data sensing process. The work in \cite{294} proposes a cluster approach that divides sensor devices into different groups. Each cluster controller ensures the privacy of its sensor group, associated with a neural network that classifies the collected data and extracts useful features for data learning. Such data processing can be helped by using MEC services that not only provides low-latency computation but also enhances privacy for physical sensor devices [16]. A DL scheme for privacy protection and computation services for physical sensors can be seen in Fig.~\ref{Fig:DL_MEC_Privacy}.
	\item User location privacy: An example is in \cite{295} that introduces a privacy preservation scheme to help users to manage their location information. The key objective is to enable users to obtain the privacy preferences based on their choice factors such as QoS, trust levels. \textcolor{black}{Based on that, a k-learning model is built that can make decisions for users in preserving location privacy. However, the statistical analysis of learning latency is missing.}  
	The authors in \cite{296} are interested in privacy in mobile social networks. They first perform a real experiment to collect user location information in a campus network and investigate the possibility of location privacy leakage. Then, a privacy management scheme is designed to provide privacy control for mobile users. This model is empowered by a ML-based decision tree algorithm that learns the user privacy preference and classifies location contexts in the social ecosystem. 
	
	At present, the location of mobile devices is mostly identified by GPS technology, but this platform remains inherent limitations in terms of energy overheads and reliance on the third party (e.g., service providers), which poses user information at privacy risks. A new method presented in \cite{297} can achieve a user privacy location guarantee with energy savings. By using the features of current location that represents the privacy location preference, an ML approach is applied to learn the privacy model, aiming to classify privacy preferences depending on whether the user want or not want to share its location information based on their activities. It is noting that capturing features only relies on sensors available on smart phones without using GPS that is power saving and eliminates the need of third party. To preserve user location privacy in social networks, the work in \cite{299} jointly takes context factors, e.g., occasion and time, and individual factors into considerations to establish a privacy location preservation protocol. A ML-based regression model is involved to evaluate the privacy levels based on such features, showing that the occasion factor dominates in determining the privacy degree during the location sharing. 
	
\end{itemize}
\begin{figure}
	\centering
	\includegraphics [width=0.95\linewidth]{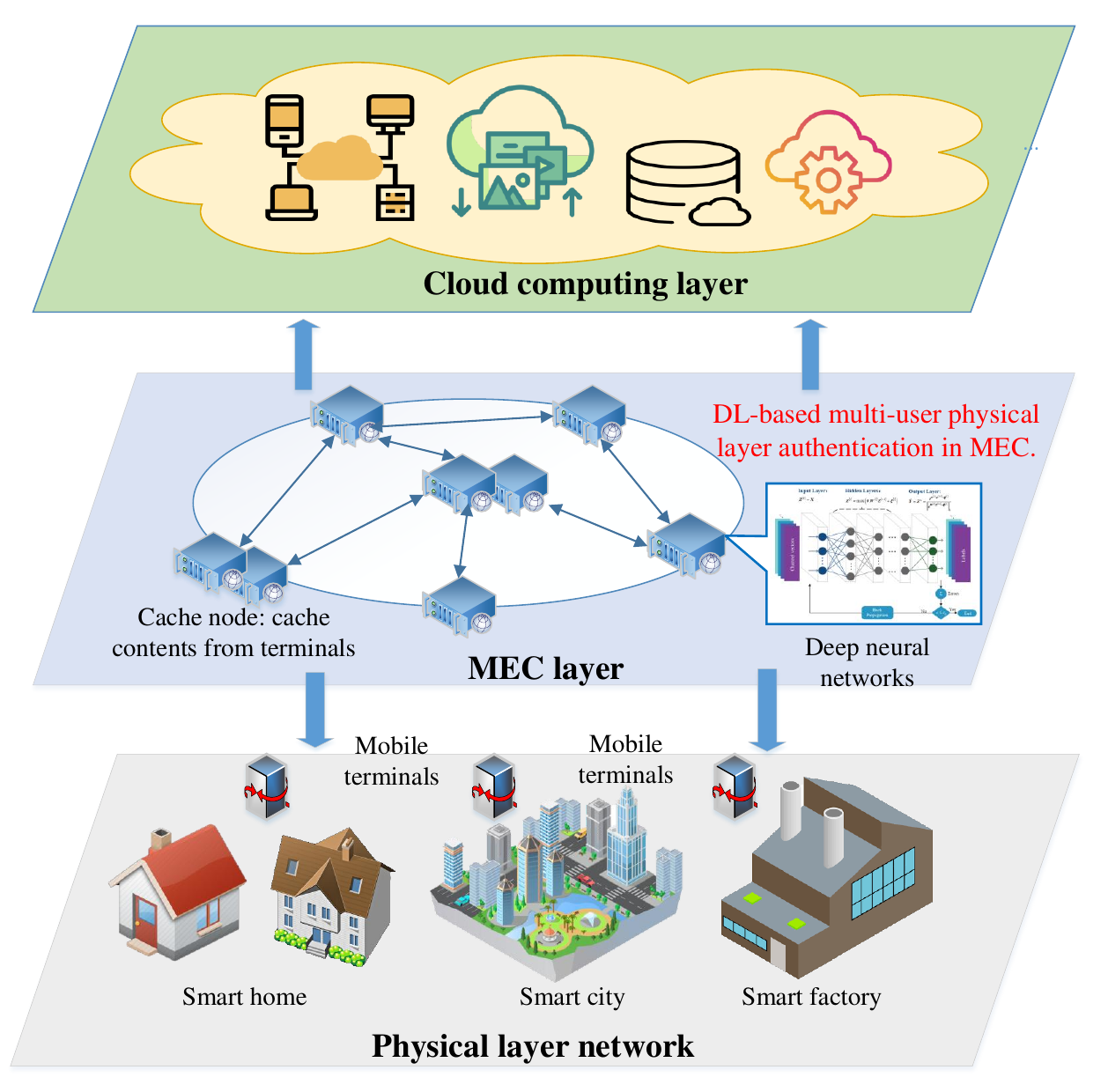}
	\caption{DL for MEC privacy preservation. }
	\label{Fig:DL_MEC_Privacy}
	\vspace{-0.15in}
\end{figure}

\subsection{	AI Data Security }
In addition to privacy protection, AI also has the great potential for enhancing security for wireless networks. We here present how AI can help to protect data security.

\textbf{Problem formulation: } Data security is the top research concern in wireless network design. A wireless network system is unable to work well and robust under the presence of security threats and attacks. The openness and dynamicity of wireless networks makes them prone to external data threats, while unauthorized users can access the resource of the network. 

\textbf{Drawbacks of conventional methods: } Wireless networks have gained increasing influences on modern life that makes cyber security an significant research area. Despite installed defensive security layers, the wireless networks inherit cyber attacks from the IT environments. Many security models have been proposed, mainly using middle boxes such as Firewall, Antivirus and Intrusion Detection Systems (IDS) \cite{500}. These solutions aims to control the traffic flow to a network based on predefined security rules, or scan the system for detecting malicious activities. However, many traditional security mechanisms still suffer from a high false alarm rate, ignoring actual serious threats while generating alerts for non-threatening situations. Another issue of the existing security systems is the impossibility of detecting unknown attacks. In the context of increasingly complex mobile networks, the human-device environments change over time, sophisticated attacks emerge constantly. How to develop a flexible, smart and reliable security solution to cope with various attacks and threats for future wireless networks is of paramount importance for network operators. 

\textbf{Unique advantages of using wireless AI techniques: }
AI has been emerged as a very promising tool to construct intelligent security models. Thanks to the ability of learning from massive datasets, AI can recognize and detect attack variants and security threats, including unknown ones in an automatic and dynamic fashion without relying on domain knowledge. Also, by integrating with on-device learning algorithms, it is possible for AI to recognize the unique features of a certain user (e.g., through AI classifiers), aiming for the user classification against threats.

\textbf{Recent works: }We here introduce some important security services that AI can offer, including user authentication and attack detection.
\begin{itemize}
	\item AI for user authentication: 
	Some recent works are interested in exploiting AI for user authentication in wireless networks. An example in \cite{300} introduces a user authentication model using millimetre sensors. Each radar sensor is equipped with four receive antennas and two transmit antennas. Signals that re reflected from a part of human body are analyzed and extracted to produce a feature dataset that is then fed to a ML-based random forest classifier. \textcolor{black}{Through training and learning, the classifier can recognize the biometric properties of any individuals with a 98\% accuracy and high identification rate. In the future, this model should be extended to make it well adaptive to a more general network setting, e.g., IoT user authentication.} In \cite{301}, a user authentication scheme is proposed based on CSI measurements working for both stationary and mobile users. For the former case, the authors build a user profile with considerations of spoofing attacks, while for the later case, the correlations of CSI measures are taken to implement user authentication checking. Specially, an authenticator is created that runs an ML algorithm to learn from the CSI profiles and determine whether the CSI value measured in the coming packet matches with the one created in the profile storage. If they match, the authentication is successful, otherwise that user is denied for further processing.
	
	The Wi-Fi signals can be exploited to build datasets of learning for user authentication. The authors in \cite{302} try the first time for a new user authentication system using human gait biometrics from Wi-Fi signals. Here, CSI value samples are collected from Wi-Fi devices. The core component of the scheme is a 23-layer DNN network that is modelled to extract the important salient gait features collected from gait biometrics in the CSI measurement dataset. The DNN acts as a classifier to extract unique features of a user and identify that user among the mobile user set. As a different viewpoint of user authentication, the work in \cite{305} analyzes CSI data for user authentication in wireless networks. First, a CNN classifier is built to learn the extracted features from CSI dataset such that pooling layers are utilized to simplify the computation process for faster learning. To analyze CSI features, an RNN network is then designed, including some recurrent layers and a fully connected layer where user recognition is implemented. The combination of CNNs and RNN models leads to a new scheme, called CRNN, aiming to use the capability of information representation of CNN and information modelling of RNNs. \textcolor{black}{Despite complex CNN layer settings and high learning time, the rate for user authentication is improved  significantly,  with 99.7\%  accuracy.}
	\item AI for attack detection: 
	The work in \cite{503} introduces a scalable IDS for classifying and recognizing unknown and unpredictable cyberattacks. The IDS dataset collected from a large-scale network-level and host-level events is processed and learned by a DNN architecture that can classify the network behaviours and intrusions. Compared to commercial IDSs that mainly rely on physical measurements or defined thresholds on feature sets to realize the intrusions, learning can assist flexible IDS without any further system configurations, and work well under varying network conditions such as network traffic, user behaviour, data patterns. \textcolor{black}{The remaining issue of this detection system is the complex multi-layer DNN learning procedure that results in high computational latency.}
	Another solution in \cite{504} formulates an intrusion detection system using RNNs. The main focus is on evaluating the detection rate and false positive rate. The intrusion detection can be expressed equivalently to a multi-class classification problem for recognizing whether the traffic behaviour in the network is normal or not. The key motivation here is to enhance the accuracy of the classifiers so that intrusive network behaviours can be detected and classified efficiently. Motivated by this, a RNN-based classifier is leveraged to extract abnormal traffic events from the network features using the well-known NSL-KDD dataset. Specially, the attacks can be divided into four main types, including DoS (Denial of Service attacks), R2L (Root to Local attacks), U2R (User to Root attack), and Probe (Probing attacks). The results of deep training and learning show a much better classification accuracy of RNNs in comparison with ML techniques.
	
	In addition to that, wireless networks also have to solve spoofing attacks where an attack claims to be a legitimate node to gain malicious access or perform denial of service attacks. Conventional security solutions, such as digital signatures, cannot fully address the spoofing issues, especially in the dynamic environment with unknown channel parameters. Recently, AI gains interest in spoofing detection tasks. As an example, the work in \cite{509} exploits the potential of ML to detect spoofing attacks in wireless sensor networks. By analyzing the correlation of signal strength on the mobile radio channel, the learning algorithm receives signal strength samples as the input data for abnormal behaviour analysis. To enhance the detection accuracy, the collected samples can be categorized into small windows that may include the attacking nodes and legitimate nodes. Then, a collaborative learning scheme with k-NN and k-means is applied to identify the spoofing attacks in the dense networks. Meanwhile, the authors in \cite{510} use RL to implement a spoofing detection model from the PHY layer perspective for wireless networks. The key concept is to exploit the CSI of radio packets to detect spoofing vulnerabilities. The spoofing authentication process is formulated as a zero-sum game including a receiver and spoofers. Here, the receiver specifies the test threshold in PHY spoofing detection, while spoofers select attack frequency to optimize its utility with respect to Bayesian risks. Specially, the threshold selection is achieved by trial and error process using Q-learning without knowing CSI such as channel variations.  
\end{itemize}

\subsection{Lessons Learned} The main lessons acquired from the survey on the	Data-provenance AI domain are highlighted as the following.
\begin{itemize}
	\item	Recent studies also raise the discussion on the AI adoption for data-driven applications during the data generation process. They mainly concentrate on four specific domains, namely AI data compression, data clustering, data privacy, and data security. As an example, AI has been applied for supporting sensor data aggregation and fusion, by classifying useful information from the collected data samples \cite{279}. Also, the classification of AI helps monitor the data aggregation process, aiming to analyze the characteristics of all data and detect data attacks \cite{281}. 
	\item	At the network edge, usage privacy and data security protection are the most important services that should be provided. From the perspective of data generation process, the applications of AI in data privacy are reflected in two key domains, including data sensing privacy and user location privacy. Currently, AI-based data classification algorithms can be implemented on each local device. This solution not only distributes learning workload over the network of devices, but also keeps data private due to no need to transfer data on the network. Interestingly, recent studies in \cite{295}, \cite{296} show that ML can help learn the user privacy preference and classify location contexts in the social ecosystem.
	\item	In addition to privacy protection, AI also has the great potential for enhancing security for wireless networks. In the literature, AI mainly provide some important security services, including user authentication and access control. In practical wireless networks, the authentication between two users on the physical network layer based on traditional approaches is challenging due to the time variance of channels and mobility of users. AI may be an ideal option to recognize legitimate users from attackers by training the channel data and perform classification. 
	\item \textcolor{black}{ In general, in the Data-provenance AI domain, some ML algorithms such as SVM \cite{284} and random forest \cite{300} achieve superior performances in terms of high learning accuracy, low computational latency, and low training time. However, in complex network scenarios with huge datasets, advanced ML techniques can achieve better performances in learning data and extracting data features, which thus improves the learning efficiency, such as CNNs in user authentication \cite{305} and RNNs in intrusion detection \cite{504}.}
\end{itemize}
\textcolor{black}{In summary, we shows the classification of the Data-provenance AI domains in Fig.~\ref{Fig:05TaxonomyData-provenance} and summarize the AI techniques used along with their pros and cons in the taxonomy Table~\ref{Table:Data-provenance_AI}. }
\begin{figure}
	\centering
	\includegraphics [width=0.95\linewidth]{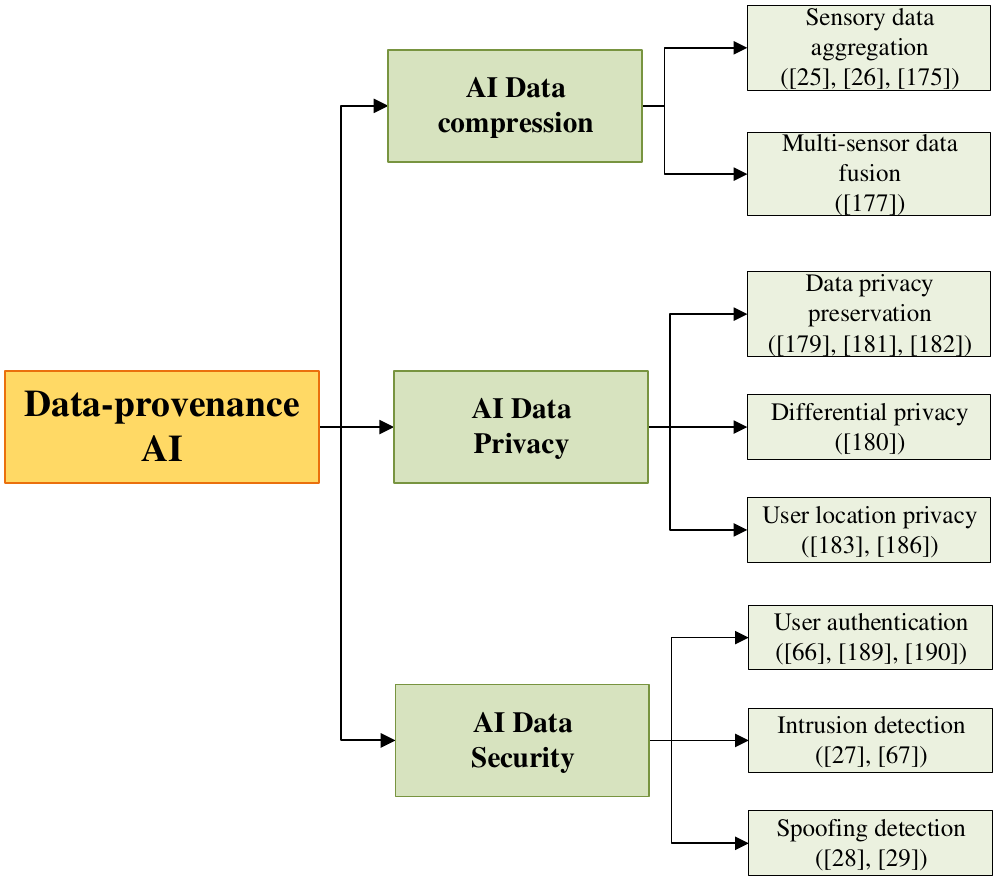}
	\caption{\textcolor{black}{Summary of Data-provenance AI domain.}}
	\label{Fig:05TaxonomyData-provenance}
	\vspace{-0.1in}
\end{figure}

\begin{table*}
	\centering
	\caption{{\color{black}Taxonomy of Data-provenance AI use cases.  }}
	{\color{black}
	\label{Table:Data-provenance_AI}
	\begin{tabular}{|P{0.4cm}|P{0.4cm}|P{1.7cm}|P{1.3cm}|P{1.2cm}|P{1cm}|P{1cm}|P{1cm}|P{1.3cm}|P{2.2cm}|P{2cm}|}
		\hline
		\textbf{Issue}& 	
		\textbf{Ref.} &	
		\textbf{Use case}&	
		\textbf{AI Algorithm}& 
		\textbf{Complexity}& 	
		\textbf{Training Time }&
		\textbf{Training Data}&
		\textbf{Accuracy}&
		\textbf{Network}&
		\textbf{Pros}&
		\textbf{Cos}
		\\
		\hline
		\parbox[t]{1.5cm}{\multirow{11}{*}{\rotatebox[origin=c]{90}{AI Data compression}}} & 
		\cite{279}&	Sensory data aggregation&	Supervised ML&	Fair&	Low&	Low&	High
		(99.8\%)&	Sensor networks&	Achieve above 99\% intrusion detection rate&	Lack of energy consumption analysis
		\\ \cline{2-11}&
		\cite{281}&	Sensory data aggregation&	Bayesian belief network&	Low&	Low&	Low&	High
		(91\%)&	Sensor networks&	Simple data sensing and learning approach&	Vulnerable data aggregation with attacks
		\\ \cline{2-11}&
		\cite{282}&	Data aggregation&	DL autoencoder&	Fair&	Fair&	Fair&	Fair&	Sensor networks&	Improved sensor privacy &Complex DL structure
		\\ \cline{2-11}&
		\cite{284}&	Multi-sensor data fusion&	SVM and KNN&	Low&	Low&	Low&	High
		(91.3\%)&
		Sensor networks&High classification accuracy& No practical sensor implementation
		\\ \cline{2-11}
		\hline
		\parbox[t]{1.5cm}{\multirow{15}{*}{\rotatebox[origin=c]{90}{AI Data Privacy}}}& 
		\cite{290}&	Data privacy preservation&	Neural networks&	Low&	Low&	Low&	-&	Mobile networks&	Provide trust for mobile sensing systems&	No learning accuracy evaluation
		\\ \cline{2-11}&
		\cite{291}&	Differential privacy&	Neural networks&	Low&	Low&	Fair&	High&	Fog computing&	Less learning complexity&Need practical model evaluation 
		\\ \cline{2-11}&
		\cite{293}&	Privacy awareness&	K-mean clustering&	Low&	Low&	Low&	-&	Wireless network&	High privacy for learning; low computing cost&	Complex system settings
		\\ \cline{2-11}&
		\cite{294}&	Data privacy&	Neural networks&	Fair&	Low&	Low&	-&	IoT networks&	High data collection privacy&	No realistic tests for evaluation 
		\\ \cline{2-11}&
		\cite{295}&	User location privacy&	k-learning &	Low&	Low&	Low&	Fair&	Wireless network&	Achieve different user privacy &	Lack of latency investigation 
		\\ \cline{2-11}&
		\cite{299}&	Privacy location preservation&
		Online ML&	Fair&	Low&	Low&	High&	Wireless network&	High location prediction accuracy &	No experiment analysis
		\\ \cline{2-11}
		\hline
		\parbox[t]{1.5cm}{\multirow{17}{*}{\rotatebox[origin=c]{90}{AI Data Security}}}& 
		\cite{300}&	User authentication&	Random forest &	Low&	Low&	Low&	High (98\%)&	Sensor networks&	Fast signal response; high identification rate&	Only use for a specific application
		\\ \cline{2-11}& 
		\cite{302}&	User authentication&	Multi-layer DNN&	Fair&	Fair&	Fair&	High (87.8\%)&	Wi-Fi networks&	Practical implementation&	Complex learning setup 
		\\ \cline{2-11}&
		\cite{305}&	User authentication&	CNN&	High&	High&	Fair&	High (99.7\%)&	Wireless networks&	Friendly to user networks &	Complex CNN layer settings
		\\ \cline{2-11}&
		\cite{503}&	Intrusion detection&	DNN&	High&	Fair&	Fair&	-&	Wireless networks&	Scalable intrusion detection&Complex model update
		\\ \cline{2-11}&
		\cite{504}&	Intrusion detection&	RNN&	Fair&	Low&	Low&	High (99.8\%)&	Wireless networks&	Fast learning rate&	Simple learning procedure
		\\ \cline{2-11}&
		\cite{509}	&Spoofing detection&	k-NN and k-means&	Low&	Fair&	Low&	High (90\%)&	Sensor networks&	Simple training; high attack rate&	Hard to implement
		\\ \cline{2-11}&
		\cite{510}	&Spoofing detection&	RL&	Low&	Low&	Low&	Fair&	IoT networks&	Efficient spoofing detection&	No practical IoT implementation
		\\ \cline{2-11}
		\hline
	\end{tabular}}
\end{table*}

\section{Research Challenges and Future Directions}
\label{Sec:Challenges}
	Different approaches surveyed in this paper clearly show that AI plays a key role in the evolution of wireless communication networks with various applications and services. However, the throughout review on the use of AI in wireless networks also reveals several critical research challenges and open issues that should be considered carefully during the system design. Further, some future directions are also highlighted. 
\subsection{Research Challenges}
We analyze some important challenges in wireless AI from three main aspects: security threats, AI limitations, and complexity of AI wireless communications.
\subsubsection{Security Threats}
Security is a major concern in wireless AI systems. Data collected from sensors and IoT devices are trained to build the model that takes actions for the involved wireless applications, such as decision making for channel allocation, user selection, or spectrum sharing. Adversaries can inject false data or adversarial sample inputs which makes AI learning invalid. Attackers can also attempt to manipulate the collected data, tamper with the model and modify the outputs \cite{332}. For instance, in reinforcement learning, the learning environment may be tampered with an attacker so that the agent senses incorrectly the information input data. Another concern can stem from the manipulation on hardware implementation by modifying hardware settings or re-configuring system learning parameters. The work in \cite{333} investigated the multi-factor adversarial attack issues in DNNs with input perturbation and DNN model-reshaping risks which lead to wrong data classification and then damages the reliability of the DNN system. 

\textcolor{black}{In these contexts, how to ensure high security for wireless AI systems is highly important. The authors in \cite{challenge1} propose a defense technique against adversaries in learning dataset samples in DNN algorithms. The learning datasets are verified to detect any adversarial samples through a testing process with a classifier that can recognize and remove the unwanted noise or adversarial data. Another solution is in \cite{challenge2} where a security protection scheme is integrated with the DL training algorithms for attack detection. A new training strategy with a learner is proposed with a backpropagation algorithm to train the benign examples against each training adversarial example.}
{\color{black}\subsubsection{	Privacy Issues}
Privacy is another growing concern for AI systems. In fact, the popularity of searchable data repositories in centralized AI architectures, e.g., AI functions in the cloud, has introduced new risks of data leakage caused by untrusted third parties or central authority. Private information in the datasets such as healthcare data, user addresses, or financial information may become the target of data attacks \cite{privacy1}. Moreover, in the context of big data, the ease of access to large datasets and computational power (e.g., GPUs) to ubiquitous users and organizations poses serious privacy concerns for AI model architectures, e.g., data loss or parameter modification, as AI models such as DNNs are used in different aspects of our lives. How to ensure high privacy preservation without compromising the training performance is an important issue to be considered when integrating AI into wireless networks. 

A possible solution is to make the AI functions distributed, by using the FL concept that enables AI model learning at user devices without compromising data privacy \cite{privacy2}. The information transferred in FL contains minimal updates to improve a particular ML model while the raw datasets are not required by the aggregator, making the training data including sensitive user information safer against data threats. Another promising approach to enhance AI privacy is  differential privacy which brings attractive properties such as privacy preservation, security, randomization, and stability \cite{privacy3}. Several attempts have been taken to apply differential privacy to AI/ML, in order to ensure high privacy of training datasets, by combining its strict mathematical proof with flexible composition theorems. For example, differential privacy mechanisms can be applied at the input layer, hidden layers or output layer of a DNN architecture, by adding noise to the gradient of neural layers to achieve protection for training datasets and prevent an adversary from extracting  accurate personal information of the training data in the output \cite{privacy3}. These solutions not only provide stability for model optimization but also protect data privacy during the training and learning process.}

\subsubsection{	Limitations of AI Algorithms}
\textcolor{black}{While AI shows impressive results in wireless communication applications, its models still have performance limitations.  A critical issue is the high cost of training AI models. Indeed, it is not uncommon to take some days for AI training on mobile devices or even edge devices with limited computation resources. The computational complexity and long training time will hinder the application of AI in practical wireless networks. Thus, optimizing data training and learning is of paramount importance for AI wireless systems.} A more challenge is the scarcity of real datasets available for wireless communications necessary to for the AI training, i.e. in supervised learning \cite{336}. In many wireless scenarios where only a small dataset is not sufficient for the training which degrades the performance of AI model in terms of low classification or prediction accuracy, and high model overfitting, and thus the confidence of AI model may not be guaranteed. This drawback would be an obstacle for deployment of AI in real-world wireless applications since the feasibility of AI should be demonstrated in testbeds or experiments in the real environments.

\textcolor{black}{Some solutions have been proposed to improve the performance of AI algorithms. For instance, an improved DNN model called DeepRebirth is introduced in \cite{challenge3} for mobile AI platforms. A streamlined slimming architecture is designed to merge the consecutive non-tensor and tensor layer vertically which can significantly accelerate the model execution and also greatly reduce the run-time memory cost since the slimmed model architecture contains fewer hidden layers. Through simulations, the proposed DeepRebirth model is able to achieve 3x speed-up and 2.5x memory saving on GoogLeNet with only 0.4\% drop on top-5 accuracy on the ImageNet validation dataset. In terms of wireless datasets, the collection of data from testbeds and experiments is an important step for facilitating wireless AI research. Some wireless AI data sources such as the massive MIMO dataset in \cite{challenge4} and the IoT trace dataset in \cite{challenge5} would help researchers implement and evaluate their learning and validation models. }

\subsubsection{Complexity of AI Wireless Networks}
The complexity may stem from wireless data and network architecture. In fact, in AI wireless communication networks, the data is produced from both application traffic and information of devices and objects. Considering the complex wireless network with various data resources, it is challenging to extract data needed for the AI training of a specific task. \textcolor{black}{In cognitive network scenarios, for example, how to label massive learning data for training AI models is also highly challenging. Moreover, in the big data era, the data sources generated from wireless systems usually refer to extremely large, heterogeneous and complex (semi-structured and unstructured) data-sets, which may not be handled by the conventional data processing methodologies \cite{challenge6}. Therefore, developing adaptive AI models that enable flexible data learning in dynamic and large-scale network scenarios is highly desirable.} From the network architecture perspective, the traditional protocol of wireless communication is based on a multi-layer network architecture. However, currently it is difficult to determine which and how many layer should be configured to run AI functions and deploy AI technologies in devices. Therefore, it is essential to consider intelligent protocol architectures with the simplified and adaptive capability according to AI functions and requirements \cite{337}.

\textcolor{black}{Recently, some approaches have been presented to solve complexity issues in wireless AI networks. As an example, the study in \cite{challenge7} suggests a proactive architecture that exploits historical data using machine learning for prediction of complex IoT data streams. In particular, an adaptive prediction algorithm called adaptive moving window regression is designed specific to time-varying IoT data which enables adaptive data learning with the dynamics of data streams in realistic wireless networks, e.g., vehicular control systems. Further, a flexible DNN architecture is designed in \cite{challenge8} that enables automatic learning of features from simple wireless signal representations, without requiring design of hand-crafted expert features like higher order cyclic moments. As a result, the complex multi-stage machine learning processing pipelines can be eliminated, and the wireless signal classifiers can be trained effectively in one end-to-end step. This work is expected to open a door for designing intelligent AI functions in complex and dynamic wireless networks.}

\subsection{Future Directions }
We discuss some of the future research directions on wireless AI motivated from the recent success in the field. 
\subsubsection{Specialized AI Architecture for Wireless Networks}
The AI architecture has significant impacts on the overall performance of the wireless AI systems, while the current AI model design is still simple. Many AI architectures proposed in \cite{21}, \cite{22}, \cite{24}, \cite{26} are mostly relied on conventional algorithms with learning iterations, data compression, decoding techniques. The performance of wireless AI algorithms can be improved thanks to the recent AI advances with sub-block partition solutions for neural networks \cite{703} or data augmentation for DL \cite{339}. However, these AI approaches suffer from dimensionality problems when applying to wireless applications due to the high complexity of the wireless systems, and data training would be inefficient when the network dimension is increased with the increase of mobile devices and traffic volumes. Therefore, developing adaptive AI architectures specialized in wireless communication scenarios is the key for empowering future intelligent applications. As suggested by \cite{340}, by adapting with the dynamicity of wireless communications, AI can model effectively the wireless systems and discover the knowledge of data generated from network traffic and distributed devices. Besides, motivated by \cite{24}, \cite{26}, AI functions should be implemented and deployed in the wireless network in a distributed manner. This architecture not only influences the wireless data transmission infrastructure, but also modernizes the way that wireless communications are organized and managed.

Furthermore, with the ambition of deploying AI on mobile devices/edge devices to extend the network intelligence to local clients, developing specialized hardware/chipsets for AI is vitally important. Recent advances in hardware designs for high-bandwidth CPUs and specialized AI accelerators open up new opportunities to the development of AI devices at the edge of the network \cite{342}. Motivated by this, recently Facebook bring ML inference to the mobile devices and edge platforms \cite{344} thanks to specialized AI functions with high data training speed and low energy costs. The work in \cite{345} also introduces a DL inference runtime model running on embedded mobile devices. This scheme is able to support data training and MIMO modelling using a hardware controller empowered by mobile CPUs with low latency and power consumption with bandwidth savings.
\subsubsection{From Simulation to Implementation}
Reviewing the literature, most of the AI algorithms conceived for wireless communication networks are in their simulation stages. Thus, the researchers and practitioners needs to further improve AI functions before they are realized in real-world wireless networks. We here propose some solutions that can be considered in practical wireless AI design. 

First, real-world datasets collected from physical communication systems should be made available for ready AI function implementation. The sets of AI data generated from simulation environments such as deepMIMO dataset \cite{346} can contain the important network features to perform AI functions, but they are unable to reflect exactly the characteristics of a real wireless system. \textcolor{black}{ Moreover, real data collected from mobile networks in civil and social networks can be valuable to wireless AI research. For example, the dataset in \cite{challenge9} provided by Shanghai Telecom that contains 7.2 million network access records of 9481 mobile phones using 3233 base stations can be useful to train and test the wireless AI models. } 

Second, as discussed in the previous section, designing specialized hardware/chipsets for AI is extremely important to deploy AI applications on edge/mobile devices. These new hardware architectures should provide functions specializing AI to support on-device training, modelling and classification at the edge of the network. \textcolor{black}{For example, the recent experiments presented in \cite{challenge10} show the promising performance of AI-specific mobile hardware, including Intel and Nvidia platforms using the standard TensorFlow library, in terms of reduced data training latency and improved learning accuracy.} 

Third, the multidimensional assessment for wireless AI applications in various scenarios is also vitally necessary to realize the feasibility of the AI adoption in practice. By comparing with simulation performances, the designers can adjust the model parameters and architecture settings so that AI can well fit the real channel statistical characteristics.

\subsubsection{	AI for 5G and Beyond}
The heterogeneous nature of 5G and beyond wireless networks features multi-access ultra-dense networks with high data transmission capability and service provisions. AI has been regarded as the key enabler to drive the 5G networks by providing a wide range of services, such as intelligent resource allocation, network traffic prediction, and smart system management. For example, the integration of AI enables network operators to realize intelligent services, such as autonomous data collection across heterogeneous access networks, dynamic network slicing to provide ubiquitous local applications for mobile users, and service provisioning management \cite{346}. Wireless AI also acts as AI-as-a-service that can reap the potentials of Massive MIMO, a key 5G technology. For example, AI can be used to predict the user distribution in MIMO 5G cells, provide MIMO channel estimation, and optimize massive MIMO beamforming \cite{347}. With the rapid development of edge 5G services, AI would potentially transform the way edge devices perform 5G functionalities. In fact, the network management and computation optimization of mobile edge computing (MEC) would be simplified by using intelligent AI software through learning from the historical data. Lightweight AI engines at the 5G devices can perform real-time mobile computing and decision making without the reliance of remote cloud servers.

Beyond the fifth-generation (B5G) networks, or so-called 6G, will appear to provide superior performance to 5G and meet the increasingly service requirements of future wireless networks in the 2030s. The 6G networks are envisioned to provide new human-centric values with transformative services, such as quantum communications, wearable computing, autonomous driving, virtuality, space-air-ground networks, and the Internet of Nano-Things \cite{348}.  AI technologies are expected to play an indispensable role in facilitating end-to-end design and optimization of the data processing in 6G wireless networks. The future cellular networks will be much more complex and dynamic, and the number of parameters to be controlled will increase rapidly. In such scenarios, AI can provide smart solutions by its self-learning and online optimization capabilities to support advanced data sensing, data communication, and signal processing \cite{350}. These benefits that AI bring about would accelerate the deployment of future 6G networks and enable innovative wireless services. In summary, the research challenges and future directions are highlighted in Table~\ref{Table:Challenge}. 

The application of AI in wireless is still in its inception and will quickly mature in the coming years for providing smarter wireless network services. The network topology model and wireless communication architecture will be more complex with the dynamicity of data traffic and mobile devices. AI is expected to play a key role in realizing the intelligence of the future wireless services and allow for a shift from human-based network management to automatic and autonomous network operations. Therefore, the integration of AI in wireless networks has reshaped and transformed the current service provisions with high performance. This detailed survey is expected to pave a way for innovative research and solutions for empowering the next generation of wireless AI. 

	\begin{table*}
	\centering
	\caption{{\color{black}Summary of research challenges and future directions for Wireless AI. } }
	\label{Table:Challenge}
	{\color{black}
	\begin{tabular}{|P{3cm}|P{7cm}|P{7cm}|}
		\hline
		\multicolumn{3}{|c|}{\cellcolor{blue!25}\textbf{Research challenges for Wireless AI}}\\
		\hline
		\centering \textbf{Issue}& 	
		\centering \textbf{Description}& 
		\textbf{Possible solutions}	
		\\
		\hline
		\multirow{7}{1.5cm}{Security threats} & 
		\begin{itemize}
			\item Security is a major concern in wireless AI systems. Adversaries can inject false data or adversarial sample inputs which makes AI learning invalid.
			\item Attackers also manipulate the collected data, tamper with the model and modify the outputs \cite{332}.
		\end{itemize}& 
		\begin{itemize}
		\item 	Defence techniques are proposed to detect any adversarial samples in DNN learning \cite{challenge1}.
		\item	Security protection solutions are integrated with the DL training algorithms for attack detection \cite{challenge2}.
		\end{itemize}
		\\ \cline{2-3}
		\hline
		\multirow{7}{1.5cm}{Privacy issues} & 
		\begin{itemize}
			\item 	Privacy is another growing concern for AI systems, caused by centralized AI architectures \cite{privacy1}.
			\item	The ease of access to large datasets and computational power (e.g., GPUs) in AI learning has posed serious data privacy concerns. 
		\end{itemize}& 
		\begin{itemize}
			\item 	The use of FL enables AI model learning at user devices without compromising data privacy. \cite{privacy2}.
			\item	Differential privacy mechanisms are also very useful to achieve protection for training datasets \cite{privacy3}. 
		\end{itemize}
		\\ \cline{2-3}
		\hline
		
		\multirow{7}{1.5cm}{Limitations of AI algorithms} & 
		\begin{itemize}
			\item 	A critical issue is the high cost of training the AI model, i.e., computational complexity and long training time.
			\item	Another limitation is the scarcity of real datasets available for AI wireless implementation.
		\end{itemize}& 
		\begin{itemize}
			\item 	Improving AI models with efficient learning architecture are necessary, i.e., accelerated DNN model \cite{challenge3}. 
			\item 	More wireless AI data sources should be collected, e.g., the MIMO dataset \cite{challenge4} and the IoT trace dataset \cite{challenge5}. 
		\end{itemize}
		\\ \cline{2-3}
		\hline
		
		\multirow{7}{1.5cm}{Complexity of AI wireless networks} & 
		\begin{itemize}
			\item 	In the big data era, the data sources generated from wireless systems usually refers to extremely large, heterogeneous and complex datasets \cite{challenge6}.
			\item	The complex architecture in wireless network makes it challenging to design efficient wireless AI functions.
			
		\end{itemize}& 
		\begin{itemize}
			\item 	A proactive architecture is proposed which exploits historical data using machine learning for prediction of complex IoT data streams \cite{challenge7}.
			\item	A flexible DNN architecture is designed in \cite{challenge8} that enables feature learning from wireless signal representations. 
			
		\end{itemize}
		\\ \cline{2-3}
		\hline
		\hline
		\multicolumn{3}{|c|}{\cellcolor{blue!25}\textbf{Future directions for Wireless AI}}\\
		\hline
		\centering \textbf{Issue}& 	
		\centering \textbf{Description}& 
		\textbf{Possible solutions}	
		\\
		\hline
		\multirow{7}{1.5cm}{Specialized AI architecture for wireless networks} & 
		\begin{itemize}
			\item The current AI model design is still simple that needs to be improved. 
			\item	The AI learning may be inefficient due to the network complexity and the increased network dimension with the increase of mobile devices and traffic volumes.
		\end{itemize}& 
		\begin{itemize}
			\item 		Developing adaptive AI architectures specialized in wireless communication scenarios is the key for empowering future intelligent applications \cite{340}.
			\item	AI functions should be implemented and deployed in the wireless network in a distributed manner for privacy and reduced network traffic \cite{26}. 
			
		\end{itemize}
		\\ \cline{2-3}
		\hline
		
		\multirow{7}{1.5cm}{From simulation to implementation} & 
		\begin{itemize}
			\item 	Real-world datasets collected from physical communication systems should be made available for ready AI function implementation.
			\item 	Designing specialized hardware/chipsets for AI is extremely important for running AI functions. 
			\item 	It is necessary for  multidimensional assessment of wireless AI functions to make them suitable for practical network scenarios.
			
		\end{itemize}& 
		\begin{itemize}
			\item 	The real data collected from mobile networks in civil and social networks can be valuable to wireless AI research \cite{challenge9}.
			\item	Some hardware platforms including Intel and Nvidia have been provided using the standard TensorFlow library for mobile AI deployment \cite{challenge10}. 
			\item	The designers should perform both simulations and practical implementations to adjust the AI model parameters and architectures based on real-world settings. 
			
		\end{itemize}
		\\ \cline{2-3}
		\hline
		
		\multirow{7}{1.5cm}{AI for 5G and beyond} & 
		\begin{itemize}
			\item 	AI should enable 5G network operators to realize intelligent services and provisions of ubiquitous local applications for mobile users \cite{346}.
			\item 	6G networks are envisioned to provide new human-centric values with transformative services, such as quantum communications, autonomous driving, virtuality, space-air-ground networks \cite{348}. 
			
		\end{itemize}& 
		\begin{itemize}
			\item 	AI can act as AI-as-a-service that can realize the potential of massive MIMO and transform the way edge devices perform 5G functionalities \cite{347}. 
			\item 	AI technologies can play an indispensable role in facilitating end-to-end design and optimization of the data processing in 6G wireless networks.	
		\end{itemize}
		\\ \cline{2-3}
		\hline

	\end{tabular}}
\end{table*}
\section{Conclusions}
\label{Sec:Conclusions}
	In this paper, we have presented a state-of-the-art survey on the emerging Wireless AI in communication networks. We have first proposed a novel Wireless AI architecture from the data-driven perspective with a focus on five main themes in wireless networks, namely {Sensing AI, Network Device AI, Access AI, User Device AI and Data-provenance AI}. Then, we have focused on analyzing the use of AI approaches in each data-driven AI theme by surveying the recent research efforts in the literature. For each domain, we have performed a holistic discussion on the Wireless AI applications in different network scenarios, and the key lessons learned from the survey have been derived. Based on the holistic survey, we have pointed out the important research challenges to be considered carefully for further investigation on the AI applications in wireless networks. Finally, we have outlined potential future research directions towards AI 5G wireless networks and beyond. {We hope that this article stimulates more interest in this promising area, and encourages more research efforts toward the full realization of Wireless AI. }

\balance
\bibliography{Ref1}
\bibliographystyle{IEEEtran}

\end{document}